%% file: main.tex
\documentclass[a4paper,11pt]{article}
\pdfoutput=1
\usepackage{jheppub}
\usepackage[T1]{fontenc}
\usepackage[utf8]{inputenc}

\usepackage[separate-uncertainty=true]{siunitx}
\usepackage{booktabs}
\usepackage{subcaption}
\usepackage{multirow}
\usepackage{slashed}
\usepackage{bbm}
\usepackage{cancel,array}
\usepackage{csquotes}
\usepackage[textsize=scriptsize]{todonotes}

\usepackage{tikz}
\usepackage{import}
\usepackage{bm}
\usepackage{arydshln}
\usetikzlibrary{calc}
\usetikzlibrary{decorations.pathreplacing}
\usetikzlibrary{decorations.markings}
\usepackage{xcolor}

\newcommand{\dd}{\mathrm{d}}
\newcommand{\ii}{\mathrm{i}}
\newcommand{\e}{\mathrm{e}}
\newcommand{\tr}{\mathrm{tr}}

\newcommand{\pf}{\mathrm{Pf}}
\newcommand{\T}{\mathrm{T}}

\newcommand{\g}{\gamma}
\newcommand{\id}{\mathbbm{1}}
\newcommand{\A}{\mathcal{A}}
\newcommand{\C}{\mathcal{C}}
\newcommand{\D}{\mathcal{D}}

\newcommand{\U}{\mathcal{U}}
\newcommand{\V}{\mathcal{V}}
\newcommand{\ie}{i.e.\ }
\newcommand{\eg}{e.g.\ }

\newcommand{\aEta}{{\text{a-}\eta^\prime}}
\newcommand{\aF}{{\text{a-}f_0}}
\newcommand{\aPi}{{\text{a-}\pi}}
\newcommand{\aA}{{\text{a-}a}}

\newcommand{\gluinoglue}{{\tilde{\mathrm{g}}\mathrm{g}}}
\newcommand{\SYM}{\mbox{$\mathcal{N}=1$}~SYM }
\newcommand{\slat}{\mbox{$8^3\times16$}}
\newcommand{\mlat}{\mbox{$16^3\times32$}}
\newcommand{\llat}{\mbox{$24^3\times48$}}
\newcommand{\pdiff}[2]{\frac{\partial #1}{\partial #2}}
\def\ring{\mathaccent"7017}
\definecolor{pi1}{HTML}{A6CEE3}
\definecolor{pi0}{HTML}{1F78B4}
\definecolor{gg1}{HTML}{B2DF8A}
\definecolor{gg0}{HTML}{33A02C}
\definecolor{a1}{HTML}{FB9A99}
\definecolor{a0}{HTML}{E31A1C}
\definecolor{0pp1}{HTML}{FDBF6F}
\definecolor{0pp0}{HTML}{FF7F00}
\definecolor{0mp1}{HTML}{cab2d6}
\definecolor{0mp0}{HTML}{6a3d9a}
\definecolor{0deg}{HTML}{808080}
\definecolor{45deg}{HTML}{FF00FF}
\definecolor{90deg}{HTML}{ffd92f}

\title{\boldmath $\mathcal{N}=1$ Super-Yang-Mills theory on the lattice with twisted mass fermions}

\author{Marc Steinhauser,}
\author{Andr\'e Sternbeck,}
\author{Bj\"orn Wellegehausen,}
\author{Andreas Wipf}
\affiliation{Friedrich Schiller University Jena, Max-Wien-Platz 1, 07743 Jena, Germany}

\emailAdd{marc.steinhauser@uni-jena.de}
\emailAdd{andre.sternbeck@uni-jena.de}
\emailAdd{bjoern.wellegehausen@uni-jena.de}
\emailAdd{wipf@tpi.uni-jena.de}

\abstract{
	Super-Yang-Mills theory (SYM)
	is a central building block for supersymmetric extensions 
	of the Standard Model of particle physics. Whereas the weakly coupled 
	subsector of the latter can be treated within a perturbative setting,
	the strongly coupled subsector must be dealt with a non-perturbative
	approach. Such an approach is provided by the lattice formulation. 
	Unfortunately a lattice regularization breaks supersymmetry and consequently 
	the mass degeneracy within a supermultiplet. In this article
	we investigate the properties of $\mathcal{N}=1$ supersymmetric SU(3) Yang-Mills theory with a lattice Wilson Dirac operator 
	with an additional parity mass, similar as in twisted mass lattice QCD.
	We show that a special $45^\circ$ twist effectively removes the mass splitting of the chiral partners. Thus, at finite lattice spacing both chiral and supersymmetry 
	are enhanced resulting in an improved continuum extrapolation.
    Furthermore, we show that for the non-interacting theory at $45^\circ$ 
    twist discretization errors of order $\mathcal{O}(a)$ are suppressed, 
    suggesting that the same happens for the interacting theory as well.
    As an aside, we demonstrate that the DD$\alpha$AMG multigrid 
    algorithm accelerates the 
    inversion of the Wilson Dirac operator considerably. On a $16^3\times 32$ lattice, speed-up factors 
    of up to 20 are reached if commonly used algorithms are replaced by the DD$\alpha$AMG.
}

\arxivnumber{2010.00946}

\begin{document} 
\maketitle

\input{introduction}

\input{basics}
\input{analytical_investigations}
\input{numerical_investigations}
\input{summary}

\newpage
\input{appendix}

\bibliographystyle{JHEP}
\bibliography{references.bib}

\end{document}

%% file: introduction.tex
\section{Introduction}\label{sec:introduction}
The standard model~(SM) of particle physics very successfully
describes all processes mediated by the electromagnetic, weak and strong forces
-- but several open questions remain unanswered. 
For example, the Higgs boson with mass
$m_\text{H}=(125.18\pm0.16)\text{GeV}$~\cite{PDG18} is
unreasonably light since the mass is quadratically divergent 
and a mass of the order of the Planck mass is expected.
The situation improves considerably in a supersymmetric
theory, where every bosonic particle has a fermionic superpartner 
with the same quantum numbers (besides the spin) and vice versa.
In a supersymmetric standard model a small Higgs mass is 
easier to accommodate since in leading order 
bosonic and fermionic divergences cancel 
and there is no quadratic divergence~\cite{witten_dynamical_1981,dimopoulos_softly_1981}.
Another urgent problem of modern physics is the
large amount of dark matter seen in our universe.
It outweighs the visible matter by
a factor of six, making up about 27 percent of the 
universe. Supersymmetric models naturally provide a dark-matter 
candidate, the so-called lightest supersymmetric particle~(LSP).
This particle is stable and can not decay if $R$-parity is conserved~\cite{dimopoulos_softly_1981,ellis_supersymmetric_1984}.

A straightforward extension of the SM is the minimal 
supersymmetric standard model (MSSM).
The present work deals with non-perturbative phenomena 
of the strongly coupled subsector of the MSSM which
is $\mathcal{N}=1$ Super-Yang-Mills~(SYM) theory with gauge group~\mbox{SU(3)}.
It is the supersymmetric extension of pure Yang-Mills~(YM) theory 
describing gluons in interaction with their superpartners, 
the so-called gluinos. As members of the same $\mathcal{N}=1$ 
vector super-multiplet the gluons and gluinos are
(in perturbation theory) massless.
Both are in the adjoint representation of the gauge
group SU(3) and on-shell the degrees of freedom match.
The latter statement holds true since the gluinos
are Majorana fermions. The theory is asymptotically
free and shows confinement, similar to QCD.

Our analytical and numerical investigations 
aim for a better understanding of the low-energy properties
of this confining theory. Unfortunately almost all
lattice regularizations break supersymmetry explicitly
and as a result of this breaking one observes a mass-splitting 
within a given supermultiplet.
In the present work we shall present a novel lattice formulation 
which considerably reduces the mass-splitting of the chiral 
partners in the Veneziano-Yankielowicz supermultiplet
of $\mathcal{N}=1$ SYM. As a result the difficult fine-tuning problem 
to the chiral and supersymmetric continuum limit is less severe.

Early analytic studies of supersymmetric lattice systems 
go back to Dondi and \mbox{Nicolai}~\cite{Dondi:1976tx}, who studied the
discretized Wess-Zumino model. Subsequently the restoration of supersymmetry in the continuum limit and the spectrum of
particles have been studied for these Yukawa-type lattice models~\cite{Catterall:2001fr,Bergner:2007pu,Kastner:2008zc,Kanamori:2007yx,Steinhauer:2014oda} or
in related supersymmetric non-linear sigma-models~
\cite{Flore:2012xj}.
Early simulations of four-dimensional \SYM theory with quenched 
fermions were performed in~\cite{Koutsoumbas:1996kz,donini_towards_1998}.
Clearly, dynamical fermions are an integral part
in any supersymmetric field theory and
the inclusion of light dynamical fermions in simulations
is essential. 

Extensive investigations and simulations of $\mathcal{N}=1$
SYM with gauge group SU(2) and with dynamical fermions 
were performed by the DESY-M\"unster collaboration during
the past $20$ years. In~\cite{Kirchner:1998mp} the
chiral symmetry breaking was investigated and two ground states 
have been spotted.
A comprehensive lattice study including 
the mass spectrum was first presented in~\cite{Campos:1999du} 
and concluded with~\cite{Bergner:2015adz}.
Later, those results were refined with the help of a 
variational analysis~\cite{Ali:2019gzj}.
Ward Identities were exploited in~\cite{Farchioni:2001wx} to 
determine the gluino mass as well as the mixing coefficient 
of the supercurrent.
An investigation of the theory at finite temperature revealed that deconfinement and chiral symmetry restoration occur at the same temperature~\cite{Bergner:2014saa}.
This insight was confirmed recently using the gradient flow~\cite{Bergner:2019dim}.
The lattice studies are supplemented with an one-loop 
calculation of the supersymmetric Ward identities~
\cite{Farchioni:2001yr}, the analysis of the adjoint pion within
partially quenched chiral perturbation theory~\cite{Munster:2014cja} 
and the perturbative calculation of the clover coefficient~
\cite{Musberg:2013foa}.
More recently the spectrum of the low lying 
bound states~\cite{Ali:2019agk} and 
supersymmetric Ward identities~\cite{Ali:2018fbq}
have been calculated for $\mathcal{N}=1$ SYM with gauge 
group SU(3). Besides these studies with Wilson fermions, 
first investigations and simulations with domain wall fermions
and overlap have been presented in~
\cite{Neuberger:1997bg,Kaplan:1999jn,Giedt:2008xm} 
and~\cite{Kim:2011fw,Ali:2020sbi}. With
Ginsparg-Wilson fermions no fine-tuning should be
necessary to end up with a supersymmetric continuum theory~\cite{Fleming:2000fa}.

A dimensional reduction of \SYM theory from $d=4$ to $d=2$ 
spacetime dimensions leads to the $\mathcal{N}=(2,2)$ SYM theory
and the two theories have supermultiplets of identical 
length. The mass spectrum of the reduced theory~\cite{August:2018esp}, the Ward identities~\cite{Kadoh:2009rw}, 
the dynamical breaking of supersymmetry~
\cite{Catterall:2017xox} and the large $N$ behavior~\cite{Hanada:2009hq}
were investigated in detail.
Certain field theories with extended supersymmetry can be formulated such 
that some (nilpotent) supersymmetry transformations are 
preserved exactly on the lattice~\cite{Catterall:2009it}.
In this context the four-dimensional $\mathcal{N}=4$ SYM theory was studied e.g.\ 
in~\cite{Schaich:2016jus} and its cousin, the 
two-dimensional $\mathcal{N}=(8,8)$ SYM in~\cite{Giguere:2015cga}.

In the present work we propose and carefully study a deformation of the \SYM lattice 
action by twisting the mass term. We will argue by analytic
and numeric means that this twisting leads to a sizable reduction of the
mass splitting (caused by a breaking of supersymmetry by
lattice artifacts) within the Veneziano-Yankielowicz supermultiplet.
Actually, the concept of a twisted mass was first introduced to 
lattice QCD in~\cite{Frezzotti:2000nk} with the aim to remove 
exceptional configurations. Later, \mbox{$\mathcal{O}(a)$} improvement at
(maximal) twisting angle~$\pm\pi/2$ was recognized as particularly
interesting for measuring physical quantities~\cite{Frezzotti:2003ni}. 
Also a study of the two-dimensional 
Wess-Zumino model with a twisted lattice action revealed a
dramatic suppression of the discretization errors 
for an optimal twist angle~\cite{Bergner:2007pu}.
In the  present work the IR improvement of the mass degeneracy in the Veneziano-Yankielowicz supermultiplet at optimal twist angle~$\pi/4$ is crucial. At the same time the  \mbox{$\mathcal{O}(a)$} discretization errors are reduced at this twist angle.

This paper is structured as follows:
In the following section we summarize basic facts about
\SYM theory in the continuum and on 
the lattice, which are relevant for our work.
Section~\ref{sec:AnaInvest} further elaborates on some aspects in more detail analytically. 
The results of our numerical calculations are
presented in section~\ref{sec:NumInvest}. Our conclusion and a summary is given in section~\ref{sec:Summary}.

%% file: basics.tex
\section{Basics}\label{sec:basics}
In this section we recall relevant facts 
about \SYM theory and thereby fix our notation.
In section~\ref{ch:SYM_cont} the continuum formulation, 
symmetries and effective field theory predictions are addressed.
Afterwards, the Wilson Dirac operator with twisted mass term is introduced (section~\ref{ch:LatticeFormulation}) and the 
main differences to the standard formulation are discussed (section~\ref{ch:DiracOpProp}).
In section~\ref{ch:latt_observables} we finally introduce all lattice observables whose numerical results are discussed 
then in section~\ref{sec:NumInvest}.

\subsection{$\mathcal{N}=1$ Super-Yang-Mills theory in the continuum}\label{ch:SYM_cont}
In Minkowski spacetime the on-shell action of \SYM theory reads
\begin{equation}
	S^\textrm{M}_\textrm{SYM} = \int \dd^4 x ~ \tr\left( -\frac{1}{4} F_{\mu\nu} F^{\mu\nu} + \frac{\ii}{2} \bar{\lambda} \slashed{D} \lambda - \frac{m}{2} \bar{\lambda}\lambda \right)\,,
	\label{eq:ContAction}
\end{equation}
and looks similar to the action of Quantum Chromodynamics (QCD) with 
a single flavor. In the supersymmetric theory the fermion 
and gauge boson are members of the same vector supermultiplet such 
that the former (called gluino) is 
described by a Majorana field $\lambda(x)$ and transforms in the same
adjoint representation as the gauge potential $A_\mu(x)$.
This way, fermionic and bosonic degrees of freedom match as dictated by
supersymmetry. The supersymmetry transformations are further discussed 
in section~\ref{ch:SusyTrafos}. 

The action in eq.~\eqref{eq:ContAction} contains a finite gluino mass $m$ which breaks supersymmetry softly.
On the lattice this mass is fine-tuned such that after continuum extrapolation
a supersymmetric limit is reached which 
at the same time is chirally symmetric.

At high energies or high temperatures, \SYM can 
be considered as a gas of free gluons and gluinos.
More interestingly, at low energies it is a confining theory similar 
to non-supersymmetric gauge theories and has a rich spectrum 
of low lying color-singlet bound states.
This spectrum has been investigated with
the method of effective field theory based on the 
theory's symmetries and applying anomaly  matching. 
Three different types of bound states are expected to arise: 
pure glueballs, pure meson-like gluinoballs and gluino-glueballs.

Supersymmetry arranges these bound states in 
supermultiplets of $\mathcal{N}=1$ supersymmetry.
As long as supersymmetry is unbroken, the states within a supermultiplet have equal mass. Veneziano and 
Yankielowicz predicted a chiral supermultiplet~\cite{Veneziano8206}
of bound states listed in table~\ref{tab:VYmultiplet}.
The names of the particles are chosen in analogy to QCD, with the 
prefix \enquote{a-} indicating the adjoint representation. As usual,
the quantum numbers $J^{PC}$ specify the total angular momentum~$J$, the parity~$P$ and the charge conjugation~$C$.

Subsequently  Farrar, Gabadadze and Schwetz suggested the existence of a second supermultiplet~\cite{Farrar9711} consisting of
the particles listed in table~\ref{tab:FGSmultiplet}.
Based on symmetry arguments they
suggested the more general effective Lagrangian
\begin{equation}\label{FGS1}
	\mathcal{L}^\textrm{eff} = \frac{1}{\alpha} (S^\dagger S)^{1/3}\Big|_D + \gamma \left[ \left\{S\,\log\left(\frac{S}{\mu^3}\right)-S\right\}\Big|_F + \text{h.c.} \right] + \frac{1}{\delta} \left( - \frac{U^2}{(S^\dagger S)^{1/3}} \right)\Big|_D
\end{equation}
with chiral superfield $S$, real tensor superfield $U$, dynamically generated scale $\mu$ and 
further low-energy constants $\alpha$, $\gamma$ and $\delta$.\footnote{In~\cite{Farrar:1998rm}, the same authors suggest an alternative formulation with two chiral superfields.}
In the limit $\delta\to\infty$ the effective action of Veneziano 
and Yankielowicz is recovered.
The effective Lagrangian~(\ref{FGS1}) describes
propagating massive fields,
for example the scalar and pseudoscalar glueball.
The physical states will be mixtures of states from these 
two multiplets with equal quantum numbers~\cite{Farrar9711}.

\begin{table}[tbp]
	\centering
	\renewcommand{\arraystretch}{1.2}
	\begin{tabular}{!{\vrule width 1pt}llll!{\vrule width 1pt}} \noalign{\hrule height 1pt}
		1 bosonic scalar & $s=1,~l=1,~0^{++}$ & gluinoball & $\text{a-}f_0\sim\bar{\lambda}\lambda$ \\
		1 bosonic pseudoscalar & $s=0,~l=0,~0^{-+}$ & gluinoball & $\text{a-}\eta^\prime\sim\bar{\lambda}\gamma_5\lambda$\hspace*{14mm} \\
		1 majorana-type & $s=\frac{1}{2},~l=1,~\frac{1}{2}^{\ii +\phantom{-()}}$ & gluino-glueball & $~\,\gluinoglue\,\sim F_{\mu\nu}\Sigma^{\mu\nu}\lambda$\\\noalign{\hrule height 1pt}
	\end{tabular}
	\caption{Veneziano-Yankielowicz supermultiplet.}
	\label{tab:VYmultiplet}
\end{table}
\begin{table}[tbp]
	\centering
	\renewcommand{\arraystretch}{1.2}
	\begin{tabular}{!{\vrule width 1pt}llll!{\vrule width 1pt}} \noalign{\hrule height 1pt}
		1 bosonic scalar & $s=0,~l=0,~0^{++}$ & glueball & $0^{++}\sim F_{\mu\nu}F^{\mu\nu}$ \\
		1 bosonic pseudoscalar & $s=1,~l=1,~0^{-+}$ & glueball & $0^{-+}\sim \epsilon_{\mu\nu\rho\sigma}F^{\mu\nu}F^{\rho\sigma}$ \\
		1 majorana-type & $s=\frac{1}{2},~l=0,~\frac{1}{2}^{(-\ii) +}$ & gluino-glueball & $\,~\gluinoglue~\sim F_{\mu\nu}\Sigma^{\mu\nu}\lambda$\\\noalign{\hrule height 1pt}
	\end{tabular}
	\caption{Farrar-Gabadadze-Schwetz supermultiplet.}
	\label{tab:FGSmultiplet}
\end{table}

The chiral symmetry of \SYM theory has a different breaking pattern compared to QCD. For vanishing gluino mass and gauge group SU($N_\text{c}$) the classical theory
has a global U$(1)_\text{A}$ symmetry\footnote{Usually
the angle of the chiral rotation is $\alpha$. We chose
$\alpha/2$ since in section~\ref{ch:ChiralTrafosLatticeExpValues}
the bilinear condensates are investigated and with our
choice they transform with the angle $\alpha$.}
\mbox{$\lambda \mapsto \e^{\ii\alpha\gamma_5/2}\lambda$}.
The axial anomaly reduces this U$(1)_\text{A}$ to the
discrete subgroup~$\mathbb{Z}_{2 N_\text{c}}$,
\begin{equation}
	\lambda \mapsto \e^{ 2\pi\ii n\gamma_5/2 N_\text{c}}\lambda ~~~\text{with}~~~ n\in\{1,\ldots,2 N_\text{c} \}\,.
	\label{eq:chiralSymmetry}
\end{equation}
A gluino condensate \mbox{$\langle \bar{\lambda}\lambda \rangle\neq 0$} spontaneously 
breaks this remnant symmetry further to a $\mathbb{Z}_2$~symmetry. 
Therefore $N_\text{c}$ physically equivalent vacua are expected.

To construct the lattice formulation one first switches from Minkowski 
to Euclidean theory~\cite{jaffe_euclidean_1985}.
In Euclidean spacetime 
the continuum on-shell action has the form
\begin{equation}
	S^\textrm{E}_\textrm{SYM} = \int \dd^4 x ~ \tr\left( \frac{1}{4} F_{\mu\nu} F^{\mu\nu} + \frac{1}{2} \bar{\lambda} \slashed{D} \lambda + \frac{m}{2} \bar{\lambda}\lambda \right)\,.
	\label{eq:EuclideanAction}
\end{equation}
This continuum action is the point of departure for the lattices studies
presented below.

\subsection{$\mathcal{N}=1$ Super-Yang-Mills theory on the lattice}\label{ch:SYM_latt}
To study the mass spectrum and in particular the
confinement of color charges, a non-perturbative 
method is required. We choose the ab-initio lattice method
although it breaks supersymmetry 
explicitly.\footnote{For \SYM there is no partially supersymmetric
formulation as for the theory with $32$ supercharges.}
Different lattice formulations are feasible, depending on the 
discretization of the continuum action and in particular on the 
choice of lattice fermions.

In the present work we shall use the lattice formulation
with Wilson fermions introduced by Curci and Veneziano~\cite{Curci8612}.
At finite lattice spacing, supersymmetry and chiral symmetry are 
broken simultaneously by the discretization and Wilson term.
This breaking leads to a relevant counter-term, which is proportional to the 
gluino mass term. To compensate this, an explicit gluino mass term is 
added and fine-tuned such that the (renormalized) gluino becomes massless 
in the continuum limit.
Since the gluino mass term is the only relevant operator,
supersymmetry and chiral symmetry will be restored in the continuum limit.

Unfortunately, confinement prevents the direct 
measurement of the gluino. Here we follow Veneziano and Yankielowicz 
who proposed to monitor instead the (unphysical) adjoint pion mass, defined in a partially quenched approximation, similarly as in 
1-flavor QCD~\cite{Veneziano8206}.
Its mass squared
\begin{equation}
	m_{\text{a-}\pi}^2\propto m^\text{R}
	\label{eq:mPi2}
\end{equation}
is proportional to the physical gluino mass, which can be calculated in partially quenched chiral perturbation theory~\cite{Munster:2014cja}.
This quantity requires only low statistics and is easy to compute.
By fine-tuning to the critical gluino mass $m_\text{crit}$ we are able to recover in the continuum limit simultaneously supersymmetry as well as chiral symmetry.

In contrast to QCD, where the Dirac fermions give rise to a
fermion determinant, in \SYM theory the Pfaffian of the Dirac operator enters the path integral after integrating out the Majorana fermions.
Since the Pfaffian is proportional to the square root of the determinant, the \emph{rational} hybrid Monte Carlo algorithm (RHMC)~\cite{Kennedy9809} is used in our simulations. 

\subsubsection{Lattice formulation}\label{ch:LatticeFormulation}

Different lattice formulations of a continuum theory vary in their discretization errors and how fast the correct continuum limit is reached.
In our simulations the gauge part of the lattice action \mbox{$S_\text{lat}=S_\text{g}+S_\text{f}$} is given by the Symanzik-improved L\"uscher-Weisz action
\begin{equation}
	S_\text{g}[\mathcal{U}] = \frac{\beta}{3}\left( \frac{5}{3} \sum_\square \tr(\mathbbm{1}-\text{Re}\, \mathcal{U}_\square) - \frac{1}{12} \sum_{\square\square} \tr( \mathbbm{1} - \text{Re}\, \mathcal{U}_{\square\square} ) \right)\,,
	\label{eq:BosonicLatticeAction}
\end{equation}
and the action for the Majorana field (the gluino part)
\begin{equation}
	S_\text{f}[\lambda,\bar{\lambda},\mathcal{U}] = a^4 \sum_{x,y\in\Lambda} \bar{\lambda}(x)D_\text{W}(x,y) \lambda(y)
	\label{eq:FermionicLatticeAction}
\end{equation}
contains the Wilson Dirac operator with an additional twisted mass term,
\begin{align}
	D_\text{W}^{\text{mtw}}(x,y)\!&=\!(4+m+\ii m_5\gamma_5)\delta_{x,y} - \frac{1}{2}\!\sum_{\mu=\pm1}^{\pm4}\left( \mathbbm{1}-\gamma_\mu \right)\!\mathcal{V}_\mu(x)\,\delta_{x+\hat{\mu},y}\,.
	\label{eq:DiracOperatorWithTwist}
\end{align}
Here the gauge links $\mathcal{V}_\mu(x)$ are in the adjoint representation. They are constructed from the gauge link $\mathcal{U}_\mu(x)$ in the fundamental representation and the generators, $T^a$, of the Lie algebra using
the relation 
\begin{equation}
	\left[ \mathcal{V}_\mu(x)\right] ^{ab}\equiv 2\,\tr\left[\mathcal{U}_\mu^\dagger(x)T^a\,\mathcal{U}_\mu(x) T^b \right]\,.
\end{equation}
Furthermore, we define \mbox{$\gamma_{-\mu}\equiv-\gamma_\mu$} and \mbox{$\mathcal{V}_{-\mu}(x)\equiv\mathcal{V}_\mu^\dagger (x-\hat{\mu})$} for simplicity.

At finite lattice spacing (with or without twisted mass term) supersymmetry and chiral symmetry are explicitly broken and only a fine-tuning of the gluino mass, \mbox{$m\to m_\text{crit}(\beta)$}
while taking the limit $\beta\to\infty$, assures a simultaneous restoration of both symmetries in the continuum limit.
After the critical point at \mbox{$(m_\text{crit},m_{5}=0)$} is determined via a parameter scan, the mass parameters can also be specified by their distance \mbox{$\delta m = m-m_\text{crit}$}, \mbox{$\delta m_5 = m_5$} with respect to the critical point. A useful variant are polar coordinates
centered at the critical point with distance $M$ from this point and twist angle $\alpha$,
\begin{equation}
	m-m_\text{crit}=M\cos\alpha\quad\text{and}\quad m_5=M\sin\alpha\,.
	\label{eq:PolarMass}
\end{equation}
We added a parity-breaking mass term \mbox{$\ii m_5\gamma_5\delta_{x,y}$} 
to the Wilson Dirac operator  
\begin{equation}
	D_\text{W}(x,y)\!=\!(4+m)\delta_{x,y} - \frac{1}{2}\!\sum_{\mu=\pm1}^{\pm4}\left( \mathbbm{1}-\gamma_\mu \right)\!\mathcal{V}_\mu(x)\,\delta_{x+\hat{\mu},y}\,,
	\label{eq:DiracOperator}
\end{equation} 
to reduce the explicit susy-breaking by lattice artifacts in the two-point functions
of the supermultiplet partners. This term is similar as for twisted-mass lattice QCD but for 
one Majorana fermion flavor\footnote{In contrast to 2-flavor twisted-mass QCD, where the 
twist term contains the Pauli matrix $\tau_3$, \SYM theory contains only one flavor and 
thus $\tau_3$ is absent.}. A special feature of \SYM motivates it: If we had twisted not only 
the mass but also the Wilson term (which becomes an irrelevant term in the continuum) we 
would have a double-twisted Wilson Dirac operator
\begin{align}
	D_\text{W}^{\text{dtw}}(x,y) 
	&\equiv \left(4\,\e^{\ii\varphi\g_5}+M\e^{\ii\alpha\g_5}\right)\delta_{x,y} - \frac{1}{2} \sum_{\mu=\pm 1}^{\pm 4} \left( \id \,\e^{\ii\varphi\g_5} - \g_\mu \right) \mathcal{V}_\mu(x) \delta_{x+\hat{\mu},y}.
	\label{eq:DiracOperatorWithDoubleTwist}
\end{align}
For identical twist angles $\varphi=\alpha$, the standard and double-twisted Wilson Dirac operators 
are related by a chiral rotation,
\begin{equation}
  \e^{\ii\alpha\gamma_5/2} D_\text{W} \e^{\ii\alpha\gamma_5/2} = D_\text{W}^\text{dtw}\,.
\end{equation}
The chiral rotation can be undone by a variable transformation of the Majorana fields
\begin{equation}
	\lambda\mapsto\e^{\ii \alpha\gamma_5/2}\lambda,\qquad \bar{\lambda}\mapsto\bar{\lambda}\,\e^{\ii \alpha\gamma_5/2}\,,
	\label{eq:ChiralRotation}
\end{equation}
and, if no anomaly enters through the measure, we obtain for Grassmann integrals of Majorana bilinears (i.e., the scalar and pseudo-scalar bilinears)
\begin{align}
	\int \D\lambda \,\e^{-\lambda^\T \C D^{\text{dtw}}_\text{W} \lambda} 
	\begin{pmatrix}
		\bar{\lambda}_x\lambda_x\\ 
		\ii\bar{\lambda}_x\gamma_5\lambda_x
	\end{pmatrix}
	=
    \int \D\lambda \,\e^{-\lambda^\T \C D_\text{W} \lambda} 
	\begin{pmatrix}
	\cos \alpha & ~-\sin \alpha \\ 
	\sin \alpha & ~~~\cos \alpha
	\end{pmatrix} 
	\begin{pmatrix}
		\bar{\lambda}_x\lambda_x\\ 
		\ii\bar{\lambda}_x\gamma_5\lambda_x
	\end{pmatrix}\,.
	\label{eq:BilinearDoublet}
\end{align}
At twist angles \mbox{$\alpha=\varphi=\pi/4$}, the chiral and parity condensate \mbox{$\langle\bar{\lambda}\lambda\rangle$} and \mbox{$\langle\bar{\lambda}\gamma_5\lambda\rangle$} thus have equivalent magnitudes. In addition, two-point correlators of 
adjoint mesonic states are mass degenerated by construction and
their operator basis can be combined by an
arbitrary rotation\footnote{See for example~\cite{Farrar9711}, where the $\aF$ and $\aEta$ are described by one common complex field~$A$ of the chiral multiplet~$S$.}.
This means that the double-twisted formulation with \mbox{$\alpha=\varphi=\pi/4$} 
has a continuum limit with mass-degenerated scalar and pseudoscalar mesonic states.
Actually we shall see below that the mass degeneracy is seen at finite lattice spacing 
even for the twisted-mass Wilson Dirac operator, $D^{\text{mtw}}_\text{W}$, that is the 
operator $D^\text{dtw}_\text{W}$ with \mbox{$\varphi=0$} and \mbox{$\alpha=\pi/4$}.
The chiral and the parity condensates differ, though.

Before proceeding with the properties of the twisted Dirac operator, a few notes are in order: 
Whereas in twisted-mass lattice QCD simulations the twisted basis is rotated back 
to the physical basis for the calculation of observables, we
interpret the $m_5$-mass term as a deformation which vanishes in the 
chiral limit \mbox{$m\rightarrow m_\text{crit},~m_5\rightarrow0$}.\footnote{In contrast to QCD, where the quark masses are tuned to obtain the physical
meson masses, the bare mass of the fermionic gluino $m$ is tuned via eq.~\eqref{eq:mPi2} to the point, where the renormalized gluino becomes massless, \mbox{$m^\text{R}=0$}, in the continuum limit. At finite lattice spacing, this point is characterized by a minimal adjoint pion mass, which can not fall below the lattice cut-off. Hence all numerical values of $m_\text{crit}$ mentioned in this paper are determined numerically this way.}
In section~\ref{ch:ChiralTrafosLatticeExpValues} the correlators of the chiral partners $\aEta$ and $\aF$ are studied analytically and in section~\ref{ch:ParameterScan} different \enquote{directions} in the \mbox{$(m,m_5)$}-plane for the extrapolation to the critical point are analyzed numerically. Both investigations reveal an optimal twist angle, for which the chiral partners have equal masses.
This reduces the breaking of chirality and supersymmetry at finite lattice spacing
considerably.

In contrast to Ginsparg-Wilson fermions, 
which preserve a variant of chiral symmetry even at finite lattice spacing~\cite{Fleming:2000fa}, twisted fermions still break 
chiral symmetry. But since our main focus is on
spectroscopy, the mass-degeneracy of chiral partners provides 
an improvement for the extrapolation to the critical point. 
In addition, the twisted formulation has the same computational costs as
Wilson fermions which are much smaller than for Ginsparg-Wilson fermions with
the overlap Dirac operator~\cite{Neuberger:1997bg,Ali:2020sbi} or the
domain-wall formalism~\cite{Nishimura:1997vg,Kaplan:1999jn}.

A similar twist was used in~\cite{Bergner:2007pu} for the supersymmetric 
Wess-Zumino model in two dimensions. There, a modified Wilson term 
was tuned such that the discretization errors in the eigenvalues of 
the free lattice Dirac operator are reduced to $\mathcal{O}(a^4)$. For the \SYM theory, we perform an analogous calculation for the twisted Wilson Dirac operator in section~\ref{ch:EigValues}.
As one option, we will also increase the freedom of finding a suitable action
further by choosing the twist angles $\alpha$, $\varphi$ entering
$D_\text{W}^\text{dtw}$ independently.
Then no direct connection between the 
action and the observables exist anymore, 
but $\mathcal{O}(a)$ improvement may be possible.

\subsubsection{Properties of the Wilson Dirac operator}\label{ch:DiracOpProp}
In table~\ref{tab:PropertiesDiracOperator} we compare the relevant properties of the Wilson Dirac operator with and without mass twist.
Most differences result from the loss of $\gamma_5$-hermiticity 
when a mass twist is added and only a modified $\gamma_5$-hermiticity 
involving $\pm m_5$ holds.
As a consequence, the complex eigenvalues do not come 
in complex-conjugated pairs and the determinant as well as the Pfaffian may have non-zero imaginary parts. Nevertheless, we shall demonstrate
in section~\ref{ch:Pfaffian} that only a very mild sign problem emerges.
As we have seen for 
the particular choice \mbox{$\alpha=\varphi$} in the double-twisted Wilson Dirac 
operator \eqref{eq:DiracOperatorWithDoubleTwist}, the chiral phase can be removed by 
a change of variables and  therefore the Pfaffian becomes real again.

\begin{table}[t]
	\scriptsize
	\renewcommand{\arraystretch}{1.3}
	\begin{tabular}{!{\vrule width 1pt}l l l!{\vrule width 1pt}}
	\noalign{\hrule height 1pt} 
	 & $D_\mathrm{W}$ of eq.~\eqref{eq:DiracOperator} & $D_\mathrm{W}^\mathrm{mtw}$ of eq.~\eqref{eq:DiracOperatorWithTwist} \\ 
	\noalign{\hrule height 1pt}
	\multirow{2}{*}{$\gamma_5$-hermiticity} & $\big(\gamma_5 D_\mathrm{W}\big)^\dagger = \gamma_5 D_\mathrm{W}$ & $\big(\gamma_5 D_\mathrm{W}^\mathrm{mtw}(m_5)\big)^\dagger = \gamma_5 D_\mathrm{W}^\mathrm{mtw}(-m_5) $ \\ 
	 & $\big(D_\mathrm{W}^{-1}\big)^\dagger = \gamma_5 D_\mathrm{W}^{-1} \gamma_5$ & $\big((D_\mathrm{W}^\mathrm{mtw})^{-1}\big)^\dagger \!\!=\! \big(D_\mathrm{W}^\mathrm{mtw}\!+\!2\ii m_5\gamma_5\big)\!\cdot\!\big(\gamma_5 D_\mathrm{W}^\mathrm{mtw} \gamma_5 D_\mathrm{W}^\mathrm{mtw} \!+\! 4 m_5^2\big)^{-1}$ \\*[1ex]
	$\C$-antisymmetry & $\big(\C D_\mathrm{W}\big)^\T = -\C D_\mathrm{W}$ & $\big(\C D_\mathrm{W}^\mathrm{mtw}\big)^\T = -\C D_\mathrm{W}^\mathrm{mtw}$ \\*[1ex]
	\multirow{2}{*}{eigenvalues} & \parbox[t]{3.3cm}{double degenerated in \\ complex conjugated pairs} & \multirow{2}{*}{complex} \\*[3ex]
	det & $\mathbb{R}^+$ & $\mathbb{C}$ \\ 
	Pf & $\mathbb{R}~~$ & $\mathbb{C}$ \\*[1ex]
	\noalign{\hrule height 1pt}
	\end{tabular} 
	\caption{Properties of the untwisted and twisted Wilson Dirac operator.}
	\label{tab:PropertiesDiracOperator}
\end{table}

\subsubsection{Lattice observables}\label{ch:latt_observables}
The simulations are performed with the action
$S[\U,\lambda]=S_\text{B}[\U]+S_\text{F}[\U,\lambda]$,
where the L\"uscher-Weisz action $S_\text{B}[\U]$ was
defined in eq.~\eqref{eq:BosonicLatticeAction} and 
the fermionic action is given by
\begin{equation}
S_\text{F}[\U,\lambda]=\frac{1}{2} \tr(\lambda^\T \C D[\U]\lambda)=\frac{1}{2} \tr(\lambda^\T \tilde{D}[\U]\lambda)\,,
\end{equation}
with Wilson Dirac 
operator $D[\U]$ without twist~\eqref{eq:DiracOperator} 
or with twisted mass~\eqref{eq:DiracOperatorWithTwist}.
The effective action after integrating out the 
Majorana fermions is \mbox{$S_\text{eff}[\U]=S_\text{B}[\U]-\log(\pf(\tilde{D}[\U]))$}.

For hadron spectroscopy, interpolating
lattice operators for the particles of interest
are required. The interpolating operators for mesons
are bilinears of the form
\begin{equation}
O(x)=\bar{\lambda}(x)\Gamma\lambda(x)\,.
\end{equation}
Specifically the interpolating operators for the adjoint 
mesonic states $\aEta$ and $\aF$ are~\cite{Curci8612}
\begin{equation}
	O_{\aEta}(x) = \bar{\lambda}(x) \ii\gamma_5 \lambda(x)\qquad\text{and}\qquad O_{\aF}(x) = \bar{\lambda}(x) \lambda(x)\,.
	\label{eq:MesonicOperators}
\end{equation}
After integrating over the fermion field the
correlators of these bilinears  are given by
gauge averages of products of the fermion lattice propagator 
\begin{equation}
	G_{xy}=\langle x\vert D^{-1}\vert y\rangle\,.
\end{equation}
In particular, the correlators between the source at
\mbox{$y=(0,\vec{y})$} and the sink at \mbox{$x=(t,\vec{x})$} contain 
connected and disconnected contributions
\begin{align}
	C(t) &= \big\langle O(t,\vec{p}=\vec{0})\, O^\dagger (0,\vec{p}=\vec{0}) \big\rangle= \frac{1}{|\Lambda_3|^2} \sum_{\vec{x},\vec{y}\in\Lambda_3} \big\langle O(t,\vec{x})\, O^\dagger(0,\vec{y}) \big\rangle\nonumber\\
	&= \frac{1}{|\Lambda_3|^2} \sum_{\vec{x},\vec{y}\in\Lambda_3} \big\langle \tr( \Gamma G_{xx} )\, \tr( \Gamma G_{yy} ) \big\rangle_\mathcal{U} - \frac{2}{|\Lambda_3|^2} \sum_{\vec{x},\vec{y}\in \Lambda_3} \big\langle \tr( \Gamma G_{xy} \Gamma G_{yx}) \big\rangle_\mathcal{U}\,,
	\label{eq:MesonicCorrelator}
\end{align}
where \mbox{$\Gamma\in\{\id_4,\gamma_5\}$}.
For the connected\footnote{We encounter a misuse of language. Here \enquote{connected} is understood in the sense of QFT calculations, where $W=\ln Z$ is used to compute connected Feynman diagrams approaching zero at large spatial separation. This must not be confused with the term
\enquote{connected} to distinguish between contributions like the last term of eq.~\eqref{eq:MesonicCorrelator} compared to the \enquote{disconnected} contributions of the first term.} two-point correlator, the contribution of the 
position-independent vacuum expectation value
\begin{equation}
	\frac{1}{|\Lambda_3|^2}\sum_{\vec{x},\vec{y}\in\Lambda_3} \big\langle \tr( \Gamma G_{xx} ) \big\rangle_\mathcal{U} \big\langle \tr( \Gamma G_{yy} ) \big\rangle_\mathcal{U}
	\label{eq:VacuumContribution}
\end{equation}
must be subtracted from the correlator in eq.~\eqref{eq:MesonicCorrelator}~\cite{Knechtli:2017sna}.
Instead of fitting the constant vacuum contribution~\eqref{eq:VacuumContribution}, it is beneficial to calculate the large cancellations between \mbox{$\langle \tr( \Gamma G_{xx} ) \tr( \Gamma G_{yy} ) \rangle$} and \mbox{$\langle \tr( \Gamma G_{xx} ) \rangle \langle \tr( \Gamma G_{yy} ) \rangle$} numerically.
This procedure is further stabilized when $y$ is consistently described by point sources and $x$ is averaged over the whole lattice with the stochastic estimator technique.
In parameter sets with small ensemble sizes these signals are too noisy and we use instead the (unphysical) correlators
\begin{align}
	C_\aPi(t)=\frac{2}{|\Lambda_3|^2} \sum_{\vec{x},\vec{y}\in \Lambda_3} \big\langle \tr( \gamma_5 G_{xy} \gamma_5 G_{yx}) \big\rangle_\mathcal{U}~~\text{and}~~ C_\aA(t)=\frac{2}{|\Lambda_3|^2} \sum_{\vec{x},\vec{y}\in \Lambda_3} \big\langle \tr(  G_{xy}  G_{yx}) \big\rangle_\mathcal{U}\,,
	\label{eq:ConnectedCorrelator}
\end{align}
which contain just the connected contributions\footnote{The connected 
correlators \eqref{eq:ConnectedCorrelator} as two-flavor states do not allow any vacuum contribution as in \eqref{eq:VacuumContribution}.}.

In \SYM theory there exist also mixed states containing bosonic and fermionic building blocks.
To measure the gluino-glueballs we define the 
interpolating operator\footnote{The trace runs only over the color degrees of freedom and the indices $i,j$ run only over the spatial directions to avoid any contributions of multiple time-slices~\cite{donini_towards_1998}.}
\begin{equation}
	[O_\gluinoglue(x)]_\alpha = [\Sigma_{ij}]_{\alpha\beta} \, \tr_\text{c}\left( F^{ij}(x)\lambda^\beta(x) \right)
	\label{eq:GluniGlueOperator}
\end{equation}
with \mbox{$\Sigma_{ij}\equiv[\gamma_i,\gamma_j]$} and
the spatial clover plaquette \mbox{$F_{ij}(x)$}.
Then, the corresponding correlator with source at $y$ and sink at $x$ including a matrix $\Gamma$ to contract the indices is
\begin{align}
C_\gluinoglue(x,y) &= \big\langle \Gamma^{\mu\delta} [O_\gluinoglue(x)]_\mu\, [\bar{O}_\gluinoglue(y)]_\delta\big\rangle \nonumber \\
& = -\left\langle [\Gamma^\T]^{\delta\mu} [\Sigma_{ij}]_{\mu\beta} \, \tr_\text{c}\big( F^{ij}(x) T^a \big) 
(G_{xy})^{\beta\rho}_{ab}\, \tr_\text{c}\big( F^{lm}(y) T^b \big) [\Sigma_{lm}]_{\rho\delta} \right\rangle_\mathcal{U}\,.
\end{align}
The gluino-glue correlator has a time-symmetric and a time-antisymmetric component.
By expanding the correlator in the spinor-space of complex \mbox{$4\times4$}~matrices, those are identified as the components of \mbox{$\Gamma=\gamma_4$} and \mbox{$\Gamma=\id_4$} respectively.
In our simulations those two variants as well as the combinations \mbox{$\Gamma=\frac{1}{2}(\id_4\pm\gamma_4)$} are measured.
It is reported that the antisymmetric component has a longer plateau in the effective mass and thus should be preferred for the determination of the ground state mass.
On the other hand, the symmetric component is expected to have a better signal for the excited states~\cite{Demmouche:2010sf,KuberskiMaster}.
Although this correlator has no disconnected contribution, it 
requires high statistics because of sizable gauge field fluctuations.

Besides those states with gluino content, there exist glueballs states in the Farrar-Gabadadze-Schwetz~(FGS) supermultiplet.
In the continuum, bosonic states transform under
tensor representations of the rotation group SO(3), but the lattice discretization breaks this symmetry to the 
finite cubic group.
With the help of the irreducible representations of the cubic
symmetry group the eigenstates can be classified and a 
restoration of the rotation group in the continuum limit 
can be achieved~\cite{HeitgerDiss}.
For the scalar glueball~\mbox{$F_{\mu\nu} F^{\mu\nu}$} with quantum numbers \mbox{$J^{PC}=0^{++}$} we use the interpolating operator~\cite{Berg:1982kp}
\begin{equation}
	O_{0^{++}}(x) = \text{Re} \Big( \tr \big( \U_{12}(x) + \U_{23}(x) + \U_{31}(x) \big)\Big)\,.
	\label{eq:ScalarGlueballOperator}
\end{equation}
The pseudoscalar glueball~\mbox{$\tilde{F}_{\mu\nu} F^{\mu\nu}$} with quantum numbers  \mbox{$J^{PC}=0^{-+}$} can be measured with the operator
\begin{equation}
	O_{0^{-+}}(x) = \text{Re} \sum_R \left( \tr\big(\mathcal{W}(\mathfrak{C}_R)\big) - \tr\big(\mathcal{W}(\mathrm{P}\mathfrak{C}_R)\big) \right)
	\label{eq:PseudoscalarGlueballOperator}
\end{equation}
using a standard loop along the curve~$\mathfrak{C}$ shown in
figure~\ref{fig:PseudoscalarGlueball}.
The sum extends over all rotations in the cubic group
and the path~$\mathfrak{C}_R$ is obtained by acting
with the rotation~$R$ on the standard loop.
The Wilson loops $\mathcal{W}$ are evaluated along
the path~$\mathfrak{C}_R$ and their reflections $\mathrm{P}\mathfrak{C}_R$. 
\begin{figure}[h]
	\centering
	\begin{tikzpicture}[cm={1,0,-1,-1,(0,0)},y=-4.5mm,x=1cm,z=-1cm,>=latex,very thick,scale=0.6]
	\newcommand\lengthA{3}
	\newcommand\lengthB{3}
	\newcommand\lengthC{3}
	\coordinate (c1) at (0,0,0);
	\coordinate (c2) at (\lengthA,0,0);
	\coordinate (c3) at (\lengthA,\lengthB,0);
	\coordinate (c4) at (2*\lengthA,\lengthB,0);
	\coordinate (c5) at (2*\lengthA,\lengthB,\lengthC);
	\coordinate (c6) at (\lengthA,\lengthB,\lengthC);
	\coordinate (c7) at (0,\lengthB,\lengthC);
	\coordinate (c8) at (0,\lengthB,0);
	\draw[->] (c1) -- (c2);
	\draw[->] (c2) -- (c3);
	\draw[->] (c3) -- (c4);
	\draw[->] (c4) -- (c5);
	\draw[->] (c5) -- (c6);
	\draw[->] (c6) -- (c7);
	\draw[->] (c7) -- (c8);
	\draw[->] (c8) -- (c1);
	\end{tikzpicture}
	\caption{Example for the three-dimensional path~$\mathfrak{C}$ used in the simulation for the pseudoscalar glueball~$0^{-+}$. This shape is rotated by the 24 elements of the cubic group.}
	\label{fig:PseudoscalarGlueball}
\end{figure}

Further fermionic observables of interest are the chiral condensate
\begin{equation}
	\Sigma = \frac{1}{V} \pdiff{\ln(Z)}{m} = -\frac{1}{2V} \sum_{x\in\Lambda} \langle \bar{\lambda}(x) \lambda(x) \rangle = \frac{1}{V} \sum_{x\in\Lambda} \langle\tr\, G_{xx}\rangle_\text{eff}\,,
	\label{eq:ChiralCondensate}
\end{equation}
which signals the spontaneous breaking of the remnant 
chiral symmetry (see section~\ref{ch:SYM_cont}) and the parity condensate
\begin{equation}
	\Sigma^\text{p} = -\frac{1}{2V} \sum_{x\in\Lambda} \langle \ii\bar{\lambda}(x) \gamma_5 \lambda(x) \rangle = \frac{\ii}{V} \sum_{x\in\Lambda} \langle\tr\, \gamma_5 G_{xx}\rangle_\text{eff}\,.
	\label{eq:ParityCondensate}
\end{equation}
Note that the chiral condensate \eqref{eq:ChiralCondensate} needs an additive renormalization and when parity is broken explicitly, the parity condensate needs it as well.

%% file: analytical_investigations.tex
\section{Analytical investigations}\label{sec:AnaInvest}
We begin our analytical investigations in subsection~\ref{ch:ChiralTrafosLatticeExpValues} with a discussion of expectation values 
of twisted lattice observables and will see that the twist angle \mbox{$\alpha=45^\circ$} is special.
Then we check in section~\ref{ch:SusyTrafosLatticeExpValues} that 
the chiral deformation has no influence on the supersymmetry
transformations and the supermultiplets.
Finally we study in section~\ref{ch:EigValues} the spectral properties of the free Wilson Dirac operator with a twist and find a reduction of $\mathcal{O}(a)$ discretization effects.

\subsection{Chiral transformations of fermionic observables}\label{ch:ChiralTrafosLatticeExpValues}
In section~\ref{ch:LatticeFormulation} we have argued that the twisted Wilson Dirac operator corresponds to a situation with rotated bilinears, see eq.~\eqref{eq:BilinearDoublet}.
Here we show this equivalence on the level of correlation functions for the mesonic states and the gluino-glue at $45^\circ$-twist.

To this end, we combine the Hermitean scalar and pseudoscalar bilinear of the doublet~\eqref{eq:BilinearDoublet} in a linear combination:
\begin{equation}
	\begin{aligned}
		O_{a,b}(x)=a\bar{\lambda}_x\lambda_x+b\ii\bar{\lambda}_x\gamma_5\lambda_x=
		O_{a,b}^\dagger(x)\,.
	\end{aligned}
	\label{eq:genericOp}
\end{equation}
Here we assumed that $a,b$ are real, which is the case for the mesonic states under investigation.
Without twist the operators for $\aF$ and $\aEta$ (compare to eq.~\eqref{eq:MesonicOperators}) have the form
\begin{align}
\text{a-}f_0:\quad\phantom{\ii\gamma_5\e^{\ii\alpha\gamma_5}} \bar{\lambda}_x\lambda_x &= O_{1,0}(x)\,, \\
\text{a-}\eta^\prime:\quad\phantom{\e^{\ii\alpha\gamma_5}} \ii\bar{\lambda}_x\gamma_5\lambda_x &= O_{0,1}(x)\,.
\intertext{Adding a chiral rotation as in eq.~\eqref{eq:ChiralRotation} to the spinors, those bilinears become}
\text{a-}f_0:\quad\phantom{\ii\gamma_5} \bar{\lambda}_x\e^{\ii\alpha\gamma_5}\lambda_x &= O_{\cos(\alpha),\sin(\alpha)}(x)\,,\label{eq:fRot}\\ 
\text{a-}\eta^\prime:\quad \ii\bar{\lambda}_x\gamma_5\e^{\ii\alpha\gamma_5}\lambda_x &=  O_{-\sin(\alpha),\cos(\alpha)}(x)\,.\label{eq:etaRot}
\end{align}
Then, we can calculate the (general) expectation values
\begin{align}
	M_{a,b}(x,x^\prime )\equiv
	\langle O_{a,b}(x)\,O^\dagger_{a,b}(x^\prime ) \rangle_\text{F} =& a^2\langle \bar{\lambda}_x\lambda_x \bar{\lambda}_{x^\prime}\lambda_{x^\prime}\rangle_\text{F} - b^2\langle \bar{\lambda}_x\gamma_5\lambda_x \bar{\lambda}_{x^\prime}\gamma_5\lambda_{x^\prime}\rangle_\text{F} \\ 
	&+  a b\ii\langle \bar{\lambda}_x\lambda_x \bar{\lambda}_{x^\prime}\gamma_5\lambda_{x^\prime}\rangle_\text{F} +
	a b\ii\langle \bar{\lambda}_x\gamma_5\lambda_x \bar{\lambda}_{x^\prime}\lambda_{x^\prime}\rangle_\text{F}\,.\nonumber
	\label{twist1d}
\end{align}
The two terms in the last row have negative parity and thus must vanish.
This can be seen explicitly, because the Green's function with parity transformed gauge field configuration $\mathcal{U}^P$ is related to the Green's function with the original configuration $\mathcal{U}$ as follows
\begin{equation}
G(\mathcal{U}^P;t,\vec{x};t',\vec{x}^\prime)=\gamma_0 G(\mathcal{U};t,-\vec{x};t',-\vec{x}^\prime)\gamma_0\,.
\label{eq:Gparity}
\end{equation}
For our parity-invariant theory\footnote{At the moment, the twist is only on the level of the observable and the action consists of the parity-invariant Wilson Dirac fermion action and for example the Wilson gauge action.}, $\mathcal{U}$ and $\mathcal{U}^P$ have equal weight, such that indeed
\begin{align}
C_{\aF,\aEta}(t) =& \frac{1}{|\Lambda_3|} \sum_{\vec{x}} \langle \bar{\lambda} (t,\vec{x}) \lambda(t,\vec{x})\, \bar{\lambda} (0,\vec{0}) \gamma_5 \lambda(0,\vec{0}) \rangle \nonumber \\
=& \frac{1}{|\Lambda_3|} \sum_{\vec{x}} \Big\langle \tr(G(t,\vec{x};t,\vec{x}))\,\tr(\gamma_5 G(0,\vec{0};0,\vec{0})) - 2\, \tr(G(t,\vec{x};0,\vec{0}) \gamma_5 G(0,\vec{0};t,\vec{x}))\Big\rangle_\mathcal{U} \nonumber \\
	=& - C_{\aF,\aEta}(t)\,,
	\end{align}
	\ie \mbox{$C_{\aF,\aEta}(t)$} vanishes.
Thus, we get the expectation values
\begin{align}
	\big\langle O_{\text{a-}f_0}(x) \,O^\dagger_{\text{a-}f_0}(x^\prime)\big\rangle(\alpha) 
	&=\big\langle M_{\cos(\alpha),\sin(\alpha)}(x,x^\prime)\big\rangle_\mathcal{U}\,,\nonumber\\
	\big\langle O_{\text{a-}\eta^\prime}(x) \,O^\dagger_{\text{a-}\eta^\prime}(x^\prime) \big\rangle(\alpha) &=
	\big\langle M_{-\sin(\alpha),\cos(\alpha)}(x,x^\prime)\big\rangle_\mathcal{U}
\end{align}
and we see immediately that for the angle $\alpha=45^\circ$,
\begin{equation}
	\big\langle O_{\text{a-}f_0}(x) \,O^\dagger_{\text{a-}f_0}(x^\prime) \big\rangle(45^\circ) 
	=\big\langle O_{\text{a-}\eta^\prime}(x) \,O^\dagger_{\text{a-}\eta^\prime}(x^\prime) \big\rangle(45^\circ)\,.
	\label{45grad}
\end{equation}
The two mesons in the supermultiplet have identical correlators and thus the same mass.

In section~\ref{ch:ParameterScan} this mass-degeneracy 
on the lattice is verified, although at finite 
lattice spacing supersymmetry and chiral symmetry are broken. 
Actually, in the simulations we did not chirally rotate 
the fermion field in the observables  (as we did in our 
analytic analysis) but instead used the Wilson Dirac operator with
twisted mass term \eqref{eq:DiracOperatorWithTwist}.
We have argued that (up to a twist of the irrelevant Wilson term) 
this is equivalent to twisting the field in the observables.

Finally, let us see how the third particle in 
the VY-supermultiplet is affected by a chiral rotation~\eqref{eq:ChiralRotation}.
The starting point is the interpolating
operator~\eqref{eq:GluniGlueOperator}
for the fermionic gluino-glue state with a twist,
\begin{equation*}
		[O_\gluinoglue(x)]_\mu = [\Sigma_{ij}]_{\mu\nu} \, \tr\left( F^{ij}(x)\,\big[\e^{\ii\alpha\gamma_5/2}\lambda(x)\big]^\nu \right)\,.
\end{equation*}
The corresponding correlator has the form
\begin{align}
	\langle \Gamma^{\mu\delta} [O_\gluinoglue(x)]_\mu [\bar{O}_\gluinoglue(y)]_\delta\rangle
	&=-\left\langle \tr\,\Gamma^\T\,F^{ij}(x)\Sigma_{ij} 
	\e^{\ii\alpha\gamma_5/2} G_{x,y}
	e^{\ii\alpha\gamma_5/2}F^{lm}(y)\Sigma_{lm}\right\rangle_\mathcal{U}\,.
	\label{eq:GluinoGlueCorrelatorWithTwist}
\end{align}
With the cyclicity of the trace one easily sees that for the antisymmetric correlator with \mbox{$\Gamma=\id_4$} a chiral phase factor $\e^{\ii\alpha\gamma_5}$ arises and for the symmetric correlator with \mbox{$\Gamma=\gamma_4$} the chiral twists cancel.

\subsection{Supersymmetry transformations of the lattice operators}\label{ch:SusyTrafosLatticeExpValues}\label{ch:SusyTrafos}
When the gluino is twisted as in eq.~\eqref{eq:ChiralRotation}, then
no additional terms arise in the supersymmetry transformations.
The only modification is an additional chiral phase factor
multiplying the spinor field $\lambda$, and this is carried through the 
whole calculation. It follows that every supermultiplets stays
intact. Without twists the off-shell supersymmetry transformations 
of the continuum theory have the simple form
\begin{align}
	\delta_\epsilon\lambda(x)&=\,\frac{1}{4}\,\Sigma_{\mu\nu}F^{\mu\nu}(x)\epsilon + \ii\mathcal{G}(x)\gamma_5\epsilon\,,&&\delta_\epsilon A_\mu(x) = \ii\bar{\epsilon}\gamma_\mu\lambda(x)\,,\nonumber\\
	\delta_\epsilon\bar{\lambda}(x)&=-\,\frac{1}{4}\,\bar{\epsilon}\,\Sigma_{\mu\nu}F^{\mu\nu}(x) + \ii\bar{\epsilon}\mathcal{G}(x)\gamma_5\,, &&\,\,\delta_\epsilon\mathcal{G}(x)\,\, = \bar{\epsilon} \gamma_5 \slashed{D} \lambda(x)\,.
	\label{eq:MinkowskiSusyTrafos}
\end{align}
Therein, $\epsilon$ is a constant Majorana-valued anticommuting parameter and $\mathcal{G}$ is an auxiliary field.
To determine the transformation of the gluino-glue
state one needs the transformation of  the field strength tensor,
\begin{align}
	\delta_\epsilon F_{\mu\nu}(x) 
	= \ii \bar{\epsilon}\, (\gamma_\nu D_\mu - \gamma_\mu D_\nu) \lambda(x)\,.
	\label{eq:deltaF}
\end{align}
The supersymmetry transformations of the composite
operators generating the VY-super\-multiplet are obtained with help 
of Fierz identities
 \cite{Fierz1937,Pal:2007dc}, derived from the general identity
 \begin{equation}
 	4\psi\bar{\chi} = -(\bar{\chi} \psi) - \gamma_\mu(\bar{\chi}\gamma^\mu \psi) + \frac{1}{2} \gamma_{\mu\nu} (\bar{\chi} \gamma^{\mu\nu} \psi) + \gamma_5\gamma_\mu (\bar{\chi} \gamma_5\gamma^\mu \psi) - \gamma_5 (\bar{\chi}\gamma_5 \psi)\,.
 \end{equation}
  One finds the transformations
\begin{align}
	\delta_\epsilon O_{\aF}(x) 
	&= -\frac{1}{2}\bar{\epsilon}\, O_{\gluinoglue}(x) + 2 \ii\bar{\epsilon}\mathcal{G}(x)\gamma_5 \lambda(x)\,, \label{eq:F0SusyTrafo}
	\\
	\delta_\epsilon O_{\aEta}(x) 
	&= -\frac{1}{2}\bar{\epsilon}\, \gamma_5 O_{\gluinoglue}(x) + 2\ii\bar{\epsilon}\mathcal{G}(x)\lambda(x)\,, \label{eq:EtaSusyTrafo}
	\\
	\delta_\epsilon \big( O_{\gluinoglue}(x) - 4\ii\mathcal{G}(x)\gamma_5 \lambda(x) \big)
	&= 2\ii\slashed{\partial} O_{\aF}(x) \epsilon + 2\ii \gamma_5 \slashed{\partial} O_{\aEta}(x) \epsilon + \ldots \,. \label{eq:GluinoGlueSusyTrafo}
\end{align}
The terms linear in the auxiliary field $\mathcal{G}$ as well as further terms indicated with the dots in eq.~\eqref{eq:GluinoGlueSusyTrafo} vanish on-shell and thus the VY-supermultiplet defines a chiral supermultiplet.

After a Wick-rotation to Euclidean spacetime, the on-shell 
supersymmetry transformation in eq.~\eqref{eq:MinkowskiSusyTrafos}
read \cite{LuckmannDiploma,KirchnerDiss,Montvay:2001aj}
\begin{align}
\delta_\epsilon A_\mu(x) = \ii\bar{\epsilon}\gamma_\mu\lambda(x),\qquad\delta_\epsilon\lambda(x)=\frac{1}{4\ii}\Sigma_{\mu\nu}F^{\mu\nu}(x)\epsilon,\qquad\delta_\epsilon\bar{\lambda}(x)=-\frac{1}{4\ii}\bar{\epsilon}\,\Sigma_{\mu\nu}F^{\mu\nu}(x)\,.
\label{eq:EuclideanSusyTrafos}
\end{align}
Although Majorana spinors in $4$-dimensional Euclidean spacetime  
cannot be defined consistently, we instead may use the consistent condition $\bar{\lambda}=\lambda^\T \C$ \cite{vanNieuwenhuizen:1996tv}. This way the same 
symmetries for the bilinears $\bar\psi \gamma^{\mu_1\dots\mu_n}\chi$ hold as in Minkowski spacetime.
The corresponding transformations of the
composite fields $O_{\aF}(x),\,O_{\aEta}(x)$ and $O_{\gluinoglue}(x)$
in Euclidean spacetime are just the Wick-rotations of the transformations (\ref{eq:F0SusyTrafo}),\,(\ref{eq:EtaSusyTrafo})
and (\ref{eq:GluinoGlueSusyTrafo}). This can be shown
explicitly by observing that the Fierz identities used to 
derive these transformations exist in Minkowski and Euclidean 
spacetime.

At finite lattice spacing supersymmetry is broken
and this will lead to additional terms in the transformation laws.
The lattice susy transformation can be formulated as \cite{KirchnerDiss,taniguchi_one_2000}
\begin{align}
	\delta_\theta \,\U_\mu(x) &= -\frac{\ii g}{2} \big(\bar{\theta}(x)\gamma_\mu\,\U_\mu(x)\lambda(x)+\bar{\theta}(x+\hat{\mu})\gamma_\mu\lambda(x+\hat{\mu})\,\U_\mu(x)\big)\\
	\delta_\theta \,\U_\mu^\dagger(x) &= \frac{\ii g}{2} \big(\bar{\theta}(x)\gamma_\mu\lambda(x)\,\U_\mu^\dagger(x)+\bar{\theta}(x+\hat{\mu})\gamma_\mu\,\U_\mu^\dagger(x)\lambda(x+\hat{\mu})\big)\\
	\delta_\theta \lambda(x) &= \frac{1}{4\ii} \Sigma_{\mu\nu} P^{\mu\nu}(x)\theta(x)\\
	\delta_\theta \bar{\lambda}(x) &= -\frac{1}{4\ii} \bar{\theta}(x)\Sigma_{\mu\nu}P^{\mu\nu}(x)
	\label{eq:LatticeSusyTrafos}
\end{align}
with clover plaquette $P^{\mu\nu}(x)$
and infinitesimal Majorana parameters $\bar{\theta}$ and $\theta$. In the continuum limit the corresponding transformations \eqref{eq:EuclideanSusyTrafos} are recovered.

\subsection{Eigenvalues of the free Wilson Dirac operator}\label{ch:EigValues}
For particular twists of the free lattice Dirac operator
in lower-dimensional Wess-Zumino models an improvement up to
order $\mathcal{O}(a^4)$ can be achieved \cite{Bergner:2007pu,KaestnerDiss}.
In order to see whether an improvement is also possible for the 
double-twisted lattice Dirac operator in supersymmetric
gauge theory we determine the eigenvalues of the operator
 $D_\text{W}^{\text{dtw}}$ (see eq.~\eqref{eq:DiracOperatorWithDoubleTwist})
for free fermions, that is for trivial link
variables \mbox{$\V_\mu=\id$}. Thus we calculate the eigenvalues and expand them in powers of
the lattice spacing $a$ to study the discretization errors.
Then, the dependence on the twist angles $\alpha,\varphi$ is analyzed to check if $\mathcal{O}(a)$ improvement 
is possible for particular choices\footnote{In our simulations,
	\mbox{$(r,r_5)=(1,0)$} resp.\ \mbox{$R=\sqrt{r^2+r_5^2}=1$} and \mbox{$\varphi=0$}
	is chosen if no other value is stated.}.
We decompose the double-twisted lattice Dirac operator 
for free fermions,
\begin{equation}
D_\text{W}^{\text{dtw}}=\gamma^\mu\ring{\partial}_\mu + M\,\e^{\ii\alpha\gamma_5} - \frac{aR}{2}\e^{\ii\varphi\gamma_5}\hat{\Delta}=
\gamma^\mu\ring{\partial}_\mu+X+\ii\gamma_5 Y\,,
\label{twdi1}
\end{equation}
which contains the naive antisymmetric lattice derivative~$\ring{\partial}_\mu$ 
and the symmetric lattice Laplacian~$\hat{\Delta}$
(we use the notation of \cite{Wipf:2013vp}).
The real operators $X,Y$ in the last decomposition are
\begin{equation}
X=M\cos\alpha-\frac{aR}{2}\hat{\Delta}\cos\varphi,\quad
Y=M\sin\alpha-\frac{aR}{2}\hat{\Delta}\sin\varphi\,.\label{twdi3}
\end{equation}
The periodic eigenfunctions are constant spinors times plane waves
on a $L^3\times T$ lattice:
\begin{equation}
\psi_p(x)=u_p\, \e^{\ii p_\mu x^\mu},\quad
p_0=\frac{2\pi}{aN_t}\left(n_0+\frac{1}{2}\right),\quad 
p_i=\frac{2\pi}{aN}\,n_i\,.\label{twdi5}
\end{equation}
Plane waves are eigenfunctions of the derivative operators
and the Laplacian,
\begin{equation}
\ring{\partial}_\mu \mapsto
\ii\ring{p}_\mu,\quad 
\ring{p}_\mu=\frac{1}{a}\,\sin(a p_\mu)\,,\qquad \hat\Delta\mapsto -\hat p_\mu\hat p^\mu,\quad 
\hat p_\mu=\frac{2}{a} \, \sin\left(\frac{a p_\mu}{2}\right)\,.
\label{twdi7}
\end{equation}
In a sector with fixed momentum the operator $X$ is a 
constant  $X_p$ which just shifts the eigenvalues 
of $D_\text{W}^{\text{dtw}}$ in eq.~\eqref{twdi1}. Hence it
suffices to determine the imaginary eigenvalues of the
$4$-dimensional anti-Hermitean matrix $\A$ in \mbox{$D_\text{W}^{\text{dtw}}=\A+X$}
for fixed momentum,
\begin{equation}
\A_p u_p=\left(\ii\gamma^\mu \ring{p}_\mu+\ii\gamma_5Y_p\right)u_p
=\ii 
\mu_p u_p,\quad \mu_p \text{ real}.\label{twdi9}
\end{equation}
Since \mbox{$\A_p \A_p^\dagger=\ring{p}^2+Y_p^2$} is a multiple of
the identity matrix we conclude that \mbox{$\mu_p^2=\ring{p}^2+Y_p^2$}.
In Euclidean spacetime there exists an 
antisymmetric charge conjugation matrix $\C_+$ with
\begin{equation}
\C_+\gamma^T_\mu \C_+^{-1}=\gamma_\mu,\quad 
\C_+\gamma^T_5\C_+^{-1}=\gamma_5\,.\label{twdi11}
\end{equation}
Taking the complex conjugate of the eigenvalue
equation~\eqref{twdi9} and acting with $\C_+$ on
this equation (and also using
that $\gamma^\mu$ and $\gamma_5$ are Hermitean)
we see that the charge conjugated constant spinor
$\C_+ u_p^*$ is a second eigenvector with
the same eigenvalue $\ii\mu_p$. Finally, since
\mbox{$\tr(\A_p)=0$} we deduce, that $\A_p$ has two eigenvalues
$\ii\mu_p$ and two eigenvalues $-\ii\mu_p$.
We conclude that for fixed $p_\mu$ the twisted Dirac operator 
\mbox{$D_\text{W}^{\text{dtw}}=\A+X$} has the double degenerate eigenvalues
\begin{equation}
\lambda_p=X_p+\ii \mu_p\quad\text{and}\quad 
\lambda_p^*=X_p-\ii\mu_p,\quad 
\mu_p=\sqrt{\ring{p}^2+Y_p^2}\,.\label{twdi13}
\end{equation}
Up to a possible sign the Pfaffian of the Dirac operator
is the square root of its determinant and hence given 
by the product of all $\vert\lambda_p\vert^2$, where
\begin{equation} 
\vert\lambda_p\vert^2=\ring{p}^2+X_p^2+Y_p^2=
\ring{p}^2+M^2+\frac{(aR)^2}{4}\hat{p}^2\hat{p}^2
+(aR)M\hat{p}^2\cos(\alpha-\varphi)\,.\label{twdi15}
\end{equation}
Inserting the small-$a$ expansions of $\hat p_\mu$ and $\ring{p}_\mu$ in eq.~\eqref{twdi7} gives rise to
\begin{equation}
\vert\lambda_p\vert^2=p^2+M^2+(aR)M p^2\cos(\alpha-\varphi)
+\frac{a^2}{12}\left(3R^2 (p_\mu p^\mu)^2-4
\sum_\mu p_\mu^4\right)+\mathcal{O}(a^3)\,.\label{twdi17}
\end{equation}
Here we see explicitly that setting \mbox{$\alpha-\varphi=90^\circ$ }
leads to an $\mathcal{O}(a)$ improvement
in the fermionic sector -- at least for free fermions.

Table~\ref{tab:eigenvalues} summarizes the values
for $\vert\lambda_p\vert^2$ and their
small-$a$ expansions for various
lattice Dirac operators considered in the present work.
Starting from the Wilson Dirac operator $D_1$ with $\mathcal{O}(a)$
discretization errors, we can remove the leading discretization
effects by choosing a $90^\circ$-twist like in $D_2$ (as in fully 
twisted lattice QCD \cite{Frezzotti:2003ni}) or by modifying the Wilson term 
like in $D_3$. In general, for free fermions $\mathcal{O}(a)$
improvement can be achieved when the mass term and Wilson 
term are orthogonal to each other, i.e. \mbox{$\alpha-\varphi=90^\circ~(\text{mod}~180^\circ)$} in~$D_4$.

The mass difference of the superpartners
$\aF$ and $\aEta$ is minimal for \mbox{$\alpha-\varphi=45^\circ$}, see figure~\ref{fig:PiACuts}. Since in the present
work our main
focus is on the restoration of supersymmetry and chirality 
we choose \mbox{$\alpha-\varphi=\alpha=45^\circ$} in our simulations.
Then there is a reduction of the 
leading order discretization errors by a factor of \mbox{$\cos(45^\circ)=1/\sqrt{2}$}\,.
\begin{table}[h]
	\centering
	\renewcommand{\arraystretch}{1.3}
	\begin{tabular}{!{\vrule width 1pt}ll!{\vrule width 1pt}}
		\noalign{\hrule height 1pt} 
		lattice Dirac operator & eigenvalues $\vert\lambda_p\vert^2$\\ \hline
		$D_1=\gamma^\mu\ring{\partial}_\mu + M - \frac{aR}{2} \hat{\Delta}$ & $p^2+M^2+aMRp^2+\mathcal{O}(a^2)$ \\[+0.25em]
		$D_2=\gamma^\mu\ring{\partial}_\mu + M - \frac{\ii a R}{2}\gamma_5\hat{\Delta}$ & $p^2+M^2+\kappa a^2\,+\mathcal{O}(a^4)$ \\
		$D_3=\gamma^\mu\ring{\partial}_\mu + M\,\e^{\ii\alpha\gamma_5} - \frac{aR}{2} \hat{\Delta}$	& $p^2+M^2+
		aMRp^2\cos(\alpha)
		+\mathcal{O}(a^2)$ \\[+0.25em]
		$D_4=\gamma^\mu\ring{\partial}_\mu + M\,\e^{\ii\alpha\gamma_5} - \frac{aR}{2}\e^{\ii\varphi\gamma_5}\hat{\Delta}$ & $p^2+M^2+
		aMRp^2\cos(\alpha-\varphi)
		+\kappa a^2+
		\mathcal{O}(a^3)
		$~~\\[+0.25em]\noalign{\hrule height 1pt} 
	\end{tabular}
	\caption{Eigenvalues $\vert\lambda_p\vert^2$ of several 
		lattice Dirac operators $D_i$,  expanded in powers of the lattice spacing $a$. We defined \mbox{$\kappa\equiv-\frac{1}{3}\sum_\mu p_\mu^4+\frac{R^2}{4}\big(p_\mu p^\mu\big)^2$}.}
	\label{tab:eigenvalues}
\end{table}

%% file: numerical_investigations.tex
\section{Numerical investigations}\label{sec:NumInvest}

In this section we present, compare and discuss our lattice 
results for \SYM theory with and  without twisted mass term. 
As demonstrated below, finite size effects are clearly visible in 
the data, while lattice spacing artifacts are more or less absent.
That means it will be beneficial to choose a slightly larger
gauge coupling in future simulations.
However, as we shall see in this chapter, for the optimal twist angle $\alpha=45^\circ$ finite size effects are less severe.
Table~\ref{tab:betaV} in appendix~\ref{ch:NumOverview} lists the lattice couplings, 
lattice sizes, mass parameters and Wilson parameters used in the
simulations.

\subsection{Scale-setting}\label{ch:ScaleSetting}

To set the scale, the Sommer parameter and QCD units are used, i.e., \mbox{$r_0=\mathrm{0.5\,fm}$}~\cite{Sommer:1993ce}. In the given context 
this is somewhat arbitrary but it allows for a direct comparison with results 
in the literature.

For our estimates of $a/r_0$, we calculate rectangular Wilson loops of different size and extract the static potential $V(R)$ for a range of spatial separations $R$. In temporal direction all loops are sufficiently large such that $V(R)$ remains stable. Furthermore, different levels of stout smearing are applied to the gauge fields (with staple weight $\rho=0.1$ \cite{Morningstar:2003gk}) and the Wilson fermion mass term is varied to allow for a safe extrapolation to the critical point, $m\to m_\text{crit}$.
For the different levels of smearing and Wilson term mass values, the results for $V(R)$ are separately fitted to 
\begin{equation}
  V(R) = V_0 + \sigma R - \frac{\alpha}{R}\,.
\end{equation}
From the fit parameters and setting 
\begin{equation}
  \frac{r_0}{a} \equiv \sqrt{\frac{1.65-\alpha}{\sigma a^2}}
\end{equation}
we obtain the lattice spacing and can extrapolate to the critical point.

\begin{figure}[tbp]
	\centering
	\begin{minipage}[c]{0.46\textwidth}
		\includegraphics{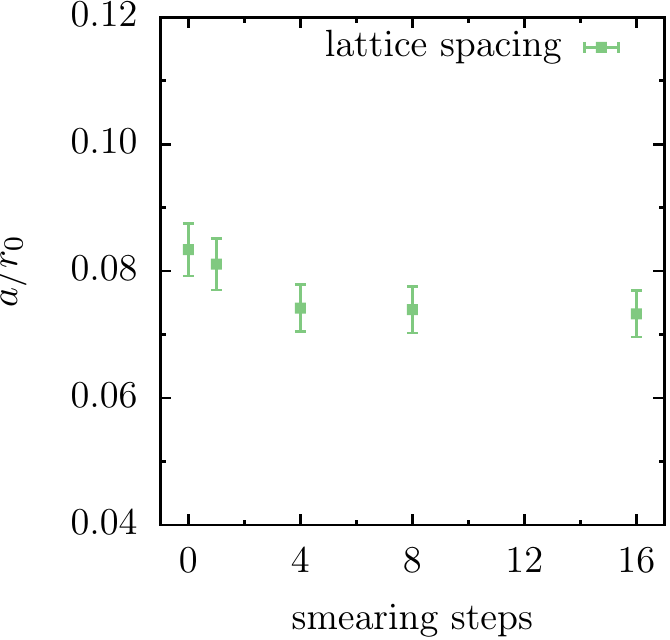}
		\caption{Dimensionless lattice spacing $a/r_0$ vs. gauge smearing steps. 
			The values include an extrapolation to the critical point at $m_\text{crit}$. To this end the parameter set (II) is fitted to $V(R)$ for different $m$, 
			see table~\ref{tab:betaV} in section~\ref{ch:NumOverview}. As long as only 
			a few gauge smearing steps are applied, all values are comparable.}
		\label{fig:ScaleSetting}
	\end{minipage}
	\hfill
	\begin{minipage}[c]{0.46\textwidth}
		\includegraphics{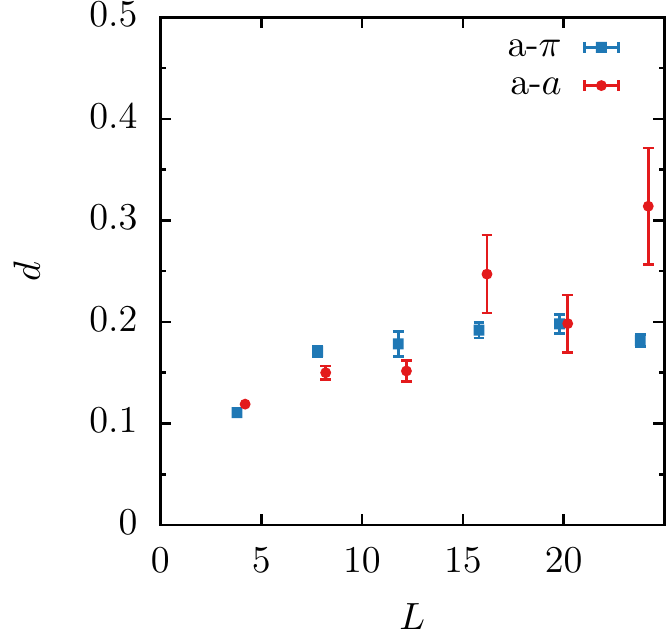}
		\caption{Dominant mass contribution of $\aPi$ and $\aA$ for different lattice sizes \mbox{$L^3\times 2L$} with $L\in\{4,8,12,16,$ $20,24\}$.
		The bare gluino mass is \mbox{$(m,m_5)=$} $(\text{-1.0506},\text{0.0})$ and 
		the lattice coupling $\beta=\textrm{5.0}$. Points are slightly displaced for better visibility.\newline}
		\label{fig:FiniteSizeScaling_1_67}
	\end{minipage}
\end{figure}

As an example, the lattice spacing for ensemble~(II) is shown in figure~\ref{fig:ScaleSetting} for different steps of stout smearing, and for $m\to m_\text{crit}$. We find that for a large number of stout smearing steps, the static potential changes its shape, but for a moderate number, as shown in figure~\ref{fig:ScaleSetting}, the lattice spacing values are all comparable. Combining the data in a linear fit leads to the lattice spacing \mbox{$a=\mathrm{(0.040\pm0.002)\,fm}$} for ensemble~(II). This translates into a spatial lattice \mbox{$aL=\mathrm{(0.64\pm0.03)\,fm}$} for this ensemble. In comparison to other lattice studies, e.g.\ \cite{Ali:2018dnd}, a box length \mbox{$aL<\textrm{1\,fm}$} appears small and finite size effects need to be carefully analyzed.
This is provided in the following section.

\subsection{Finite size analysis} \label{ch:FiniteSizeAnalysis}

\begin{figure}[tbp]
	\centering
	\includegraphics{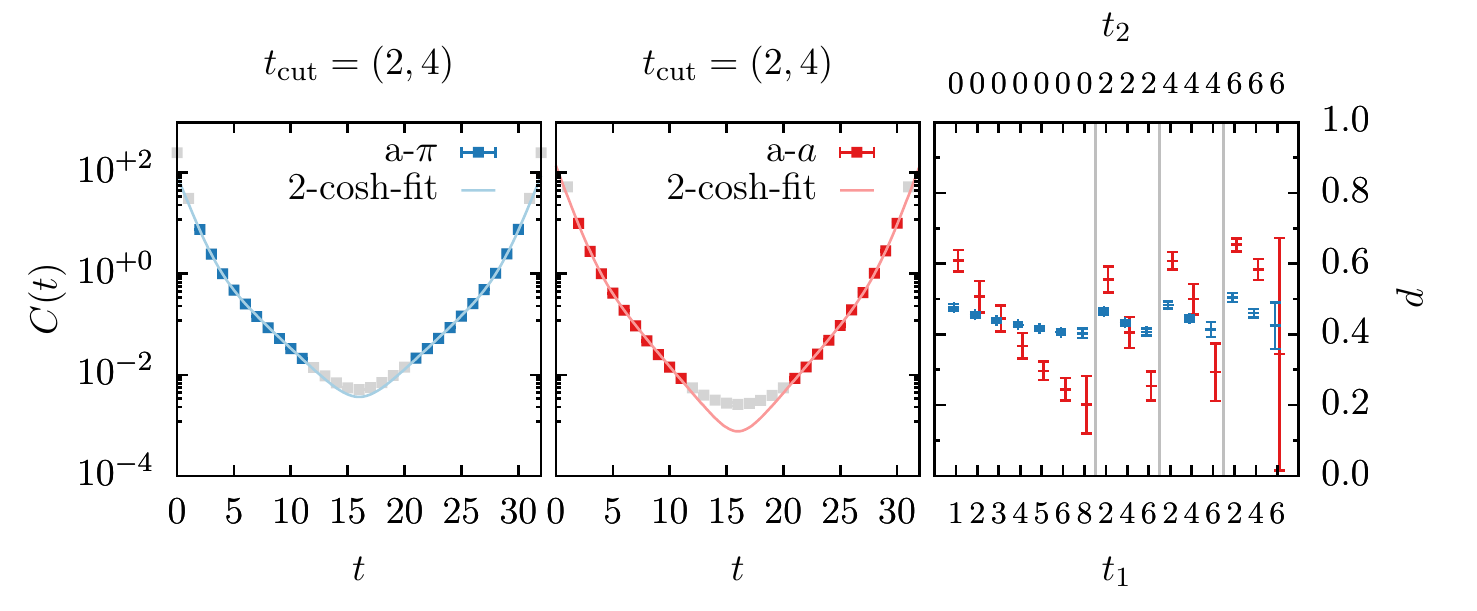}
	\caption{Left/Center: $\aPi$ resp.\ $\aA$ correlator at \mbox{$\beta=5.4$} and \mbox{$(m,m_5)=(\textrm{-0.8950},\textrm{0.0})$} with 2-cosh-fits. The gray data points are excluded from the fit to reduce contributions from higher states and to stabilize the fit. Error bars are smaller than the symbol size. Right: The dominant mass contribution $d$ from fits with different cuts $t_\text{cut}=(t_1,t_2)$. The gray vertical lines separate regions with different values of $t_2$.}
	\label{fig:PiAfit}
\end{figure}

We continue with ensemble~(V) and show data for the $\aPi$ and $\aA$ correlators for $(m,m_5)=(\textrm{-0.8950},\textrm{0.0})$ in figure~\ref{fig:PiAfit}. Looking at the left and middle panel of this figure, one clearly sees both correlators would not fit a simple cosh-like $t$-dependence. Up to $t=5$ (and $T-t=5$), contributions from higher states are significant, and the interval where a single exponential behavior dominates is rather short. To fit the $t$-dependence we therefore choose a 2-cosh ansatz
\begin{equation}
	C(t)= c_1 \cosh(d_1\big(t-T/2)\big) + c_2 \cosh\big(d_2(t-T/2)\big)
	\label{eq:fitFunction}
\end{equation}
and vary the fit ranges $t\in[t_1,T/2-t_2]$ and $t\in[T/2+t_2,T-t_1]$. Furthermore, 
we will refer to \mbox{$d=\min(d_1,d_2)$} as the \emph{dominant mass contribution}. 
It corresponds to the ground state mass on sufficiently large lattices.

As an example, the 2-cosh fits for $t_1=2$ and $t_2=4$ are included in
figure~\ref{fig:PiAfit}. Colored symbols refer to points inside the fit range, while gray
symbols to points outside. Although cutting the inner time slices is not necessary,
as we will see, it turns out to be useful nonetheless: Near the critical point, the
correlators of the connected part of the mesonic states are flat while those of the
disconnected part are dominated by statistical noise. Applying cuts on both sides of the fit
ranges stabilizes the fits and reduces the contributions of excited states.

Results for $d_{\aPi}$ and $d_{\aA}$, and for different combinations of $t_1$ and $t_2$, 
are shown in the right panel of figure~\ref{fig:PiAfit}. There, the upper $t_2$-axis 
divides the panel (vertical lines) into four domains and each domain shows $d$ 
versus $t_1$ at constant $t_2$. We see that a variation of $t_2$ has a minor 
effect on the value for $d$, whereas there is a clear linear dependence on $t_1$. 
In particular for $\aA$ this dependence is significant.

Using the same ansatz as before we can analyze $d$ as a function of $L$. 
For $\beta=5.0$ and $(m,m_5)=(\textrm{-1.0506},\textrm{0.0})$ we did simulations for $L=4,\ldots,24$
and the results for the adjoint states are shown in 
figure~\ref{fig:FiniteSizeScaling_1_67} (for $(t_1,t_2)=(2,0)$). 
For the adjoint pion, $d$ forms a plateau at approximately $d=0.2$ 
for $L\ge 16$, while for $\aA$ the situation is not as clear. 
A similar behavior is seen for $\beta=4.5$. The volume effects for $\aPi$ are mild, 
while for $\aA$ an unambiguous mass extraction is more difficult despite 
a good signal-to-noise ratio.

\begin{figure}[tbp]
	\centering
	\hspace*{-11mm}\includegraphics{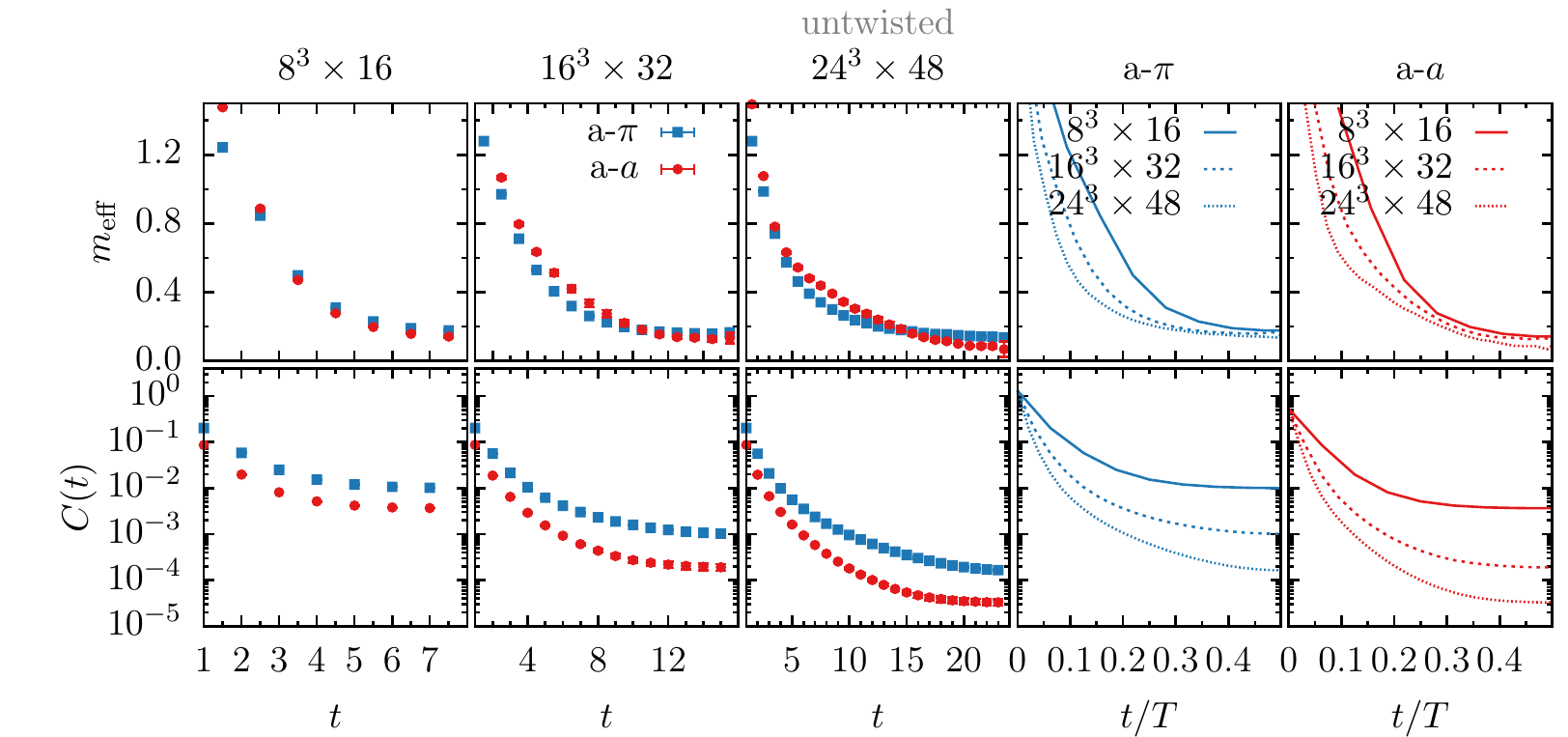}
	\caption{Top/Bottom: Effective masses/correlators of the adjoint pion and the adjoint~$a$ at fixed bare gluino mass \mbox{$m=\text{-1.0506}$} and lattice coupling \mbox{$\beta=\text{5.0}$}, without mass twist. From left to right the first three columns show results for $\aPi$ and $\aA$ from a \slat, \mlat\ and \llat~lattice. The last two compare $\aPi$ and $\aA$ for different lattice sizes. Most error bars are smaller than the symbol size. In the last two columns, data points are connected by lines and symbols and error bars are omitted for better visibility.}
	\label{fig:PiAmeffcorruntwisted81624PiA}
\end{figure}

Volume effects are also apparent in the effective mass plots. Such plots, and the corresponding $\aPi$ and $\aA$ correlators,
are shown in figure~\ref{fig:PiAmeffcorruntwisted81624PiA}, again for $\beta=5.0$,
$(m,m_5)=(\textrm{-1.0506},\textrm{0.0})$ and $L=8, 16, 24$. The first three columns compare the 
effective mass and the correlators for $\aPi$ and $\aA$ for a fixed lattice size, 
while the panels in column 4 and 5 show them separately for $\aPi$ and $\aA$ for 
different volumes and versus $t/T$. Looking at the first three upper panels in
figure~\ref{fig:PiAmeffcorruntwisted81624PiA} we notice an intersection of the 
effective mass values at a certain~$t$. The effective mass of $\aA$ falls off 
faster with $t$ than that of $\aPi$ and approaches a lower value. 
Furthermore, the deviations seem to increase with increasing volume. 
This is in contrast to common expectations,  because $\aPi$ should be the 
lighter state. Most likely it is the 
small volume ($aL<1\,\text{fm}$) which 
causes $\aA$ appearing lighter than $\aPi$.

Indications for this are also provided by the last two upper panels of
figure~\ref{fig:PiAmeffcorruntwisted81624PiA} showing the same data sets 
as the first three panels but as function of the rescaled variable
$t/T$ such that finite size effects are better visible.
We see the effective mass curves of $\aPi$ settle on the same value 
on all lattices and only the length of the plateau increases with 
lattice size. But for the $\aA$ state the effective mass seems 
not to approach a single plateau, if at all, rather $m_{\mathrm{eff}}$ gets smaller when 
increasing the lattice size which indicates an enhanced correlation length. 
This is in line with the correlator plots in the lower panels of
figure~\ref{fig:PiAmeffcorruntwisted81624PiA}. There, the $\aA$ correlator 
on the largest lattice (\llat) decays visibly faster than the $\aPi$ 
correlator for \mbox{$0<t\lesssim 15$}. For the smaller lattices this effect 
is less pronounced. Again we see that finite size effects are small for $\aPi$, 
while they are more pronounced for $\aA$. This supports our 
interpretation of the results in figure~\ref{fig:FiniteSizeScaling_1_67}. 

In summary, especially the $\aA$ state is problematic in small volumes, 
where a flat region appears in its correlator and
the extracted dominant mass contribution is underestimated.
In some distance to the critical point this lattice artifact is less 
pronounced and the mass hierarchy is as expected, $m_\aA>m_\aPi$.
But with $D^\text{mtw}$ and optimal twist angle, the correlators 
of $\aPi$ and $\aA$ have identical shapes and no observable finite volume 
artifacts remain.

\medskip
Let us recall at this point that the connected mesonic states are 
not part of the physical spectrum of the \SYM theory.  But these
auxiliary states are very useful, mainly because the signal-to-noise-ratio 
of the related correlators are much better compared to
those of the physical mesonic states with disconnected contributions. Therefore we use 
the connected mesons for ensembles with low statistics, like in this section 
or the parameter scan in the next section. In addition, the connected diagrams 
contribute to the correlators of the physical states, 
see eq.~\eqref{eq:MesonicCorrelator} and \eqref{eq:VacuumContribution}.
Hence, the connected mesonic states partly determine the behavior of the 
full physical states.

\subsection{Parameter scan}\label{ch:ParameterScan}

\begin{figure}[tbp]
	\hspace*{-6mm}\begin{minipage}[c]{0.30\textwidth}
		\includegraphics{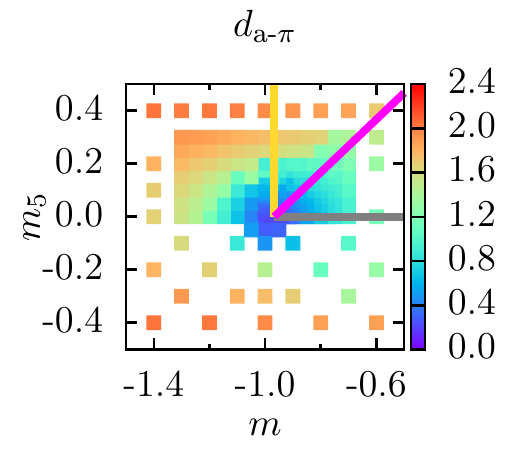}
	\end{minipage}
	\hfil
	\begin{minipage}[c]{0.30\textwidth}
		\includegraphics{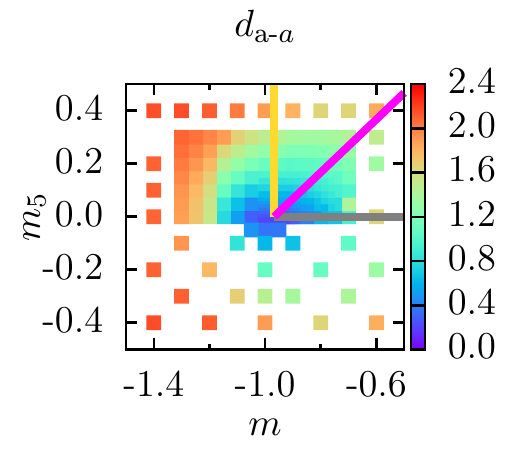}
	\end{minipage}
	\hfil
	\begin{minipage}[c]{0.30\textwidth}
		\includegraphics{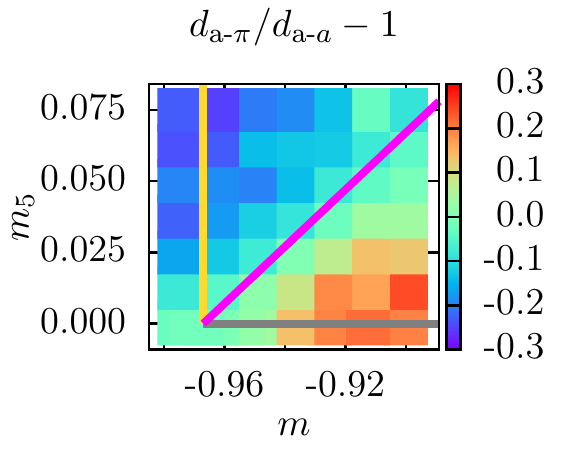}
	\end{minipage}
	\caption{Parameter scan in $m$ and $m_5$ on a \slat~lattice. In the left and middle plot the dominant mass contribution $d$ of the $\aPi$ (connected part of the $\aEta$) resp.\ $\aA$ (connected part of the $\aF$) are shown. The right plot combines those results in the subtracted ratio \mbox{$d_{\aPi}/d_{\aA}-1$}. Note the different axis ranges. The colored lines (gray, magenta and yellow) are discussed in the text.}
	\label{fig:PiA}
\end{figure}

After discussing finite size effects for the untwisted system, we now 
analyze the effect of a twisted mass term for Wilson fermions.
To this end, we calculate the dominant mass contribution of the $\aPi$ 
and $\aA$ correlators in the $(m,m_5)$ parameter space by performing a 
parameter scan. For this scan we fix the lattice coupling and size 
to \mbox{$\beta = 5.4$} and \slat\ and vary the mass parameter 
\mbox{$m \in [\textrm{-1.4}, \textrm{-0.6}]$} and the twist parameter 
\mbox{$m_5 \in [\textrm{-0.4}, \textrm{0.4}]$} around the critical point,
\mbox{$(m,m_5)=(\textrm{-0.967},\textrm{0.0})$}. Due to the
\mbox{$(m_5\leftrightarrow-m_5)$-symmetry}, fine parameter steps are 
necessary only in the upper half-plane of the parameter space; see 
left and middle plot of figure~\ref{fig:PiA}.
Every gauge ensemble consists of approximately 200 thermalized configurations 
which is sufficient for a good signal-to-noise-ratio for the correlators. To 
determine their dominant mass contribution $d$, all correlators are fitted 
to the ansatz~\eqref{eq:fitFunction} as in the previous section. On a 
rather small $\slat$  lattice, the quantity $d$ is only a rough estimate 
for the ground state mass and the results for the latter 
are more qualitative than quantitative. However, the simulation
results on a larger \mlat~lattice support our findings.

Note that we treat the twist as a deformation of 
the lattice action and do \emph{not} rotate observables back, as is done 
in twisted mass QCD. In the limit \mbox{$(m\to m_\textrm{crit}, m_5\to 0)$} 
the twisted Wilson Dirac operator $D_\text{W}^{\text{mtw}}$ \eqref{eq:DiracOperatorWithTwist} is equivalent to the Wilson Dirac 
operator $D_\text{W}$~\eqref{eq:DiracOperator} such that both operators 
correspond to the same continuum theory. But along certain paths
ending at the critical parameters (belonging to the continuum theory)
the breaking of chiral symmetry and of supersymmetry maybe suppressed.

\begin{figure}[tbp]
	\centering
	\includegraphics{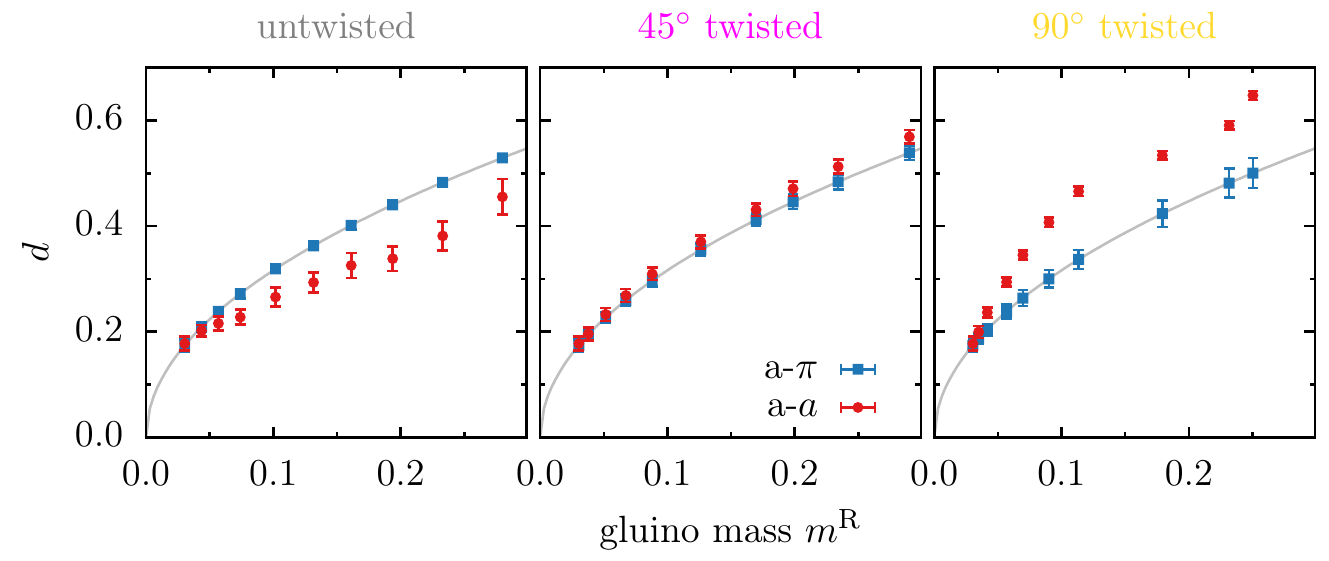}
	\caption{The three plots correspond to twist angles \mbox{$\alpha\in\{0^\circ,45^\circ,90^\circ\}$} marked in Figure~\ref{fig:PiA} with lines in gray, magenta and yellow. The same data of the \slat~lattice is used and three clearly different mass hierarchies of the dominant contribution $d$ are revealed. Some error bars are smaller than the symbol size. The gray solid lines visualize the dependence of the adjoint pion on the renormalized gluino mass, see eq.~\eqref{eq:mPi2}.}
	\label{fig:PiACuts}
\end{figure}

\begin{figure}[tbp]
	\centering
	\includegraphics{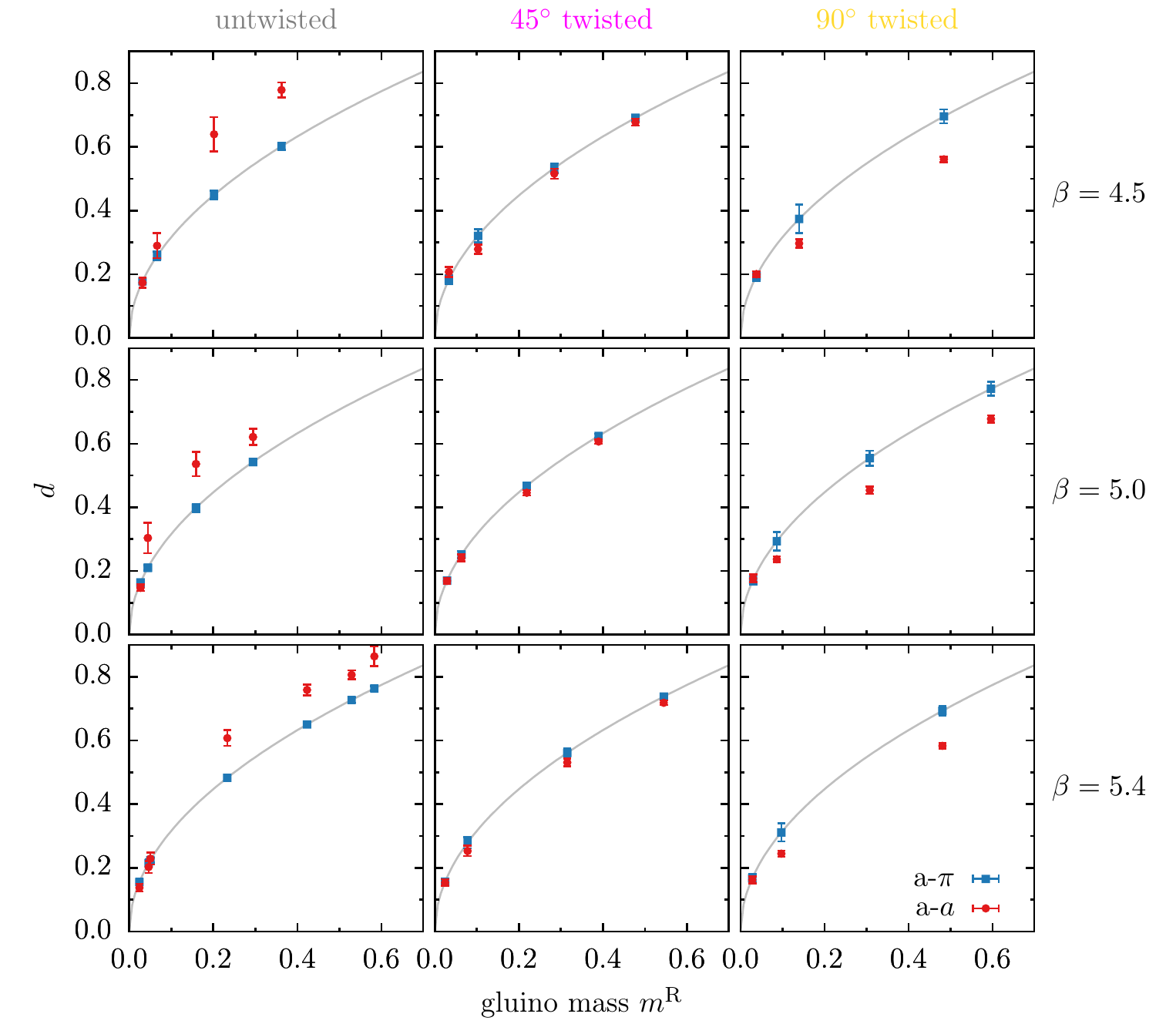}
	\caption{Connected mesons on the \mlat~lattice. From left to right the twist angle \mbox{$\alpha\in\{0^\circ,45^\circ,90^\circ\}$} rises. From top to bottom the lattice coupling \mbox{$\beta\in\{4,5.5.0,5.4\}$} increases. Some error bars are smaller than the symbol size and some data points for $\alpha=45^\circ$ lie on top of each other. The gray solid lines visualize eq.~\eqref{eq:mPi2}.}
	\label{fig:PiA16}
\end{figure}

Figure~\ref{fig:PiA} shows the dominant mass contributions. The left and center panel 
show $d_{\aPi}$ and $d_{\aA}$, respectively, while in the right panel the subtracted 
ratio \mbox{$d_{\aPi}/d_{\aA}-1$} near the critical point is shown.
Three interesting choices for the twist angle $\alpha$ are 
highlighted in these panels:
\begin{itemize}
	\item The data points for the untwisted case with \mbox{$\alpha=0^\circ$} and $m_5=0$ 
	along the gray line indicate that $d_\aPi$ is greater than $d_\aA$.
	\item For \mbox{$\alpha=45^\circ$} along the diagonal magenta line, the dominant 
	mass contributions of the chiral partners $\aPi$ and $\aA$ seem to match. 
	\item At maximal twist, i.e., \mbox{$\alpha=90^\circ$}, where the bare gluino 
	mass is kept fixed at its critical value, \mbox{$m=m_\textrm{crit}=\textrm{-0.967}$}, 
	and only the twisted mass parameter $m_5$ is varied, $d_\aA$ is greater 
	than $d_\aPi$, see vertical yellow line.
\end{itemize}
The results clearly favor a twist angle  \mbox{$\alpha=45^\circ$} 
with improved chiral properties at finite lattice spacing.
This interpretation is supported by the results shown in
figure~\ref{fig:PiACuts}, where the dominant mass contributions $d_{\aA}$ and $d_{\aPi}$ 
are shown versus the renormalized gluino mass \mbox{$m^\text{R}\propto m_{\aPi}^2$}. 
At \mbox{$\alpha=45^\circ$} the two chiral partners have equal mass
within errors. In contrast, for \mbox{$\alpha=0^\circ$} and
\mbox{$\alpha=90^\circ$} we clearly see a split of the two masses\footnote{See
appendix~\ref{ch:PionProof} for a discussion of the expected mass hierarchy and 
also compare with figure~\ref{fig:PiA16}.}.


To substantiate this observation on the small \slat~lattice, we double the 
lattice in each direction and repeat our calculation for the gauge couplings
\mbox{$\beta\in\{4.5,5.0,5.4\}$} along the three aforementioned directions 
in parameter space. For the fit we choose $t_\text{cut}=(2,4)$. 
The results are summarized in figure~\ref{fig:PiA16}. We see that
without twist the dominant $\aA$ contribution is greater than the $\aPi$ 
contribution, at \mbox{$\alpha=45^\circ$} both contributions 
are equal, and at maximal twist $\alpha=90^\circ$ the $\aA$ contribution is 
smaller than that of $\aPi$. Compared to the small volume results 
in figure~\ref{fig:PiACuts}, the mass hierarchies for $\alpha=0^\circ$ 
and $\alpha=90^\circ$ are inverted, which is a finite size effect, 
but our findings for $\alpha=45^\circ$ remain and are barely
affected by the size of the lattice.

In subsequent sections, we will therefore focus on the twist 
angle \mbox{$\alpha=45^\circ$} with improved chiral and supersymmetry
properties at finite lattice spacing. Furthermore 
from section~\ref{ch:EigValues} we know this special twist comes with 
an $\mathcal{O}(a)$ improvement at tree level which may at least reduce 
lattice spacing artifacts also at the non-perturbative level. Performing 
continuum extrapolations along the \mbox{$\alpha=45^\circ$} direction may 
thus be beneficial.

\begin{figure}[tbp]
	\centering
	\includegraphics{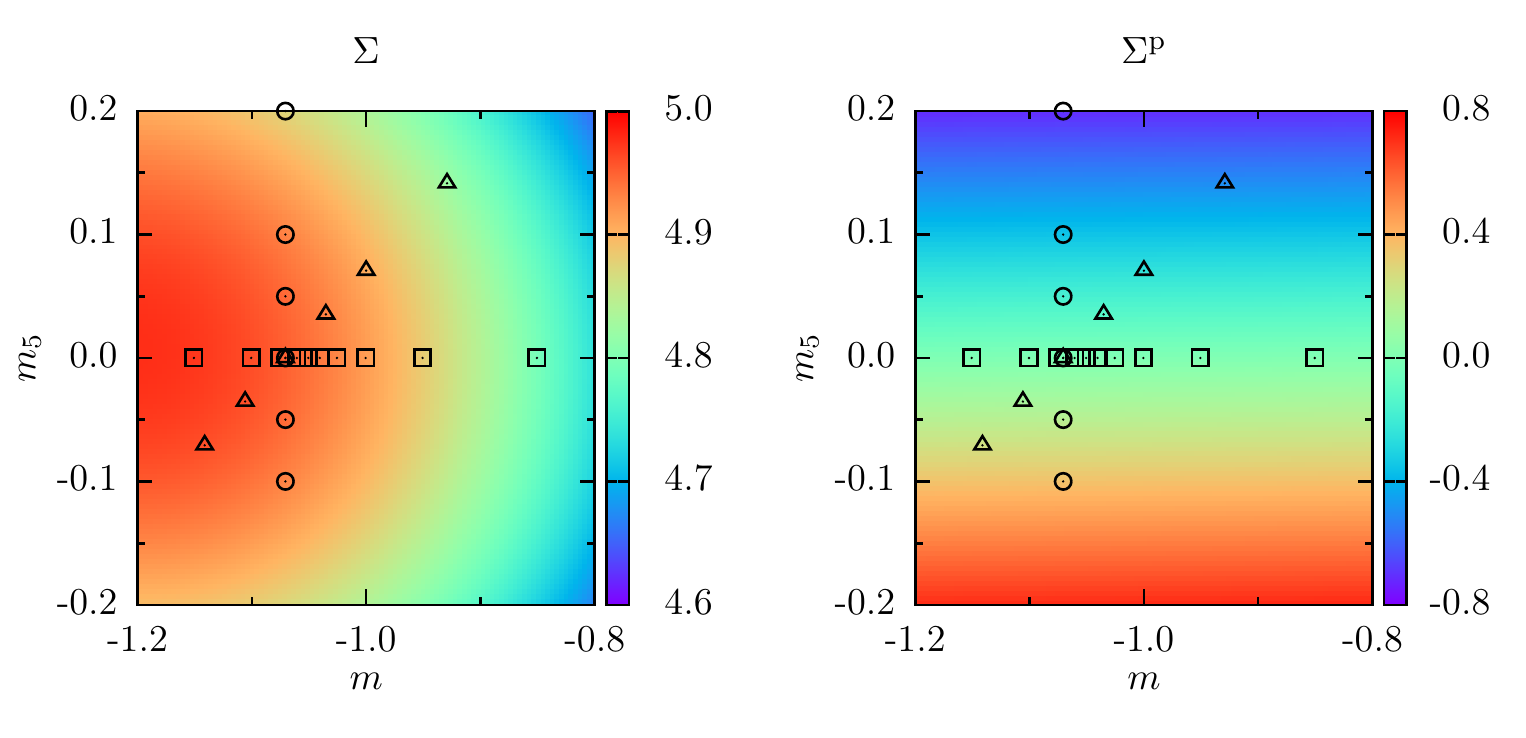}
	\caption{Left: The (unrenormalized) chiral condensate $\Sigma$ and fit to polynomial of second order. Right: The parity condensate $\Sigma^\text{p}$ and fit to a plane. The squares, triangles and circles correspond to $\alpha=0^\circ$, $45^\circ$ and $90^\circ$.}
	\label{fig:CondensatesParameterScan}
\end{figure}

\medskip
What remains is a cross-check of our findings for other observables. 
The chiral condensate \mbox{$\Sigma\sim\langle \bar{\lambda}(x)\lambda(x)\rangle$} 
and the parity condensate \mbox{$\Sigma^\text{p}\sim\langle\ii
 \bar{\lambda}(x)\gamma_5\lambda(x)\rangle$} are good candidates built
from the gluino field, see eqs.~\eqref{eq:ChiralCondensate} 
and \eqref{eq:ParityCondensate}.
A parameter scan of those condensates along the three \enquote{directions}, 
i.e., \mbox{$\alpha\in\{0^\circ,45^\circ,90^\circ\}$}, is shown in figure~\ref{fig:CondensatesParameterScan}. In the left panel we notice that 
the chiral condensate is maximal for $m_5=0$ and falls of 
as soon as $m_5\neq 0$.
Again we see a mirror symmetry in \mbox{$m_5\leftrightarrow-m_5$} as for 
the dominant mass contribution $d_\aPi$ of the adjoint pion.
The chiral condensate can be fitted well with a polynomial of second order 
while for the parity condensate a first order polynomial is sufficient.
The parity condensate is shown in the right panel of figure~\ref{fig:CondensatesParameterScan}.
Along $m_5=0$ it is zero, but if $m_5$ increases the condensate decreases 
linearly and vice versa.

Altogether we learn from figure~\ref{fig:CondensatesParameterScan}, for the condensates $\alpha=45^\circ$ is not a distinguished direction in the $(m,m_5)$ parameter plane. Only on-axis directions, that is $0^\circ$ and $90^\circ$, are special. However, we will see below that for a double-twisted Dirac operator, \mbox{$\alpha=\varphi=45^\circ$} is special also for the condensates, because then the condensates are equal (see figure~\ref{fig:CondensatesAbsDifference} and eq.~\eqref{diffw1} in section~\ref{ch:DoubleTwist}).

\subsection{Physical mesonic states}\label{ch:FullMesonic}

Up to now, only the connected contribution to the mesonic states $\aEta$ and $\aF$ has been analyzed. For a determination of their mass in the VY-supermultiplet, additional lattice calculations of the correlator's disconnected diagram is required, see eqs.~\eqref{eq:MesonicCorrelator} and \eqref{eq:VacuumContribution}. Compared to the connected contribution, the numerical effort for the disconnected part is rather large. Its magnitude is small and it comes with a large statistical uncertainty. Furthermore, there are two 
contributions: \mbox{$\langle \tr( \Gamma G_{xx} ) \tr( \Gamma G_{yy} ) \rangle_\mathcal{U}$} and \mbox{$\langle \tr( \Gamma G_{xx} )\rangle \langle \tr( \Gamma G_{yy} ) \rangle_\mathcal{U}$}, whose difference enters the correlator.  
High statistics is thus a prerequisite for a reasonable mass estimate not only for those two VY-supermultiplet partners.

For the twist angle $\alpha=45^\circ$, we have performed high-statistics calculations of both the connected and disconnected contribution for a fixed lattice coupling ($\beta=5.0$) and a single lattice size (\mlat). Thereby the mass and twisted-mass parameters were varied 
to extrapolate them afterwards towards their critical values (see table~\ref{tab:16_5.0}). 

\begin{figure}[tbp]
	\centering
	\includegraphics{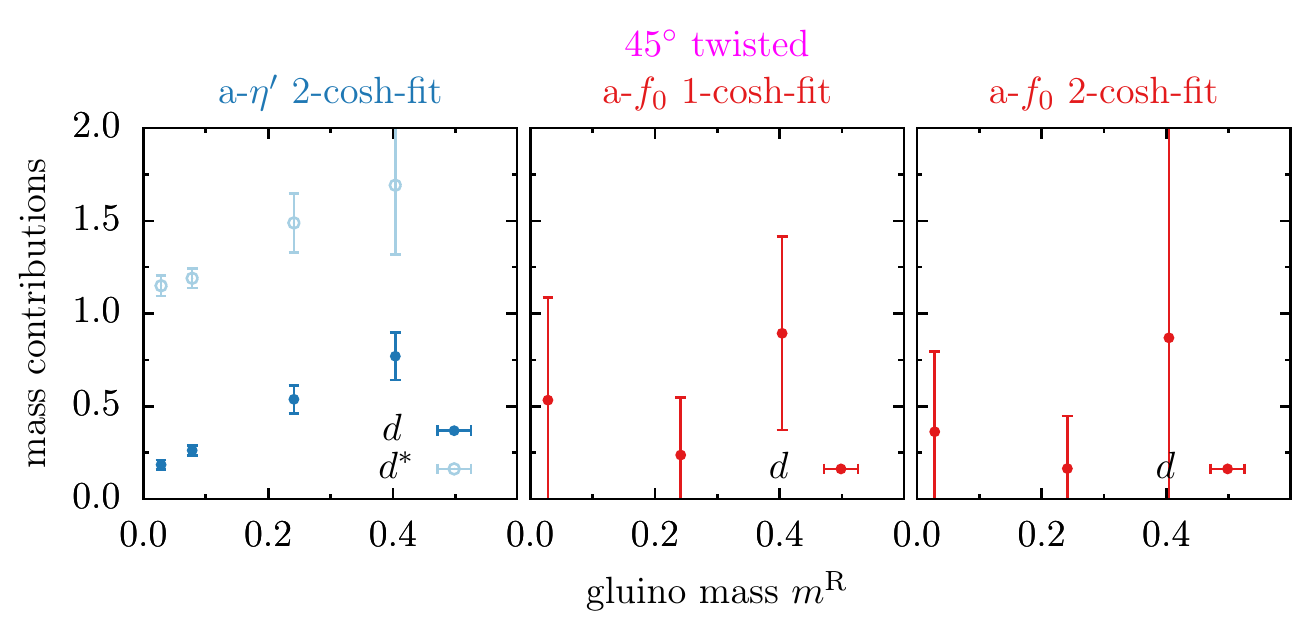}
	\caption{Masses for $\aF$ and $\aEta$ from a \mlat~lattice at $\beta=5.0$ and for a twist angle \mbox{$\alpha=45^\circ$}. Results are shown as a function of the renormalized gluino mass $m^\text{R}$. Left: lowest $d$ and next higher mass contribution $d^\ast$ of $\aEta$ extracted by a 2-cosh-fit. Center/Right: lowest contribution $d$ of $\aF$ extracted by a 1-cosh-fit/2-cosh-fit.  Two resp.\ four time slices are excluded in the correlator fit at the boundary resp.\ around the inner time slice, i.e., \mbox{$t_{\mathrm{cut}}=(2,4)$}.}
	\label{fig:EtaF16}
\end{figure}

Results for the dominant mass contribution (i.e., for the approximate ground state mass) for $\aEta$ and $\aF$ are shown in figure~\ref{fig:EtaF16} versus the renormalized gluino mass. They are obtained from fits of the lattice two-point correlators to the same 2-cosh ansatz as used above. For $\aF$, additional results from a 1-cosh fit are shown. For $\aEta$ there are also results for the next higher state, $d^*$, in figure~\ref{fig:EtaF16}.

While for $d_\aF$ the statistical fluctuations are large, both for the 2-cosh 
and 1-cosh fit, the results for $\aEta$ are much preciser such that a trend 
can be seen. For $\aEta$, $d$ and $d^*$ clearly decrease with $m^\textrm{R}$ and 
approach finite values at $m^\textrm{R}=0$. One would expect, the ground state 
mass of $\aEta$ near the critical point is approximately $0.2$ in lattice units, 
while the mass of the next higher state tends towards a value above $1$. 
For $d_\aF$, the lowest mass contribution is below $0.4$ within errors. 

Within statistical fluctuations we can hardly distinguish the correlation 
functions of the physical mesons $\aEta$, $\aF$ and from those of their partially 
quenched approximations $\aPi$ and $\aA$. Since the ground state masses of $\aPi$ and $\aA$ vanish in the chiral limit this would also be true for the physical meson masses. However
in section~\ref{ch:ChiralLimit}, we will revisit the chiral extrapolations of the would-be Goldstone bosons and physical mesons and argue that the physical masses remain massive in the chiral limit. 
In addition we include further states beside the two mesonic states considered here.

Although the $\aF$ correlator is noisy, in particular at the inner time slices, 
we try to get an approximate value for its first excited state at small $t$ 
where the signal-to-noise-ratio is better. Without knowing the exact ground 
state mass, we assume \mbox{$d_{\aF}= d_{\aEta}$} and fit 
\mbox{$C(t)-c_1\,\e^{-d t} \approx c_2\,\e^{-d^\ast t}$}. 
Repeating the same analysis with \mbox{$d_{\aF}=0.9\cdot d_{\aEta}$} 
and \mbox{$d_{\aF}=1.1\cdot d_{\aEta}$}, to account for a ground state 
mass error, we finally get \mbox{$d_{\aF}^\ast\approx \textrm{1.02}\pm\textrm{0.02}$}
at the bare mass parameter $m=\textrm{-1.0105}$.
This value is significantly lower than \mbox{$d_{\aEta}^\ast\approx \textrm{1.73}$} 
at the same parameters but still in the ballpark of allowed values, given 
all the other uncertainties and systematic errors (in particular due to 
the finite box size).

\subsection{Gluino-glue}\label{ch:GluinoGlue}

\begin{figure}[tbp]
	\centering
	\begin{minipage}[c]{0.47\textwidth}
		\hspace*{-2mm}\includegraphics{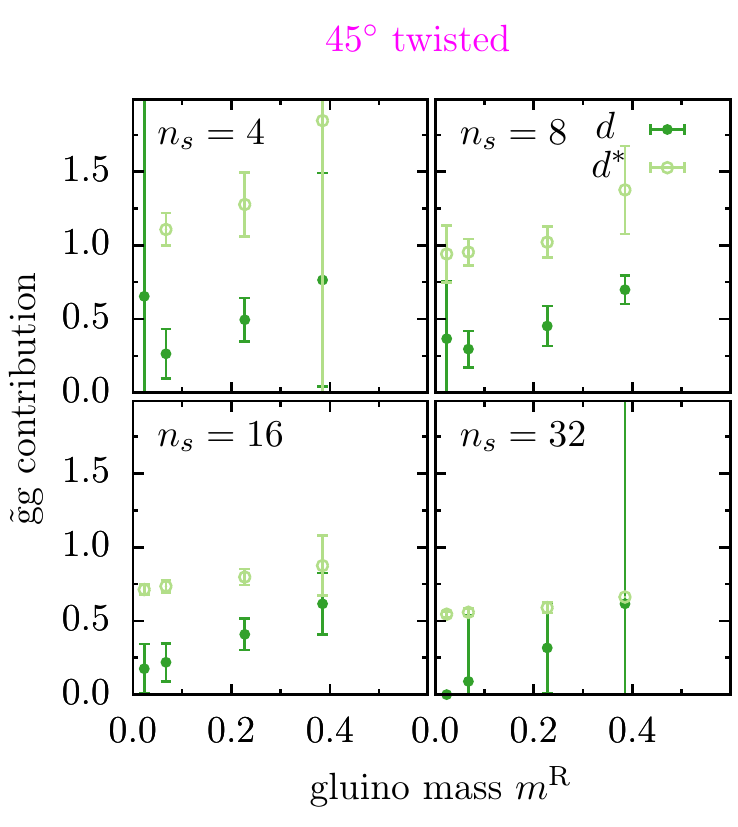}
		\caption{Symmetric gluino-glue from a \mlat~lattice and for twist angle \mbox{$\alpha=45^\circ$} versus the gluino mass $m^\text{R}$. The different panels show data for $n_s=4,8,16$ and $32$ stout smearing steps. For the fits, two time-slices have been excluded at both ends, $t_\mathrm{cut}=(2,2)$.}
		\label{fig:GluinoGlue}
	\end{minipage}
	\hfill
	\begin{minipage}[c]{0.47\textwidth}
		\includegraphics{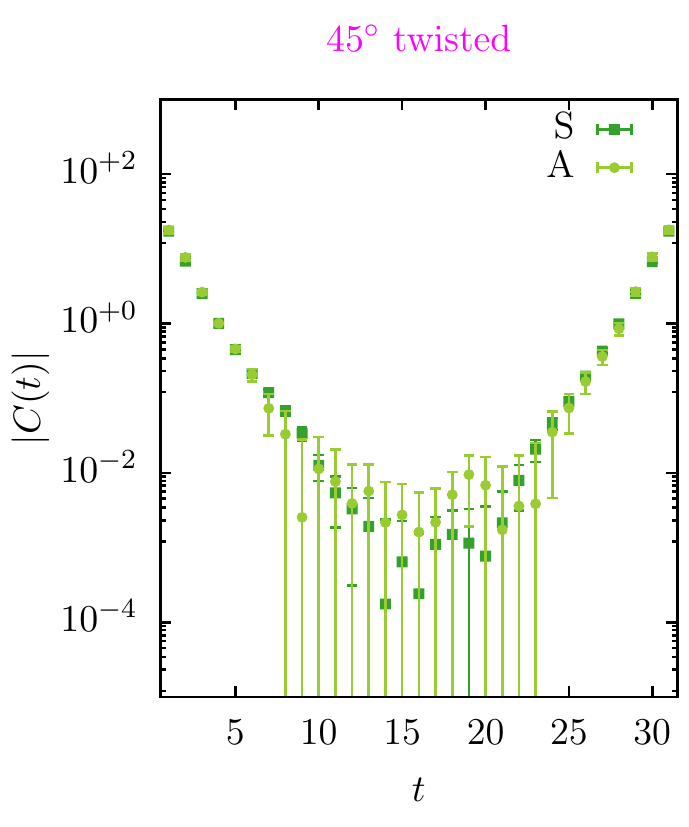}
		\caption{Comparison of time-symmetric (S) and time-antisymmetric (A) gluino-glue correlator at $m=\textrm{-1.0105}$, and for $n_s=8$ stout smearing steps. Absolute values are shown for a better comparison of the cosh- and sinh-shaped correlators.\newline}
		\label{fig:compareSA}
	\end{minipage}
\end{figure}

We continue with the third particle of the VY-super\-multiplet, the gluino-glue $\gluinoglue$. Figure~\ref{fig:GluinoGlue} shows its dominant mass contribution for different numbers of stout smearing steps, specifically for $n_s=4,8,16$ and 32. Gauge-link smearing smoothes the $t$-dependence of the correlators and suppresses contributions from excited states, if a sufficiently (but not too) large number of smearing steps are applied to the gauge links. Figure~\ref{fig:GluinoGlue} suggests that $n_s=8,\ldots,16$ smearing steps are optimal for our simulation parameters. For both cases the lowest mass contribution of the gluino-glue near the critical point is between 0.2 and 0.4 in lattice units, and between 0.7 and 0.9 for the next higher state. In comparison, fits to correlators for only $n_s=4$ smearing steps lead to higher uncertainties, while for $n_s=32$ some fits even fail.

In figure~\ref{fig:compareSA}, we compare (the absolute value of) the symmetric and antisymmetric correlators of the gluino-glue. Clearly, most data points lie on top of each other, but the noise of the antisymmetric correlator is increased at the inner time slices, where the sinh-shaped correlator crosses zero. Hence, no additional insight from the antisymmetric gluino-glue can be expected. Focusing on the cosh-shaped symmetric gluino-glue should be sufficient.

\subsection{Glueballs}\label{ch:Glueballs}
Before continuing with a chiral extrapolation of the VY-supermultiplet states in the 
next section, let us present some results for the FGS-supermultiplet. This multiplet
contains two glueballs and a further gluino-glueball, see table~\ref{tab:FGSmultiplet}. 
A lattice determination is thus numerically demanding. Enhanced gauge link 
fluctuations in the glueball interpolator fields require large ensemble 
sizes. A reasonable mass determination 
would exceed our computing time budget. Hence all results presented here 
are exploratory and preliminary.

\begin{figure}[tbp]
	\centering
	\includegraphics{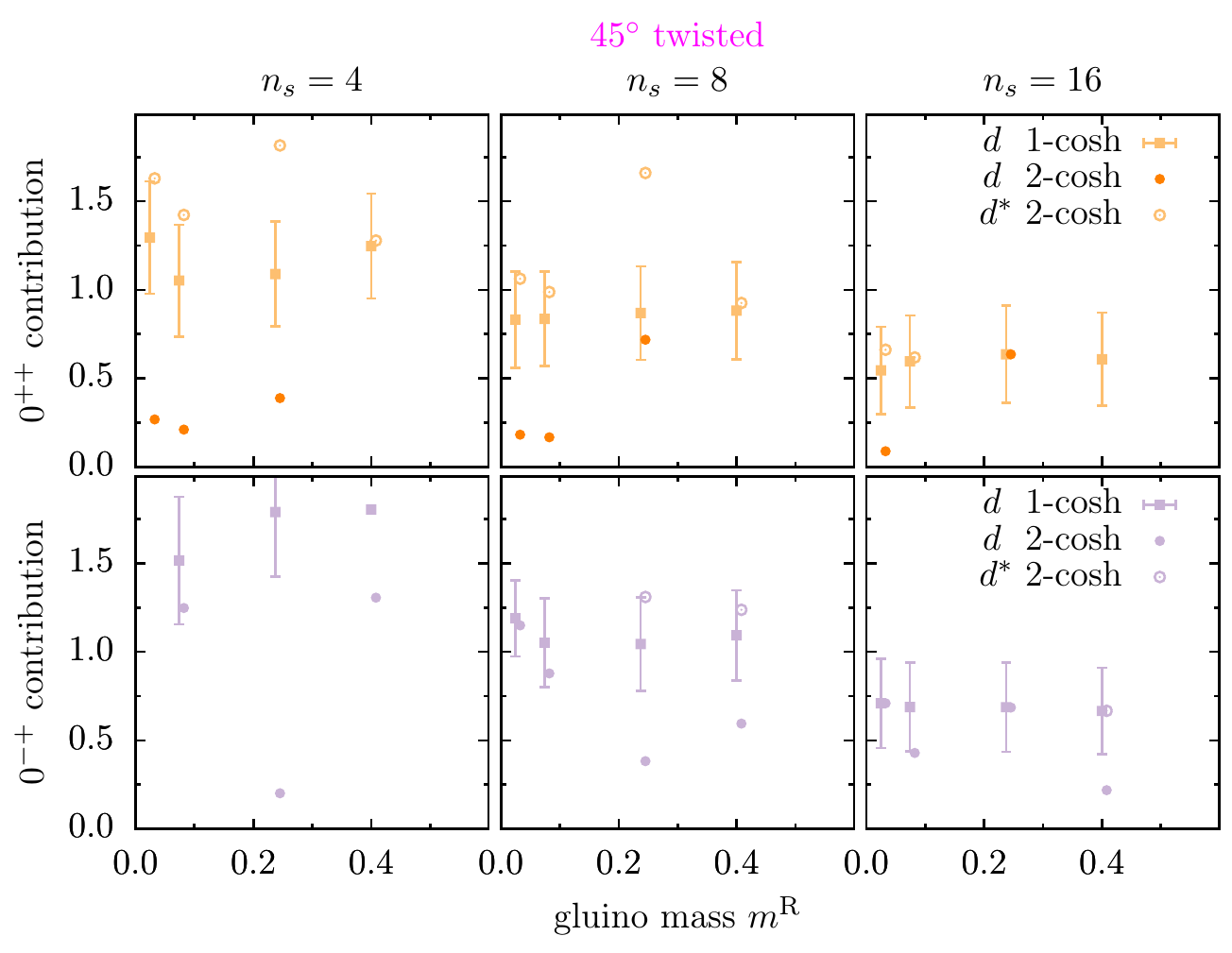}
	\caption{Top/Bottom: Scalar/Pseudoscalar glueball on the \mlat~lattice with twist angle $\alpha=45^\circ$ as a function of the gluino mass $m^\text{R}$. Data points are slightly displaced for better visibility. In the different panels, $n_s=\{4,8,16\}$ steps of stout smearing are applied to smooth the data. Both 1-cosh-fit and 2-cosh-fit results are shown for comparison. Two resp.\ four time slices are ignored in the correlator fit at the lattice boundary resp.\ in the center. For better clarity some (over-estimated) error bars are not shown.}
	\label{fig:Glueballs}
\end{figure}

Figure~\ref{fig:Glueballs} shows the dominant mass contributions for the glueballs with quantum numbers $0^{++}$ and $0^{-+}$. Within errors, $d$ does not depend on the renormalized gluino mass $m^\text{R}$. Similar holds for the next higher state of the scalar glueball. Extrapolated to the critical point, the scalar glueball is somewhat lighter than the pseudoscalar glueball, cf.\ top and bottom rows of figure~\ref{fig:Glueballs}. The extrapolated values at critical gluino mass is somewhere between 0.2 and 0.3 in lattice units. The mass of the next higher state of the scalar glueball extrapolates to a value somewhere between 0.6 and 1.3.

Comparing the three columns in figure~\ref{fig:Glueballs} we see that the number of stout smearing steps clearly affects $d$. Extrapolations to the critical mass are consistent with a horizontal line in all panels, that is a $m^\text{R}$ dependency is not resolvable, but the offset of each line depends on the number of smearing steps. For the chiral extrapolation of all multiplet states in the next section we will choose the results for $n_s=8$ stout smearing steps. Only for the lowest mass contribution of $0^{++}$, the results for $n_s=4$ are chosen, because the mass hierarchy is better seen (see top row of figure~\ref{fig:Glueballs}).

Fitting the dominant mass contribution of the $0^{-+}$, and that of the next higher state, is difficult, even with a 2-cosh-fit ansatz (see bottom row of figure~\ref{fig:Glueballs}). The small lattice volume does not allow for a reasonable determination of the ground state mass. Therefore, only one (excited) contribution has been determined with a value between 0.7 and 1.5 in lattice units, depending on the number of stout smearing steps.
In~\cite{Ali:2019gzj,Ali:2019agk} it was reported that the lowest state of the
pseudoscalar glueball is comparable with the first excited states of mesonic states 
and the gluino-glue. This agrees with our observations.

\subsection{Chiral limit}\label{ch:ChiralLimit}

To connect lattice results with the supersymmetric continuum theory, first an 
extrapolation to the critical point and then to the continuum limit should be
performed. In what follows, all previously discussed results (see
 sections~\ref{ch:FullMesonic}, \ref{ch:GluinoGlue} and \ref{ch:Glueballs}) 
will be extrapolated to the critical point where the renormalized gluino mass 
vanishes at fixed lattice spacing. In the previous sections this extrapolation 
has been discussed for the individual states already. The focus here is on a 
comparison of the extrapolated values for all supersymmetric partners of a 
multiplet, in particular if they coincide within errors.

The leading order of chiral perturbation theory suggests that the residual gluino mass $m^\text{R}$ is given by squared mass of the would-be Goldstone bosons, i.e.\ \mbox{$m^\text{R}\propto m_{\aPi}^2$}. In~\cite{Evans:1997jy,Bali:2016lvx,Aoki:2005mb,Farchioni:2007dw} it has been argued that the leading correction to non-zero meson and baryon masses in the chiral limit is also proportional to $m_{\aPi}^2$ such that we assume a linear $m^\text{R}$ dependency in the 
extrapolation to the chiral point. Hence we will obtain non-zero masses for the 
physical mesons -- in contrast to the partially quenched approximations in the chiral limit --
although at our finite values of $m^\text{R}$ the masses are hardly distinguishable.

\begin{figure}[tbp]
	\centering
	\includegraphics{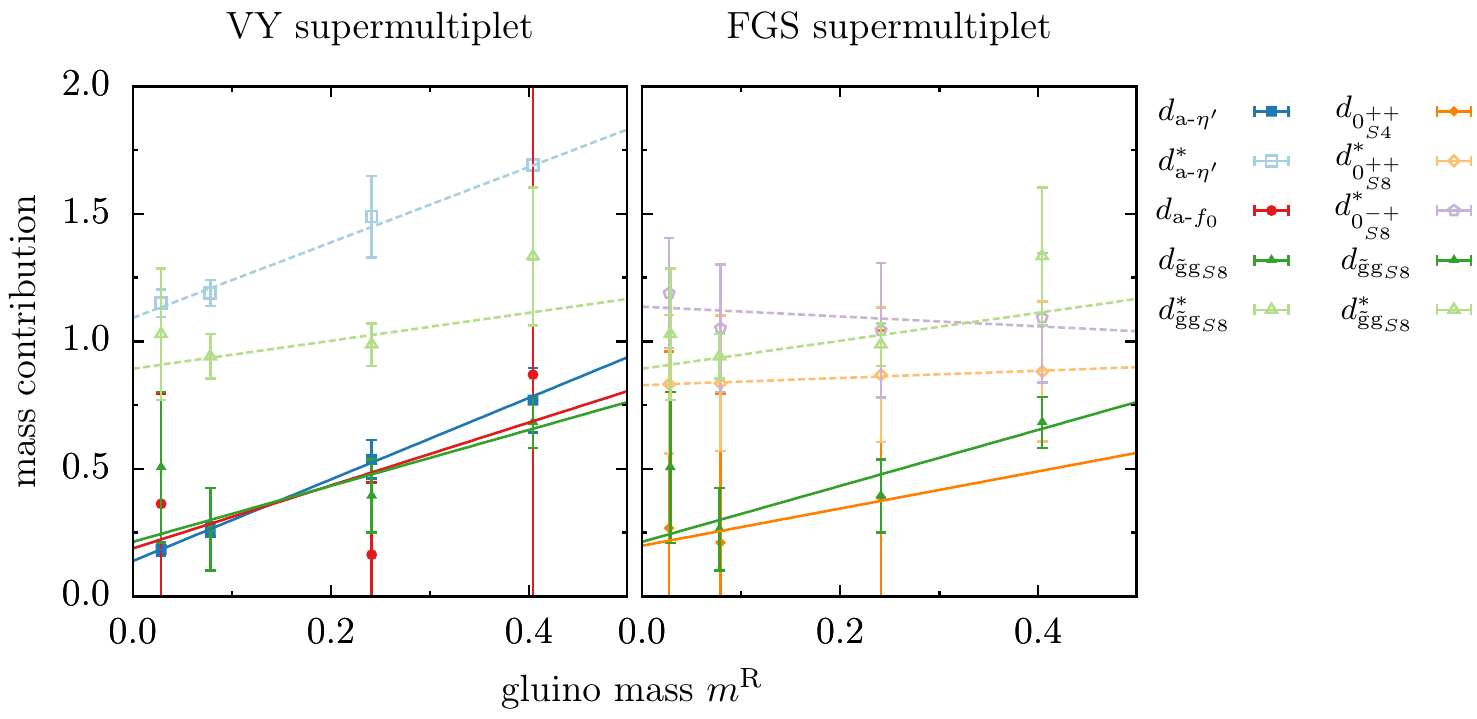}
	\caption{Left/Right: Chiral extrapolation of the VY/FGS-supermultiplet. Each particle is depicted with a different color. Solid/Dashed lines are linear fits to the lowest/next higher mass contribution. The amount of smearing steps is indicated in the indices, \eg $\gluinoglue_{S8}$ stands for 8 levels of stout smearing. For the gluino-glue the symmetric operator is considered as discussed in section~\ref{ch:GluinoGlue}. Two resp.\ four time slices are ignored in all correlator fits at the lattice boundary resp.\ in the center. All fits are 2-cosh except $0^{++}_{S8}$ and $0^{-+}_{S8}$, which are fitted to a single cosh.}
	\label{fig:ChiralExtrapolation}
\end{figure}

For the VY-supermultiplet, the linear extrapolations are depicted in the left 
panel of figure~\ref{fig:ChiralExtrapolation}, and the corresponding values 
are given in table~\ref{tab:ChiralFits}. We see that the lowest mass contributions 
of $\aEta$, $\aF$ and $\gluinoglue_{S8}$ (this index indicates the usage of 8 stout smearing steps)
are degenerated within errors. For the 
next higher state of the VY-supermultiplet, the situation is less clear. 
Nonetheless, a tendency for a mass degeneracy is seen which may be manifest 
in the continuum limit. Possibly the relatively small lattice size causes the second 
excited state to superpose with the first, resulting in larger contributions 
to $d_\aEta^\ast$ which we cannot resolve. Conversely, smearing the gluino-glue 
operator may have overly dampened the first excited state $d_\gluinoglue^\ast$ 
such that its mass is underestimated.

\begin{table}[tbp]
	\centering
	\renewcommand{\arraystretch}{1.2}
	\begin{tabular}{!{\vrule width 1pt}lllll!{\vrule width 1pt}}
		\noalign{\hrule height 1pt}  
		state & $d$ & $s$ & $d^\ast$ & $s^\ast$ \\ 
		\hline
		${\aEta}$ & $0.14 \pm 0.01$ & $1.60 \pm 0.04$ & $1.09 \pm 0.02$ & $1.48 \pm 0.23$ \\ 
		${\aF}$ & $0.19 \pm 0.12$ & $1.23 \pm 1.49$ & -- & --  \\ 
		${\gluinoglue_{S8}}$ & $0.21 \pm 0.12$ & $1.09 \pm 0.40$ & $0.89 \pm 0.09$ & $0.55 \pm 0.43$ \\ 
		${0^{++}_{S8}}$ & $0.20 \pm 0.05$ & $0.73 \pm 0.32$ & $0.83 \pm 0.05$ & $0.14 \pm 0.02$ \\ 
		${0^{-+}_{S8}}$ & -- & -- & $1.14 \pm 0.06$ & $-0.19 \pm 0.25$ \\ 
		\noalign{\hrule height 1pt}  
	\end{tabular} 
	\caption{Results of the linear fits. The lowest mass contribution~$d$, the next higher mass contribution~$d^\ast$, the corresponding slopes~$s$ resp.\ $s^\ast$ and their fit errors are rounded to 2 digits.}
	\label{tab:ChiralFits}
\end{table}

The right panel of figure~\ref{fig:ChiralExtrapolation} shows the 
extrapolation of the FGS-supermultiplet states. Looking at the ground state, 
the scalar glueball $0^{++}$ shows a clear mass degeneracy with the gluino-glue.
It is as heavy as the $\aF$ of the VY-supermultiplet, but slightly heavier 
than the $\aEta$ state. A prediction which of the two multiplets is the 
lightest in the continuum limit is not possible with the present data.
In the excited spectrum, $0^{++}_{S8}$, $\gluinoglue_{S8}$ and $0^{-+}_{S8}$ 
lie in the interval $[0.8,1.2]$.
If those states all belong to the first excitation, or if this excitation in 
fact is a superposition of all higher states, cannot be resolved. Simulations in
larger volumes are required to address this in a reasonable manner.

We identified the lowest contribution of the $0^{-+}$ glueball as its first excited state.
This is in accordance to~\cite{Bergner:2015adz}, where $m_{0^{++}}^1 \approx m_{0^{-+}}^0$ was found.
In another study~\cite{Ali:2019gzj}, results from a lattice calculation using the variational method are discussed. The authors found that the $\aEta$ and $0^{-+}$ operators do not mix in the variational basis, even though both lead to the same masses for the exited states when analyzed individually.
In~\cite{Ali:2019agk}, in which SU(3) \SYM theory has been addressed, the scalar glueball and $\aF$ interpolation operators were combined into a variational basis. Both operators showed a good overlap with the lowest state and mixing occurs.
In the pseudoscalar channel, the lowest state was dominated by the $\aEta$ operator while the signal for the $0^{-+}$ operator was comparably small.

To conclude, our spectroscopic results of the VY- and FGS-supermultiplet with the twisted Wilson Dirac operator demonstrate that a mass degeneracy of the ground states can be observed.
In future studies, the first excited states should be refined and 
with a continuum extrapolation the question, which of the 
supermultiplets is the lightest, should be addressed.

\subsection{Chiral anomaly and relevance of Wilson term} \label{ch:DoubleTwist}
\begin{figure}[tbp]
	\centering
	\hspace*{15mm}\includegraphics{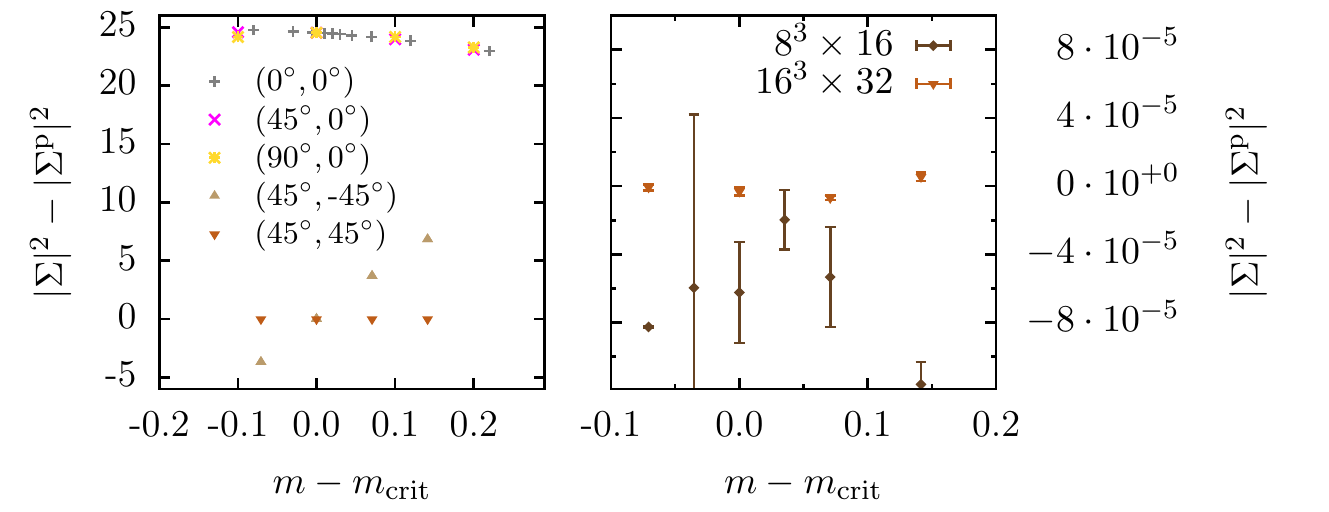}
	\caption{Difference of the absolute values squared of the chiral condensate~$\Sigma$ and the parity condensate~$\Sigma^\textrm{p}$. For $\varphi=0^\circ$ the chiral condensate dominates this value while for \mbox{$\varphi=45^\circ$} the contributions of the chiral condensate and the parity condensate are approximately the same. Left:~Different twist angles $(\alpha,\varphi)$ on the \mlat~lattice. Right:~Different lattice sizes for \mbox{$\alpha=\varphi=45^\circ$}.}
	\label{fig:CondensatesAbsDifference}
\end{figure}

Disregarding a potential anomaly due to a
non-invariance of the measure
a twist of both the mass term and Wilson term 
with the same angle
can be undone by a chiral rotation~\eqref{eq:ChiralRotation}
which rotates the interpolating  operators.
So far we have investigated mesonic correlators of the type~\mbox{$\langle \bar{\lambda}_x\Gamma^1\lambda_x \bar{\lambda}_{x^\prime}\Gamma^2\lambda_{x^\prime} \rangle$}.
Above we have compared connected and disconnected contributions to 
these correlators. Thereby one should keep in mind that
the latter depend via the condensates
very sensitive  on external conditions.
To quantify a possible anomaly and at the same time 
study the quality of the $45^\circ$-twist, we now 
consider the chiral and parity condensate, \ie condensates of type \mbox{$\sum_x \langle \bar{\lambda}_x\Gamma\lambda_x\rangle$} with \mbox{$\Gamma=1,\gamma_5$}.
Under a chiral transformation~\eqref{eq:ChiralRotation}, the 
doublet of bilinears is rotated, see~\eqref{eq:BilinearDoublet},
and so are the condensates (\ref{eq:ChiralCondensate}) and
(\ref{eq:ParityCondensate}):
\begin{align}
	\Sigma(\alpha)&=
	\cos(\alpha)\Sigma + \sin(\alpha)\Sigma^\text{p}\,,\nonumber\\
	\Sigma^\text{p}(\alpha) &=
	\cos(\alpha)\Sigma^\text{p} - \sin(\alpha) \Sigma\,.
\end{align}
Hence the sum \mbox{$|\Sigma(\alpha)|^2 + |\Sigma^\text{p}(\alpha)|^2 = |\Sigma|^2 + |\Sigma^\text{p}|^2$}
is independent of $\alpha$ and
the difference
\begin{align}
	|\Sigma(\alpha)|^2-|\Sigma^\text{p}(\alpha)|^2 =& 
	|\Sigma|^2 \big( \cos^2(\alpha) - \sin^2(\alpha) \big) - |\Sigma^\text{p}|^2 \big( \cos^2(\alpha) - \sin^2(\alpha) \big)
	\label{diffw1}
\end{align}
should be zero at \mbox{$\alpha=\pm 45^\circ$}.
For \mbox{$\varphi=0^\circ$} eq.~\eqref{diffw1} measures 
both the breaking of chirality by the measure and by the
irrelevant Wilson term. If instead the difference is measured for the double-twist
\mbox{$\alpha=\varphi=45^\circ$} then the difference is only due to
a potential non-invariance of the measure.
This way we can disentangle the breaking of chirality by the 
Wilson term and the measure.
The right panel of figure~\ref{fig:CondensatesAbsDifference}
is compatible with the chiral invariance of the latter. 
On the \slat~lattice the deviation of
\mbox{$|\Sigma|^2-|\Sigma^\text{p}|^2$} from zero is smaller than $10^{-4}$ 
and on the \mlat~lattice even below $10^{-5}$.
A possible deviation is so small that we see
no symmetry breaking induced by a non-invariant measure. In subsequent
studies a perturbative lattice calculation should support this finding.

The similarity of the condensates for a double-twist \mbox{$\alpha=\varphi=45^\circ$} 
can be used to our advantage when we analyze the physical mesonic states.
Their disconnected contributions depend on the chiral condensate 
resp.\ the parity condensate. In section~\ref{ch:ChiralTrafosLatticeExpValues} 
we argued that $\aEta$ and $\aF$ are identical when the spinors are 
rotated with $45^\circ$.
Without twisting the Wilson term, that is for \mbox{$\varphi=0^\circ$}, the numerical data presented in section~\ref{ch:ParameterScan} show that the connected part of $\aEta$ and $\aF$ 
agree.
At the same time, the chiral 
condensate~$\Sigma$ is much bigger than the parity 
condensate~\mbox{$\Sigma^\text{p}\ll 1$}, see left 
part of figure~\ref{fig:CondensatesAbsDifference}.
It follows that in the $\aF$ correlator large numbers of the order 
\mbox{$\langle \tr( \Gamma G_{xx} )\rangle \langle \tr( \Gamma G_{yy} ) 
	\rangle \sim |\Lambda| \cdot \Sigma^2 $} must be subtracted unlike for 
the $\aEta$. This explains the unequal noise in those two 
correlators at \mbox{$\alpha=45^\circ$} -- even though we would expect 
them to be equal according to section~\ref{ch:ChiralTrafosLatticeExpValues}.

Now, with a rotation of the mass term and the Wilson 
term, \ie $\alpha=\varphi=45^\circ$, also the disconnected contributions 
of those two mesonic states match.
This implies an even better degeneracy of the $\aEta$ and $\aF$.
A compromise would be the choice \mbox{$\alpha=45^\circ=-\varphi$}, where the difference of the condensates is significantly lower than in the scenario with $\varphi=0^\circ$, see left panel of figure~\ref{fig:CondensatesAbsDifference}.
Additionally this difference shrinks linearly towards the critical point and for \mbox{$\alpha-\varphi=90^\circ$} discretization improvements of $\mathcal{O}(a)$ are possible as discussed in section~\ref{ch:EigValues}.

Altogether, there are several interesting setups $(\alpha,\varphi)$ for future investigations compared to the untwisted Wilson Dirac operator, which all lead to an improved mass degeneracy of the chiral partners:
\begin{enumerate}
	\item $(45^\circ, 0^\circ)$: equal connected contributions to $\aEta$ and $\aF$, $\mathcal{O}(a)$ errors reduced.
	\item $(45^\circ, 45^\circ)$: equal connected and disconnected contributions to $\aEta$ and $\aF$. Note that this choice amounts to a redefinition of the observables.
	\item $(45^\circ, \textrm{-}45^\circ)$: equal connected contributions to $\aEta$ and $\aF$, disconnected contributions become equal as the critical point is approached, $\mathcal{O}(a)$ improvement.
\end{enumerate}

\subsection{Sign of the Pfaffian}\label{ch:Pfaffian}

In order to have a positive Boltzmann weight in the path integral, 
the Pfaffian must be positive. Otherwise our lattice calculations 
may suffer a sign problem. In the continuum, the Pfaffian of \SYM 
theory is real, but our twisted lattice Dirac operator may have
a complex Pfaffian. To check the severeness 
of that problem additional lattice calculations of the Pfaffian on 
lattices up to a size of $7^3\times14$ have been performed. 
Since the computational costs scale as $\mathcal{O}(N^3)$ and the 
memory requirement as $\mathcal{O}(N^2)$ with the size $N$ of the 
Dirac matrix, the explicit calculation of the Pfaffian with the optimized serial algorithm~\cite{Wimmer_2012} was only performed for lattice sizes from $2^3\times4$ to $7^3\times14$.

\begin{figure}[tbp]
	\centering
	\begin{minipage}[c]{0.48\textwidth}
		\hspace*{5mm}\includegraphics{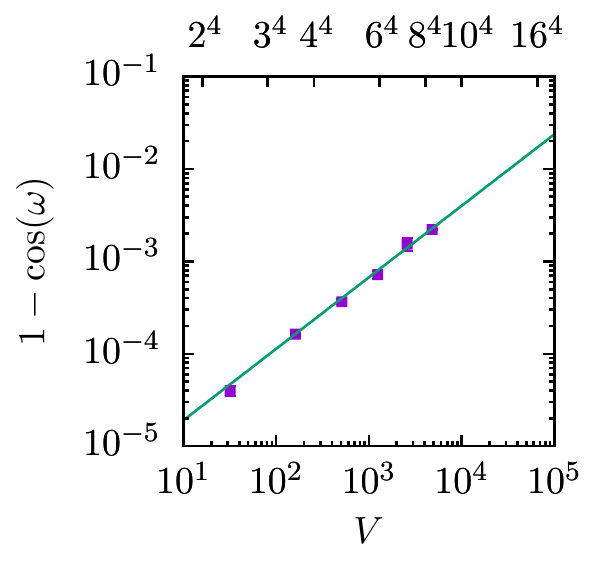}
	\end{minipage}
	\hfill
	\begin{minipage}[c]{0.48\textwidth}
		\includegraphics{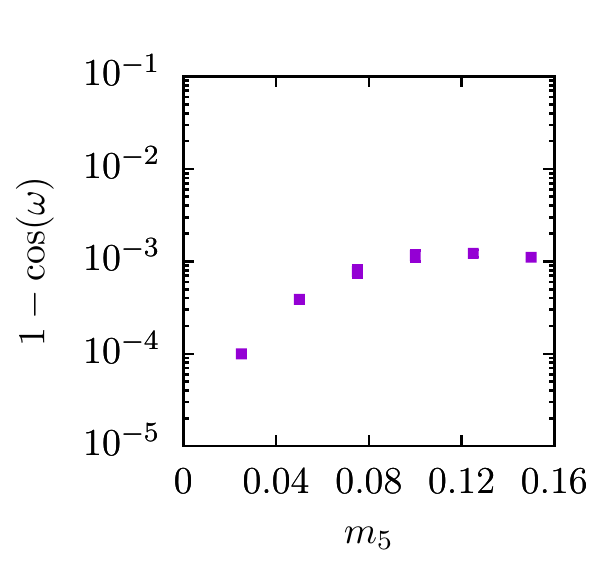}
	\end{minipage}
	\caption{Left: Phase of the Pfaffian for different lattice sizes ranging from \mbox{$2^3\times4$} to \mbox{$7^3\times14$}. The green line is an exponential fit to extrapolate the results to the lattice size \mbox{\mlat}. Right: Phase of the Pfaffian for different values of \mbox{$m_5\in[0,0.15]$}. When approaching the critical point at \mbox{$m_5=0$}, the phase of the Pfaffian decreases. Error bars are mostly smaller than the symbol size.}
	\label{fig:Pfaffian}
\end{figure}

The left panel of figure~\ref{fig:Pfaffian} shows the phase $\omega$ 
of \mbox{$\pf(\C D_\text{W}^{\text{mtw}})=|\C D_\text{W}^{\text{mtw}}|\cdot\e^{\ii\omega}$} 
for different lattice sizes and simulation parameters: \mbox{$\beta=5.4$} 
and \mbox{$(m,m_5)=(\textrm{-0.85},\textrm{0.1})$}, where \mbox{$d_\aPi\approx0.60$}. 
Extrapolated to the typical lattice size of our calculations, \mlat, 
we find the phase remains small: \mbox{$1-\cos(\omega) < \textrm{0.035}$}. 
That is, we expect no significant sign problem for our calculations. 
Furthermore we find that the phase becomes smaller towards the critical 
point, see right plot of figure~\ref{fig:Pfaffian}.

\subsection{Multigrid acceleration} \label{sec:Multigrid}
When calculating correlator functions, a large amount of computation time is spent in the inversion of the Wilson Dirac operator.
In lattice QCD, the implementation of multigrid methods has led to a significant speed-up.
Their strength is the separate treatment of high and low modes by an alternating 
application of a domain decomposition smoother and a coarse-grid correction.
We adjusted the adaptive aggregation-based domain decomposition multigrid \mbox{(DD$\alpha$AMG)} library~\cite{Alexandrou:2016izb,DDalphaAMG} to the adjoint representation of \SYM theory and used the DD$\alpha$AMG inverter when calculating correlators or condensates. This turned out as a valuable investment, because it has allowed us to significantly reduce the statistical noise for all results presented in the sections~\ref{ch:FullMesonic} to \ref{ch:ChiralLimit} by using a large number of stochastic estimators and point sources. This would have been impossible with the commonly used conjugate gradient (CG) algorithm, in particular given our limited CPU time budget. 

To illustrate the performance boost by the \mbox{DD$\alpha$AMG} inverter, we perform a benchmark study with the following setup:
Inversion precision $10^{-12}$, two multigrid levels, block size~$2^4$, mixed precision and the solver combination FGMRES with red-black Schwarz.

\begin{figure}[tbp]
	\centering
	\includegraphics{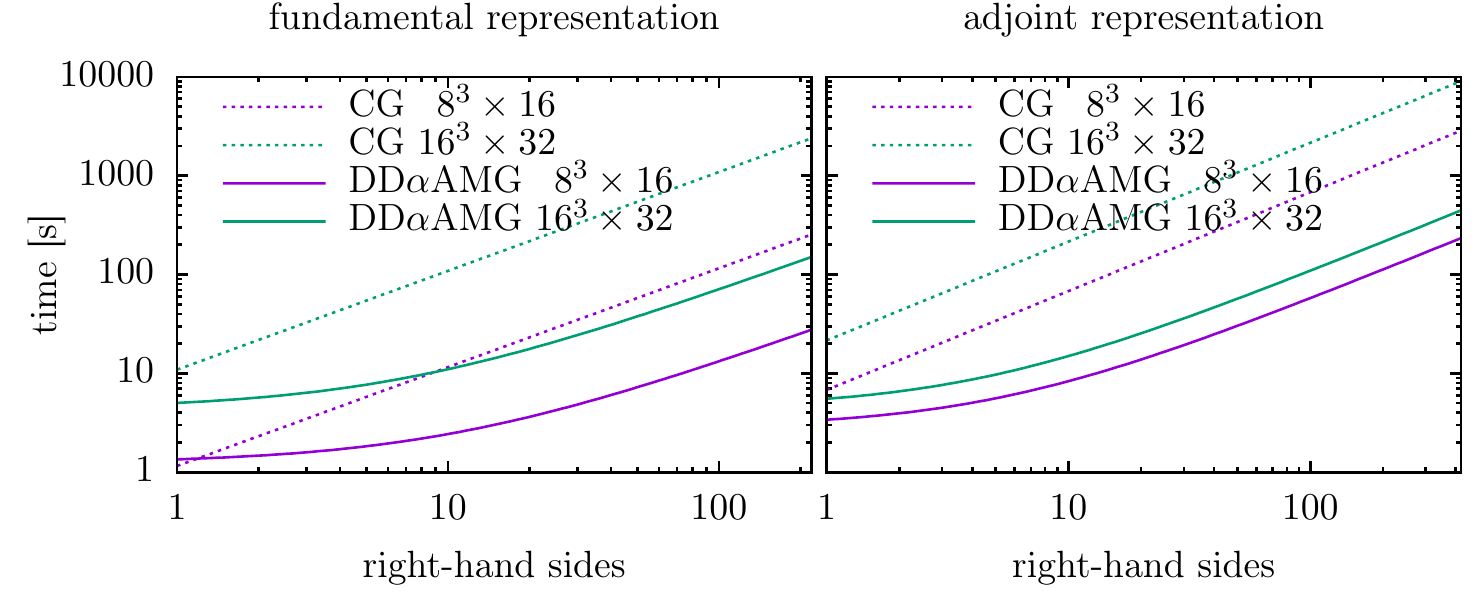}
	\caption{Left/Right: Measured time in seconds to invert the Wilson Dirac operator for different numbers of right-hand sides in the fundamental/adjoint representation of SU(3). The different colors correspond to the lattice sizes \slat\ resp.\ \mlat. Solid (dotted) lines are for the DD$\alpha$AMG (CG) inverter. Note the double-logarithmic scale.}
	\label{fig:DDbenchmarkRHScombined}
\end{figure}

Figure~\ref{fig:DDbenchmarkRHScombined} shows the timings for inversions of the Wilson Dirac operator for the CG and the \mbox{DD$\alpha$AMG} inverter. The left panel is for the Wilson Dirac operator in the fundamental representation, the right for the adjoint representation of SU(3). In both cases up to 100 stochastic estimators and 5 point sources are considered.
For comparison, the timings for two different lattice sizes, \slat\ and \mlat, are shown. We see that 
on the \mlat~lattice the DD$\alpha$AMG solver is always faster than the CG algorithm.
Only for the fundamental representation with a single right-hand side the CG solver is slightly faster. This is because of the time needed for the DD$\alpha$AMG setup. However, if many different right-hand sides are calculated 
this setup time becomes negligible.
Especially, on large lattices and for the adjoint representation the DD$\alpha$AMG algorithm yields a significant performance gain and is much faster than the CG. For this case our benchmark
study reveals a speed-up factor of 20.
Additionally, the DD$\alpha$AMG can reduce the critical slowing down 
near the critical point.

%% file: summary.tex
\section{Summary and outlook}\label{sec:Summary}

In this work we have introduced, analyzed and applied a new type of Wilson Dirac operator for 
lattice calculations of \mbox{$\mathcal{N}=1$} supersymmetric \mbox{SU(3)} Yang-Mills theory. 
Inspired by twisted-mass lattice QCD and simulations of 
lower-dimensional supersymmetric theories we have added a twisted mass term to 
the fermionic lattice action and interpreted it as a deformation whose 
parameter requires tuning.
With analytical arguments we showed that 
at $45^\circ$~twist the correlators of the chiral partners in the Veneziano-Yankielowicz supermultiplet are identical and additionally the 
discretization artifacts are reduced at tree level.
With the help of lattice simulations we have demonstrated that this 
particular twist angle leads to an improvement of the mass degeneracy of the 
mesonic chiral partners at finite lattice spacing. Consequently, 
chiral symmetry as well as supersymmetry are improved reducing the 
distance to the supersymmetric continuum limit.

In the exploratory simulations presented in this work
the lattice parameters were not optimally 
chosen such that some lattice results are
afflicted with non-negligible volume artifacts. Nonetheless, on a 
qualitative level our findings presumably will not change and we leave 
it to forthcoming lattice
studies to verify them on larger volumes. Those studies should start 
at smaller (inverse) gauge couplings $\beta$ to increase 
the physical box size. 
Depending on the available computer time, a larger lattice size 
may be helpful to reduce the statistical noise. Furthermore, a 
combination of ensembles with different couplings should enable us
to better extrapolate to the continuum limit and to determine the physical 
masses of the Veneziano-Yankielowicz and Farrar-Gabadadze-Schwetz 
supermultiplets. 

After twisting the mass term only, we also analyzed the double-twist 
scenario with a twist angle $\alpha$ for the mass term and another 
angle $\varphi$ for the Wilson term.
Preliminary results of the two condensates suggest 
that no anomaly occurs at $\alpha=\varphi=45^\circ$.
We observed that a double-twist can reduce the numerical difference 
of the disconnected contributions between the chiral 
partners $\aEta$ and $\aF$.
Optimally chosen twist angles reduce lattice artifacts such that
the double-twist approach provides a promising improvement of
lattice Super-Yang-Mills theory and could be used in future lattice simulations.

Much improvement has been achieved with an adapted DD$\alpha$AMG multigrid algorithm for fermions in the adjoint representation.
In a benchmark study, a speed-up factor of 20 has been achieved.
This way, we could reduce our computation cost considerably
and at the same time increase the number of stochastic estimators 
and point sources.

Ultimately, dynamical supersymmetric quarks (squarks) should be added 
to perform lattice studies for Supersymmetric Quantum Chromodynamics 
(aka.\ Super-QCD). First steps in that direction are presented in
\cite{Costa:2017rht,Wellegehausen:2018opt,Bergner:2018znw}. One-flavor
Super-QCD with Wilson fermions has nine relevant operators, but as 
demonstrated in \cite{Wellegehausen:2018opt}, certain properties of 
the one-loop potential of the squark field may help to
fine-tune these parameters.
With respect to the $R$-symmetry of Super-QCD, the twisted formulation provides a variety of options. One possibility is to twist only the gluinos or to twist gluinos and squarks in the same resp.\ 
the opposite direction. Upcoming numerical studies may clarify the influence of the Yukawa-type interaction between the two fermionic fields on the masses of bound states.

\acknowledgments

MS likes to thank Georg Bergner for helpful discussions, especially on the mass of the pion as presented in appendix~\ref{ch:PionProof}. The authors gratefully acknowledge the Leibniz Supercomputing Centre (LRZ, \href{http://www.lrz.de}{\url{www.lrz.de}}) for granting computer time on SuperMUC and SuperMUC-NG for this project (pr48ji).
Additional computer time on the DFG-funded Ara cluster at the Friedrich-Schiller-University Jena is acknowledged. AS acknowledges support by the BMBF under Grant No.\ 05P15SJFAA (FAIR-APPA-SPARC) and by the DFG Research Training Group GRK1523. MS and AW have been supported by  the Deutsche Forschungsgemeinschaft (DFG) under GrantNo. 406116891 within the Research Training Group RTG2522/1.

%% file: appendix.tex
\appendix
\section{Why the pion is the lightest mesonic state}\label{ch:PionProof}

For the benefit of the reader we elaborate on an argument 
put forward by Weingarten
\cite{Weingarten:1983uj} which makes clear that 
the pion is the lightest mesonic state on the 
lattice
(see the texts \cite{Kilcup:1995ww,Shuryak:2004pry}).
Clearly, if two (connected) correlators obey for large enough $x$ (where excited states do not contribute) the inequality
\begin{align}
\vert C_1(0,x)\vert > \vert C_2(0,x)\vert,\qquad x\gg1\,,
\end{align}
then the exponential decay of $C_2$ is faster and thus the ground state mass of the corresponding particle is heavier.
Starting from a generic mesonic creation and annihilation operator with mass-degenerated fermions $\psi_1$ and $\psi_2$, the mesonic correlator is
\begin{align}
C(0,x)=\langle \bar{\psi}_1(0) \Gamma \psi_2(0)\,\bar{\psi}_2(x) \tilde{\Gamma} \psi_1(x) \rangle &=
\langle \tr\big( G(0,x) \Gamma G(x,0) \tilde{\Gamma} \big) \rangle_\mathcal{U} \nonumber \\
&= \langle \tr\big( G(0,x) \Gamma \gamma_5 G^{\dagger}(0,x) \gamma_5 \tilde{\Gamma} \big) \rangle_\mathcal{U}\,,
\end{align}
where the subscript $\mathcal{U}$ indicates the average with respect to gluonic
degrees of freedom and the trace is in color and spinor space.
In the last step we used the $\gamma_5$-hermiticity
(which holds for untwisted fermions) and that the Green function can be written as
\[ 
G(x,y)=
\gamma_5\big\langle x\big\vert\gamma_5 \frac{1}{D}\gamma_5\big\vert y\big\rangle\gamma_5=
\gamma_5\big\langle x\big\vert \frac{1}{D^\dagger}\big\vert y\big\rangle\gamma_5=
\gamma_5 G^{\dagger}(y,x)\gamma_5\,,
\]
where the adjoint is in spinor and color space only. In the following
$x$ is fixed and we are dealing with a matrix problem in
color and spinor space only. We recall the Frobenius scalar product
of two matrices and the Frobenius norm of a matrix,
\[
(A,B)=\tr(A^\dagger B)\quad \text{with}\quad
\Vert A\Vert=\sqrt{(A,A)}\,.
\]
They fulfill all properties of a scalar product,
in particular
\[
\vert (A,B)\vert\leq
\|A\|\,\|B\|\,.
\]
Since the Hermitean $\gamma_5$ squares to $\id$ we have
(we set $G(0,x)=G_x$)
\begin{equation}
\big\vert C(0,x)\big\vert =\big\vert\big(G_x,
\gamma_5\tilde\Gamma G_x\Gamma\gamma_5\big)\big\vert
\leq \Vert G_x\Vert\,\Vert\gamma_5\tilde\Gamma G_x\Gamma\gamma_5\Vert=
\Vert G_x\Vert\,\Vert\tilde\Gamma G_x\Gamma\Vert\,.
\end{equation}
The inequality turns into an equality if and only
if the two arguments of
the scalar product are linearly dependent,
\begin{equation}
\gamma_5\tilde\Gamma G_x\Gamma\gamma_5=\lambda G_x\,.\label{inequality}
\end{equation}
For the $\aA$ with $\Gamma=\tilde{\Gamma}=\id_4$ the condition
(\ref{inequality}) in not fulfilled and we obtain
\begin{equation}
	\big\vert C_{\aA}(0,x)\big\vert<\Vert G_x\Vert^2\,.
\end{equation}
For the $\aPi$ with $\Gamma=\tilde{\Gamma}=\gamma_5$ the condition
(\ref{inequality}) is fulfilled and we obtain
\begin{equation}
\big\vert C_{\aPi}(0,x)\big\vert=\Vert G_x\Vert^2\,.
\end{equation}
The two last relations imply the inequality
\begin{equation}
\big\vert C_{\aA}(0,x)\big\vert<\big\vert C_{\aPi}(0,x)\big\vert\,.
\label{inequality3}
\end{equation}
In conclusion, the $\aA$ (and all other mesonic states) are heavier than the $\aPi$.
Note that this proof is only correct without twist when
the Dirac operator is $\gamma_5$-hermitean.
We also used that the expectation values $\langle\dots\rangle_\mathcal{U}$ are calculated with a 
positive measure which we do not have in case there
is a sign problem.
Finally, the conclusion about the mass-hierarchy
only holds for infinite volume, when all connected 
correlators approach zero. In a finite volume the correlators 
are cosh-shaped and (\ref{inequality3}) would not necessarily
imply $m_\aPi<m_\aA$.

\begin{figure}[tbp]
	\centering
	\includegraphics{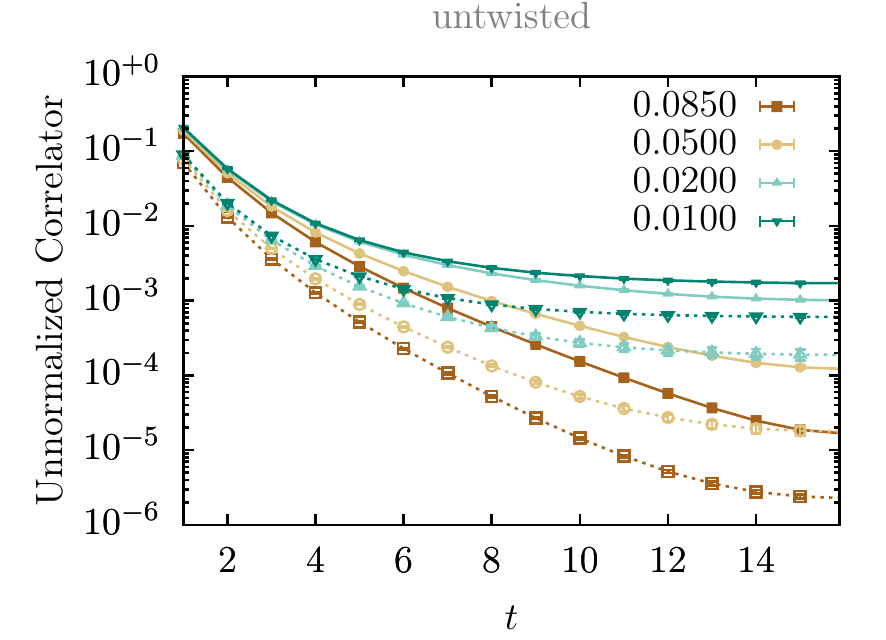}
	\caption{Correlators for our four mass parameters at \mbox{$\beta=\textrm{5.0}$} from untwisted simulations on the \mlat~lattice without normalization, see top four rows of table~\ref{tab:16_5.0}. Solid/dotted lines with filled/open markers connect the data points of $\aPi$ resp.\ $\aA$ to guide the eye. The labels indicate the distance \mbox{$|m-m_\text{crit}|$} to the critical point. Errors are mostly smaller than the marker size.}
	\label{fig:PiAcorrelators}
\end{figure}

To see whether the results in section~\ref{ch:ParameterScan} 
are in line with above inequality, we have a closer look
at the correlators of $\aPi$ and $\aA$.
This way we can check whether the unexpected mass-hierarchy originates
from problems with fitting the correlators correctly.
Figure~\ref{fig:PiAcorrelators} depicts the correlators 
of both connected mesonic states without normalization.
In full agreement with (\ref{inequality3}) we see that 
the correlator of the $\aPi$ is always above that of
$\aA$ such that the adjoint pion should be lighter.
In the range $t\in[2,12]$ the $\aA$ correlator falls off faster than the $\aPi$ correlator and thus $m_\aA>m_\aPi$, as expected.
With an appropriate fit range, the influence of excited 
states at small $t$ and the lattice artifacts around $t=T/2$ 
can be reduced.
See section~\ref{ch:ScaleSetting} for a further discussion of the mass extraction and section~\ref{ch:FiniteSizeAnalysis} for the finite size effects.
Similar observations hold for the other lattice 
gauge couplings $\beta\in\{4.5,5.4\}$.

\section{Overview of numerical data}\label{ch:NumOverview}
In table~\ref{tab:betaV}, we summarize the parameters of our simulations.
For the parameter scan on the \slat~lattice at $\beta=\textrm{5.4}$, all ensembles have around 200 configurations.
Table~\ref{tab:16} contains the values of the bare mass $m$, the twisted mass $m_5$ as well as the number of configurations for the various gauge couplings $\beta$ on the \mlat~lattice.

\begin{table}[tbp]
	\centering
	\renewcommand{\arraystretch}{1.2}
	\begin{tabular}{!{\vrule width 1pt}llllllll!{\vrule width 1pt}}
		\noalign{\hrule height 1pt} 
		ID & $\beta$ & $L^3\times T$ & $m_\text{crit}$ & $m$ & $m_5$ & $r$ & $r_5$ \\ 
		\hline 
		(I) & 4.5 & \mlat & -1.22428 & [-1.1443,\,-1.22428] & [0.0000,\,0.0800] & 1.0000 & 0.0000 \\ 
		(II) & 5.0 & \mlat & -1.0706 & [-0.9856,\,-1.0706] & [0.0000,\,0.0850] & 1.0000 & 0.0000 \\ 
		(III) & 5.0 & \mlat & -0.7570 & [-0.6156,\,-0.8277] & [-0.6156,\,-0.8277] & 0.7071 & 0.7071 \\ 
		(IV) & 5.4 & $8^3~\,\times16$ & -0.967 & [-1.4000,\,-0.6000] & [-0.4000,\,0.4000] & 1.0000 & 0.0000 \\ 
		(V) & 5.4 & \mlat & -0.9750 & [-0.8450,\,-0.9750] & [0.0000,\,0.1300] & 1.0000 & 0.0000 \\ 
		\noalign{\hrule height 1pt} 
	\end{tabular} 
	\caption{Overview of the parameter sets. Three different lattice couplings $\beta$, two different lattice volumes \mbox{$V=L^3\times T$} and two different combinations of \mbox{$(r,r_5)$} are used within this paper. For each setting, the mass parameter $m_\text{crit}$ of the critical point as well as the ranges of $m$ and $m_5$ are listed.}
	\label{tab:betaV}
\end{table}

\begin{table}[tbp]
	\centering
	\renewcommand{\arraystretch}{1.2}
	\begin{minipage}[c]{0.32\textwidth}
		\centering
		\begin{tabular}{!{\vrule width 1pt} lll !{\vrule width 1pt}}
			\noalign{\hrule height 1pt} 
			$m$ & $m_5$ & $\#$ \\ 
			\hline 
			-1.2143 & 0.0000 & 100 \\ 
			-1.2043 & 0.0000 & 100 \\ 
			-1.1743 & 0.0000 & 100 \\ 
			-1.1443 & 0.0000 & 100 \\ 
			& & \\
			-1.2172 & 0.0071 & 100 \\ 
			-1.2101 & 0.0141 & 100 \\ 
			-1.1889 & 0.0354 & 100 \\ 
			-1.1677 & 0.0566 & 100 \\ 
			& & \\
			-1.22428 & 0.0100 & 50 \\ 
			-1.22428 & 0.0200 & 50 \\ 
			-1.22428 & 0.0500 & 50 \\ 
			-1.22428 & 0.0800 & 50 \\ 
			\noalign{\hrule height 1pt}
		\end{tabular} 
		\subcaption{$\beta=4.5$}
		\label{tab:16_4.5}
	\end{minipage}
	\hfill
	\begin{minipage}[c]{0.32\textwidth}
		\centering
		\begin{tabular}{!{\vrule width 1pt} lll !{\vrule width 1pt}}
			\noalign{\hrule height 1pt} 
			$m$ & $m_5$ & $\#$ \\ 
			\hline 
			-1.0606 & 0.0000 & 200 \\ 
			-1.0506 & 0.0000 & 200 \\ 
			-1.0206 & 0.0000 & 200 \\ 
			-0.9856 & 0.0000 & 200 \\ 
			& & \\
			-1.0635 & 0.0071 & 2110 \\ 
			-1.0565 & 0.0141 & 2370 \\ 
			-1.0352 & 0.0354 & 2705 \\ 
			-1.0105 & 0.0601 & 3100 \\ 
			& & \\
			-1.0706 & 0.0100 & 50 \\ 
			-1.0706 & 0.0200 & 100 \\ 
			-1.0706 & 0.0500 & 50 \\ 
			-1.0706 & 0.0850 & 50 \\ 
			\noalign{\hrule height 1pt}
		\end{tabular} 
		\subcaption{$\beta=5.0$}
		\label{tab:16_5.0}
	\end{minipage}
	\hfill
	\begin{minipage}[c]{0.32\textwidth}
		\centering
		\begin{tabular}{!{\vrule width 1pt} lll !{\vrule width 1pt}}
			\noalign{\hrule height 1pt} 
			$m$ & $m_5$ & $\#$ \\ 
			\hline 
			-0.9650 & 0.0000 & 100 \\ 
			-0.9500 & 0.0000 & 100 \\ 
			-0.8950 & 0.0000 & 100 \\ 
			-0.8450 & 0.0000 & 100 \\ 
			& & \\
			-0.9679 & 0.0071 & 100 \\ 
			-0.9573 & 0.0177 & 100 \\ 
			-0.9184 & 0.0566 & 100 \\ 
			-0.8831 & 0.0919 & 100 \\
			& & \\
			-0.9750 & 0.0100 & 100 \\ 
			-0.9750 & 0.0250 & 100 \\ 
			-0.9750 & 0.0800 & 100 \\ 
			-0.9750 & 0.1300 & 100 \\
			\noalign{\hrule height 1pt}
		\end{tabular} 
		\subcaption{$\beta=5.4$}
		\label{tab:16_5.4}
	\end{minipage}
	\caption{Ensemble sizes on the \mlat~lattice.}
	\label{tab:16}
\end{table}

\newpage

%% file: main.bbl
\providecommand{\href}[2]{#2}\begingroup\raggedright\begin{thebibliography}{10}

\bibitem{PDG18}
{\scshape Particle Data Group} collaboration, \emph{{Review of Particle
  Physics}}, \href{https://doi.org/10.1103/PhysRevD.98.030001}{\emph{Phys.
  Rev.} {\bfseries D98} (2018) 030001}.

\bibitem{witten_dynamical_1981}
E.~Witten, \emph{Dynamical breaking of supersymmetry},
  \href{https://doi.org/10.1016/0550-3213(81)90006-7}{\emph{Nuclear Physics B}
  {\bfseries 188} (1981) 513}.

\bibitem{dimopoulos_softly_1981}
S.~Dimopoulos and H.~Georgi, \emph{Softly broken supersymmetry and {SU}(5)},
  \href{https://doi.org/10.1016/0550-3213(81)90522-8}{\emph{Nuclear Physics B}
  {\bfseries 193} (1981) 150}.

\bibitem{ellis_supersymmetric_1984}
J.~Ellis, J.~S. Hagelin, D.~V. Nanopoulos, K.~Olive and M.~Srednicki,
  \emph{Supersymmetric relics from the big bang},
  \href{https://doi.org/10.1016/0550-3213(84)90461-9}{\emph{Nuclear Physics B}
  {\bfseries 238} (1984) 453}.

\bibitem{Dondi:1976tx}
P.~Dondi and H.~Nicolai, \emph{{Lattice Supersymmetry}},
  \href{https://doi.org/10.1007/BF02730448}{\emph{Nuovo Cim. A} {\bfseries 41}
  (1977) 1}.

\bibitem{Catterall:2001fr}
S.~Catterall and S.~Karamov, \emph{{Exact lattice supersymmetry: The
  Two-dimensional N=2 Wess-Zumino model}},
  \href{https://doi.org/10.1103/PhysRevD.65.094501}{\emph{Phys. Rev. D}
  {\bfseries 65} (2002) 094501}
  [\href{https://arxiv.org/abs/hep-lat/0108024}{{\ttfamily hep-lat/0108024}}].

\bibitem{Bergner:2007pu}
G.~Bergner, T.~Kaestner, S.~Uhlmann and A.~Wipf, \emph{{Low-dimensional
  Supersymmetric Lattice Models}},
  \href{https://doi.org/10.1016/j.aop.2007.06.010}{\emph{Annals Phys.}
  {\bfseries 323} (2008) 946}
  [\href{https://arxiv.org/abs/0705.2212}{{\ttfamily 0705.2212}}].

\bibitem{Kastner:2008zc}
T.~Kastner, G.~Bergner, S.~Uhlmann, A.~Wipf and C.~Wozar,
  \emph{{Two-Dimensional Wess-Zumino Models at Intermediate Couplings}},
  \href{https://doi.org/10.1103/PhysRevD.78.095001}{\emph{Phys. Rev. D}
  {\bfseries 78} (2008) 095001}
  [\href{https://arxiv.org/abs/0807.1905}{{\ttfamily 0807.1905}}].

\bibitem{Kanamori:2007yx}
I.~Kanamori, F.~Sugino and H.~Suzuki, \emph{{Observing dynamical supersymmetry
  breaking with euclidean lattice simulations}},
  \href{https://doi.org/10.1143/PTP.119.797}{\emph{Prog. Theor. Phys.}
  {\bfseries 119} (2008) 797}
  [\href{https://arxiv.org/abs/0711.2132}{{\ttfamily 0711.2132}}].

\bibitem{Steinhauer:2014oda}
K.~Steinhauer and U.~Wenger, \emph{{Loop formulation of supersymmetric
  Yang-Mills quantum mechanics}},
  \href{https://doi.org/10.1007/JHEP12(2014)044}{\emph{JHEP} {\bfseries 12}
  (2014) 044} [\href{https://arxiv.org/abs/1410.0235}{{\ttfamily 1410.0235}}].

\bibitem{Flore:2012xj}
R.~Flore, D.~Korner, A.~Wipf and C.~Wozar, \emph{{Supersymmetric Nonlinear O(3)
  Sigma Model on the Lattice}},
  \href{https://doi.org/10.1007/JHEP11(2012)159}{\emph{JHEP} {\bfseries 11}
  (2012) 159} [\href{https://arxiv.org/abs/1207.6947}{{\ttfamily 1207.6947}}].

\bibitem{Koutsoumbas:1996kz}
G.~Koutsoumbas and I.~Montvay, \emph{{Gluinos on the lattice: Quenched
  calculations}},
  \href{https://doi.org/10.1016/S0370-2693(97)00178-0}{\emph{Phys. Lett. B}
  {\bfseries 398} (1997) 130}
  [\href{https://arxiv.org/abs/hep-lat/9612003}{{\ttfamily hep-lat/9612003}}].

\bibitem{donini_towards_1998}
A.~Donini, M.~Guagnelli, P.~Hernandez and A.~Vladikas, \emph{Towards n=1
  {superYang}-mills on the lattice},
  \href{https://doi.org/10.1016/S0550-3213(98)00166-7}{\emph{Nucl. Phys.}
  {\bfseries B523} (1998) 529}.

\bibitem{Kirchner:1998mp}
{\scshape DESY-Munster} collaboration, \emph{{Evidence for discrete chiral
  symmetry breaking in N=1 supersymmetric Yang-Mills theory}},
  \href{https://doi.org/10.1016/S0370-2693(98)01523-8}{\emph{Phys. Lett. B}
  {\bfseries 446} (1999) 209}
  [\href{https://arxiv.org/abs/hep-lat/9810062}{{\ttfamily hep-lat/9810062}}].

\bibitem{Campos:1999du}
{\scshape DESY-Munster} collaboration, \emph{{Monte Carlo simulation of SU(2)
  Yang-Mills theory with light gluinos}},
  \href{https://doi.org/10.1007/s100520050651}{\emph{Eur. Phys. J. C}
  {\bfseries 11} (1999) 507}
  [\href{https://arxiv.org/abs/hep-lat/9903014}{{\ttfamily hep-lat/9903014}}].

\bibitem{Bergner:2015adz}
G.~Bergner, P.~Giudice, G.~Münster, I.~Montvay and S.~Piemonte, \emph{{The
  light bound states of supersymmetric SU(2) Yang-Mills theory}},
  \href{https://doi.org/10.1007/JHEP03(2016)080}{\emph{JHEP} {\bfseries 03}
  (2016) 080} [\href{https://arxiv.org/abs/1512.07014}{{\ttfamily
  1512.07014}}].

\bibitem{Ali:2019gzj}
S.~Ali, G.~Bergner, H.~Gerber, S.~Kuberski, I.~Montvay, G.~Münster et~al.,
  \emph{{Variational analysis of low-lying states in supersymmetric Yang-Mills
  theory}}, \href{https://doi.org/10.1007/JHEP04(2019)150}{\emph{JHEP}
  {\bfseries 04} (2019) 150}
  [\href{https://arxiv.org/abs/1901.02416}{{\ttfamily 1901.02416}}].

\bibitem{Farchioni:2001wx}
{\scshape DESY-Munster-Roma} collaboration, \emph{{The Supersymmetric Ward
  identities on the lattice}},
  \href{https://doi.org/10.1007/s100520200898}{\emph{Eur. Phys. J. C}
  {\bfseries 23} (2002) 719}
  [\href{https://arxiv.org/abs/hep-lat/0111008}{{\ttfamily hep-lat/0111008}}].

\bibitem{Bergner:2014saa}
G.~Bergner, P.~Giudice, G.~Münster, S.~Piemonte and D.~Sandbrink, \emph{{Phase
  structure of the $ \mathcal{N}=1 $ supersymmetric Yang-Mills theory at finite
  temperature}}, \href{https://doi.org/10.1007/JHEP11(2014)049}{\emph{JHEP}
  {\bfseries 11} (2014) 049} [\href{https://arxiv.org/abs/1405.3180}{{\ttfamily
  1405.3180}}].

\bibitem{Bergner:2019dim}
G.~Bergner, C.~L\'opez and S.~Piemonte, \emph{{Study of center and chiral
  symmetry realization in thermal $\mathcal{N}=1$ super Yang-Mills theory using
  the gradient flow}},
  \href{https://doi.org/10.1103/PhysRevD.100.074501}{\emph{Phys. Rev.}
  {\bfseries D100} (2019) 074501}
  [\href{https://arxiv.org/abs/1902.08469}{{\ttfamily 1902.08469}}].

\bibitem{Farchioni:2001yr}
F.~Farchioni, A.~Feo, T.~Galla, C.~Gebert, R.~Kirchner, I.~Montvay et~al.,
  \emph{{SUSY Ward identities in 1 loop perturbation theory}},
  \href{https://doi.org/10.1016/S0920-5632(01)01892-8}{\emph{Nucl. Phys. B
  Proc. Suppl.} {\bfseries 106} (2002) 941}
  [\href{https://arxiv.org/abs/hep-lat/0110113}{{\ttfamily hep-lat/0110113}}].

\bibitem{Munster:2014cja}
G.~Münster and H.~Stüwe, \emph{{The mass of the adjoint pion in $\mathcal{N}
  =$ 1 supersymmetric Yang-Mills theory}},
  \href{https://doi.org/10.1007/JHEP05(2014)034}{\emph{JHEP} {\bfseries 05}
  (2014) 034} [\href{https://arxiv.org/abs/1402.6616}{{\ttfamily 1402.6616}}].

\bibitem{Musberg:2013foa}
S.~Musberg, G.~Münster and S.~Piemonte, \emph{{Perturbative calculation of the
  clover term for Wilson fermions in any representation of the gauge group
  SU(N)}}, \href{https://doi.org/10.1007/JHEP05(2013)143}{\emph{JHEP}
  {\bfseries 05} (2013) 143} [\href{https://arxiv.org/abs/1304.5741}{{\ttfamily
  1304.5741}}].

\bibitem{Ali:2019agk}
S.~Ali, G.~Bergner, H.~Gerber, I.~Montvay, G.~Münster, S.~Piemonte et~al.,
  \emph{{Numerical results for the lightest bound states in $\mathcal{N}=1$
  supersymmetric SU(3) Yang-Mills theory}},
  \href{https://doi.org/10.1103/PhysRevLett.122.221601}{\emph{Phys. Rev. Lett.}
  {\bfseries 122} (2019) 221601}
  [\href{https://arxiv.org/abs/1902.11127}{{\ttfamily 1902.11127}}].

\bibitem{Ali:2018fbq}
S.~Ali, H.~Gerber, I.~Montvay, G.~Münster, S.~Piemonte, P.~Scior et~al.,
  \emph{{Analysis of Ward identities in supersymmetric Yang--Mills theory}},
  \href{https://doi.org/10.1140/epjc/s10052-018-5887-9}{\emph{Eur. Phys. J. C}
  {\bfseries 78} (2018) 404}
  [\href{https://arxiv.org/abs/1802.07067}{{\ttfamily 1802.07067}}].

\bibitem{Neuberger:1997bg}
H.~Neuberger, \emph{{Vector - like gauge theories with almost massless fermions
  on the lattice}}, \href{https://doi.org/10.1103/PhysRevD.57.5417}{\emph{Phys.
  Rev. D} {\bfseries 57} (1998) 5417}
  [\href{https://arxiv.org/abs/hep-lat/9710089}{{\ttfamily hep-lat/9710089}}].

\bibitem{Kaplan:1999jn}
D.~B. Kaplan and M.~Schmaltz, \emph{{Supersymmetric Yang-Mills theories from
  domain wall fermions}}, {\emph{Chin. J. Phys.} {\bfseries 38} (2000) 543}
  [\href{https://arxiv.org/abs/hep-lat/0002030}{{\ttfamily hep-lat/0002030}}].

\bibitem{Giedt:2008xm}
J.~Giedt, R.~Brower, S.~Catterall, G.~T. Fleming and P.~Vranas, \emph{{Lattice
  super-Yang-Mills using domain wall fermions in the chiral limit}},
  \href{https://doi.org/10.1103/PhysRevD.79.025015}{\emph{Phys. Rev. D}
  {\bfseries 79} (2009) 025015}
  [\href{https://arxiv.org/abs/0810.5746}{{\ttfamily 0810.5746}}].

\bibitem{Kim:2011fw}
{\scshape JLQCD} collaboration, \emph{{Lattice study of 4d {\cal N}=1 super
  Yang-Mills theory with dynamical overlap gluino}},
  \href{https://doi.org/10.22323/1.139.0069}{\emph{PoS} {\bfseries LATTICE2011}
  (2011) 069} [\href{https://arxiv.org/abs/1111.2180}{{\ttfamily 1111.2180}}].

\bibitem{Ali:2020sbi}
S.~Ali, G.~Bergner, H.~Gerber, C.~López, I.~Montvay, G.~Münster et~al.,
  \emph{{Continuum limit of SU(3) $\mathcal{N}=1$ supersymmetric Yang-Mills
  theory and supersymmetric gauge theories on the lattice}},  in \emph{{37th
  International Symposium on Lattice Field Theory}}, 1, 2020,
  \href{https://arxiv.org/abs/2001.09682}{{\ttfamily 2001.09682}}.

\bibitem{Fleming:2000fa}
G.~T. Fleming, J.~B. Kogut and P.~M. Vranas, \emph{{SuperYang-Mills on the
  lattice with domain wall fermions}},
  \href{https://doi.org/10.1103/PhysRevD.64.034510}{\emph{Phys. Rev. D}
  {\bfseries 64} (2001) 034510}
  [\href{https://arxiv.org/abs/hep-lat/0008009}{{\ttfamily hep-lat/0008009}}].

\bibitem{August:2018esp}
D.~August, M.~Steinhauser, B.~Wellegehausen and A.~Wipf, \emph{{Mass spectrum
  of $2$-dimensional $\mathcal{N}=(2,2)$ super Yang-Mills theory on the
  lattice}}, \href{https://doi.org/10.1007/JHEP01(2019)099}{\emph{JHEP}
  {\bfseries 01} (2019) 099}
  [\href{https://arxiv.org/abs/1802.07797}{{\ttfamily 1802.07797}}].

\bibitem{Kadoh:2009rw}
D.~Kadoh and H.~Suzuki, \emph{{SUSY WT identity in a lattice formulation of 2D
  = (2,2) SYM}},
  \href{https://doi.org/10.1016/j.physletb.2009.11.028}{\emph{Phys. Lett. B}
  {\bfseries 682} (2010) 466}
  [\href{https://arxiv.org/abs/0908.2274}{{\ttfamily 0908.2274}}].

\bibitem{Catterall:2017xox}
S.~Catterall, R.~G. Jha and A.~Joseph, \emph{{Nonperturbative study of
  dynamical SUSY breaking in N=(2,2) Yang-Mills theory}},
  \href{https://doi.org/10.1103/PhysRevD.97.054504}{\emph{Phys. Rev. D}
  {\bfseries 97} (2018) 054504}
  [\href{https://arxiv.org/abs/1801.00012}{{\ttfamily 1801.00012}}].

\bibitem{Hanada:2009hq}
M.~Hanada and I.~Kanamori, \emph{{Lattice study of two-dimensional N=(2,2)
  super Yang-Mills at large-N}},
  \href{https://doi.org/10.1103/PhysRevD.80.065014}{\emph{Phys. Rev. D}
  {\bfseries 80} (2009) 065014}
  [\href{https://arxiv.org/abs/0907.4966}{{\ttfamily 0907.4966}}].

\bibitem{Catterall:2009it}
S.~Catterall, D.~B. Kaplan and M.~Unsal, \emph{{Exact lattice supersymmetry}},
  \href{https://doi.org/10.1016/j.physrep.2009.09.001}{\emph{Phys. Rept.}
  {\bfseries 484} (2009) 71} [\href{https://arxiv.org/abs/0903.4881}{{\ttfamily
  0903.4881}}].

\bibitem{Schaich:2016jus}
D.~Schaich, S.~Catterall, P.~H. Damgaard and J.~Giedt, \emph{{Latest results
  from lattice N=4 supersymmetric Yang--Mills}},
  \href{https://doi.org/10.22323/1.256.0221}{\emph{PoS} {\bfseries LATTICE2016}
  (2016) 221} [\href{https://arxiv.org/abs/1611.06561}{{\ttfamily
  1611.06561}}].

\bibitem{Giguere:2015cga}
E.~Giguère and D.~Kadoh, \emph{{Restoration of supersymmetry in
  two-dimensional SYM with sixteen supercharges on the lattice}},
  \href{https://doi.org/10.1007/JHEP05(2015)082}{\emph{JHEP} {\bfseries 05}
  (2015) 082} [\href{https://arxiv.org/abs/1503.04416}{{\ttfamily
  1503.04416}}].

\bibitem{Frezzotti:2000nk}
{\scshape Alpha} collaboration, \emph{{Lattice QCD with a chirally twisted mass
  term}}, {\emph{JHEP} {\bfseries 08} (2001) 058}
  [\href{https://arxiv.org/abs/hep-lat/0101001}{{\ttfamily hep-lat/0101001}}].

\bibitem{Frezzotti:2003ni}
R.~Frezzotti and G.~Rossi, \emph{{Chirally improving Wilson fermions. 1. O(a)
  improvement}},
  \href{https://doi.org/10.1088/1126-6708/2004/08/007}{\emph{JHEP} {\bfseries
  08} (2004) 007} [\href{https://arxiv.org/abs/hep-lat/0306014}{{\ttfamily
  hep-lat/0306014}}].

\bibitem{Veneziano8206}
G.~Veneziano and S.~Yankielowicz, \emph{{An effective Lagrangian for the pure N
  = 1 supersymmetric Yang-Mills theory}},
  \href{https://doi.org/http://dx.doi.org/10.1016/0370-2693(82)90828-0}{\emph{Physics
  Letters B} {\bfseries 113} (1982) 231 }.

\bibitem{Farrar9711}
G.~R. Farrar, G.~Gabadadze and M.~Schwetz, \emph{{On the effective action of
  N=1 supersymmetric Yang-Mills theory}},
  \href{https://doi.org/10.1103/PhysRevD.58.015009}{\emph{Phys. Rev.}
  {\bfseries D58} (1998) 015009}
  [\href{https://arxiv.org/abs/hep-th/9711166}{{\ttfamily hep-th/9711166}}].

\bibitem{Farrar:1998rm}
G.~R. Farrar, G.~Gabadadze and M.~Schwetz, \emph{{The spectrum of softly broken
  N=1 supersymmetric Yang-Mills theory}},
  \href{https://doi.org/10.1103/PhysRevD.60.035002}{\emph{Phys. Rev. D}
  {\bfseries 60} (1999) 035002}
  [\href{https://arxiv.org/abs/hep-th/9806204}{{\ttfamily hep-th/9806204}}].

\bibitem{jaffe_euclidean_1985}
A.~Jaffe, \emph{{Euclidean quantum field theory}},
  \href{https://doi.org/10.1016/0550-3213(85)90208-1}{\emph{Nuclear Physics B}
  {\bfseries 254} (1985) 31}.

\bibitem{Curci8612}
G.~Curci and G.~Veneziano, \emph{{Supersymmetry and the Lattice: A
  Reconciliation?}},
  \href{https://doi.org/10.1016/0550-3213(87)90660-2}{\emph{Nucl. Phys.}
  {\bfseries B292} (1987) 555}.

\bibitem{Kennedy9809}
A.~D. Kennedy, I.~Horvath and S.~Sint, \emph{{A new exact method for dynamical
  fermion computations with nonlocal actions}},
  \href{https://doi.org/10.1016/S0920-5632(99)85217-7}{\emph{Nucl. Phys. Proc.
  Suppl.} {\bfseries 73} (1999) 834}
  [\href{https://arxiv.org/abs/hep-lat/9809092}{{\ttfamily hep-lat/9809092}}].

\bibitem{Nishimura:1997vg}
J.~Nishimura, \emph{{Four-dimensional N=1 supersymmetric Yang-Mills theory on
  the lattice without fine tuning}},
  \href{https://doi.org/10.1016/S0370-2693(97)00674-6}{\emph{Phys. Lett. B}
  {\bfseries 406} (1997) 215}
  [\href{https://arxiv.org/abs/hep-lat/9701013}{{\ttfamily hep-lat/9701013}}].

\bibitem{Knechtli:2017sna}
F.~Knechtli, M.~Günther and M.~Peardon, \emph{{Lattice Quantum Chromodynamics:
  Practical Essentials}}, SpringerBriefs in Physics. Springer, 2017,
  \href{https://doi.org/10.1007/978-94-024-0999-4}{10.1007/978-94-024-0999-4}.

\bibitem{Demmouche:2010sf}
K.~Demmouche, F.~Farchioni, A.~Ferling, I.~Montvay, G.~Munster, E.~Scholz
  et~al., \emph{{Simulation of 4d N=1 supersymmetric Yang-Mills theory with
  Symanzik improved gauge action and stout smearing}},
  \href{https://doi.org/10.1140/epjc/s10052-010-1390-7}{\emph{Eur. Phys. J. C}
  {\bfseries 69} (2010) 147} [\href{https://arxiv.org/abs/1003.2073}{{\ttfamily
  1003.2073}}].

\bibitem{KuberskiMaster}
S.~Kuberski, ``{Bestimmung von Massen in der supersymmetrischen
  Yang-Mills-Theorie mit der Variationsmethode}.''

\bibitem{HeitgerDiss}
F.~Heitger, ``{Darstellungstheorie der kubischen Gruppe in Anwendung auf
  Operatoren der N=1 SUSY-Yang-Mills-Theorie auf dem Gitter}.''

\bibitem{Berg:1982kp}
B.~Berg and A.~Billoire, \emph{{Glueball Spectroscopy in Four-Dimensional SU(3)
  Lattice Gauge Theory. 1.}},
  \href{https://doi.org/10.1016/0550-3213(83)90620-X}{\emph{Nucl. Phys. B}
  {\bfseries 221} (1983) 109}.

\bibitem{Fierz1937}
M.~Fierz, \emph{{Zur Fermischen Theorie des $\beta$-Zerfalls}},
  \href{https://doi.org/10.1007/BF01330070}{\emph{Zeitschrift für Physik}
  {\bfseries 104} (1937) 553}.

\bibitem{Pal:2007dc}
P.~B. Pal, \emph{{Representation-independent manipulations with Dirac
  spinors}},  \href{https://arxiv.org/abs/physics/0703214}{{\ttfamily
  physics/0703214}}.

\bibitem{LuckmannDiploma}
S.~Luckmann, ``{Ward-Identitäten in der N = 1 Super-Yang-Mills-Theorie}.''

\bibitem{KirchnerDiss}
R.~Kirchner, ``{Ward Identities and Mass Spectrum of N=1 Super Yang-Mills
  Theory on the Lattice}.''

\bibitem{Montvay:2001aj}
I.~Montvay, \emph{{Supersymmetric Yang-Mills theory on the lattice}},
  \href{https://doi.org/10.1142/S0217751X0201090X}{\emph{Int. J. Mod. Phys. A}
  {\bfseries 17} (2002) 2377}
  [\href{https://arxiv.org/abs/hep-lat/0112007}{{\ttfamily hep-lat/0112007}}].

\bibitem{vanNieuwenhuizen:1996tv}
P.~van Nieuwenhuizen and A.~Waldron, \emph{{On Euclidean spinors and Wick
  rotations}}, \href{https://doi.org/10.1016/S0370-2693(96)01251-8}{\emph{Phys.
  Lett. B} {\bfseries 389} (1996) 29}
  [\href{https://arxiv.org/abs/hep-th/9608174}{{\ttfamily hep-th/9608174}}].

\bibitem{taniguchi_one_2000}
Y.~Taniguchi, \emph{One loop calculation of {SUSY} ward-takahashi identity on
  lattice with wilson fermion},
  \href{https://doi.org/10.1103/PhysRevD.63.014502}{\emph{Physical Review D}
  {\bfseries 63} (2000) 014502}
  [\href{https://arxiv.org/abs/hep-lat/9906026}{{\ttfamily hep-lat/9906026}}].

\bibitem{KaestnerDiss}
T.~Kästner, \emph{{Supersymmetry on a space-time lattice}}. Dissertation,
  Friedrich Schiller University Jena, 2008.

\bibitem{Wipf:2013vp}
A.~Wipf, \emph{{Statistical approach to quantum field theory}: {An
  introduction}}, vol.~864. Springer, Berlin, Heidelberg, 2013,
  \href{https://doi.org/10.1007/978-3-642-33105-3}{10.1007/978-3-642-33105-3}.

\bibitem{Sommer:1993ce}
R.~Sommer, \emph{{A New way to set the energy scale in lattice gauge theories
  and its applications to the static force and alpha-s in SU(2) Yang-Mills
  theory}}, \href{https://doi.org/10.1016/0550-3213(94)90473-1}{\emph{Nucl.
  Phys. B} {\bfseries 411} (1994) 839}
  [\href{https://arxiv.org/abs/hep-lat/9310022}{{\ttfamily hep-lat/9310022}}].

\bibitem{Morningstar:2003gk}
C.~Morningstar and M.~J. Peardon, \emph{{Analytic smearing of SU(3) link
  variables in lattice QCD}},
  \href{https://doi.org/10.1103/PhysRevD.69.054501}{\emph{Phys. Rev. D}
  {\bfseries 69} (2004) 054501}
  [\href{https://arxiv.org/abs/hep-lat/0311018}{{\ttfamily hep-lat/0311018}}].

\bibitem{Ali:2018dnd}
S.~Ali, G.~Bergner, H.~Gerber, P.~Giudice, I.~Montvay, G.~Münster et~al.,
  \emph{{The light bound states of $\mathcal{N}=1$ supersymmetric SU(3)
  Yang-Mills theory on the lattice}},
  \href{https://doi.org/10.1007/JHEP03(2018)113}{\emph{JHEP} {\bfseries 03}
  (2018) 113} [\href{https://arxiv.org/abs/1801.08062}{{\ttfamily
  1801.08062}}].

\bibitem{Evans:1997jy}
N.~J. Evans, S.~D. Hsu and M.~Schwetz, \emph{{Lattice tests of supersymmetric
  Yang-Mills theory?}},  \href{https://arxiv.org/abs/hep-th/9707260}{{\ttfamily
  hep-th/9707260}}.

\bibitem{Bali:2016lvx}
{\scshape RQCD} collaboration, \emph{{Direct determinations of the nucleon and
  pion $\sigma$ terms at nearly physical quark masses}},
  \href{https://doi.org/10.1103/PhysRevD.93.094504}{\emph{Phys. Rev. D}
  {\bfseries 93} (2016) 094504}
  [\href{https://arxiv.org/abs/1603.00827}{{\ttfamily 1603.00827}}].

\bibitem{Aoki:2005mb}
S.~Aoki, O.~Bar, S.~Takeda and T.~Ishikawa, \emph{{Pseudo scalar meson masses
  in Wilson chiral perturbation theory for 2+1 flavors}},
  \href{https://doi.org/10.1103/PhysRevD.73.014511}{\emph{Phys. Rev. D}
  {\bfseries 73} (2006) 014511}
  [\href{https://arxiv.org/abs/hep-lat/0509049}{{\ttfamily hep-lat/0509049}}].

\bibitem{Farchioni:2007dw}
F.~Farchioni, I.~Montvay, G.~Munster, E.~Scholz, T.~Sudmann and J.~Wuilloud,
  \emph{{Hadron masses in QCD with one quark flavour}},
  \href{https://doi.org/10.1140/epjc/s10052-007-0394-4}{\emph{Eur. Phys. J. C}
  {\bfseries 52} (2007) 305} [\href{https://arxiv.org/abs/0706.1131}{{\ttfamily
  0706.1131}}].

\bibitem{Wimmer_2012}
M.~Wimmer, \emph{Efficient numerical computation of the pfaffian for dense and
  banded skew-symmetric matrices},
  \href{https://doi.org/10.1145/2331130.2331138}{\emph{ACM Transactions on
  Mathematical Software} {\bfseries 38} (2012) 1–17}.

\bibitem{Alexandrou:2016izb}
C.~Alexandrou, S.~Bacchio, J.~Finkenrath, A.~Frommer, K.~Kahl and M.~Rottmann,
  \emph{{Adaptive Aggregation-based Domain Decomposition Multigrid for Twisted
  Mass Fermions}},
  \href{https://doi.org/10.1103/PhysRevD.94.114509}{\emph{Phys. Rev.}
  {\bfseries D94} (2016) 114509}
  [\href{https://arxiv.org/abs/1610.02370}{{\ttfamily 1610.02370}}].

\bibitem{DDalphaAMG}
{Simone Bacchio}, \emph{{DDalphaAMG library including twisted mass fermions}},
  \href{https://arxiv.org/abs/{https://github.com/sbacchio/DDalphaAMG}}{{\ttfamily
  {https://github.com/sbacchio/DDalphaAMG}}}.

\bibitem{Costa:2017rht}
M.~Costa and H.~Panagopoulos, \emph{{Supersymmetric QCD on the Lattice: An
  Exploratory Study}},
  \href{https://doi.org/10.1103/PhysRevD.96.034507}{\emph{Phys. Rev. D}
  {\bfseries 96} (2017) 034507}
  [\href{https://arxiv.org/abs/1706.05222}{{\ttfamily 1706.05222}}].

\bibitem{Wellegehausen:2018opt}
B.~Wellegehausen and A.~Wipf, \emph{{$\mathcal{N}=1$ Supersymmetric $SU(3)$
  Gauge Theory - Towards simulations of Super-QCD}},
  \href{https://doi.org/10.22323/1.334.0210}{\emph{PoS} {\bfseries LATTICE2018}
  (2018) 210} [\href{https://arxiv.org/abs/1811.01784}{{\ttfamily
  1811.01784}}].

\bibitem{Bergner:2018znw}
G.~Bergner and S.~Piemonte, \emph{{Supersymmetric and conformal theories on the
  lattice: from super Yang-Mills towards super QCD}},
  \href{https://doi.org/10.22323/1.334.0209}{\emph{PoS} {\bfseries LATTICE2018}
  (2019) 209} [\href{https://arxiv.org/abs/1811.01797}{{\ttfamily
  1811.01797}}].

\bibitem{Weingarten:1983uj}
D.~Weingarten, \emph{{Mass Inequalities for QCD}},
  \href{https://doi.org/10.1103/PhysRevLett.51.1830}{\emph{Phys. Rev. Lett.}
  {\bfseries 51} (1983) 1830}.

\bibitem{Kilcup:1995ww}
G.~Kilcup and S.~R. Sharpe, eds., \emph{{Phenomenology and lattice QCD.
  Proceedings: Uehling Summer School, Seattle, USA, Jun 21-Jul 2, 1993}}, 1995.

\bibitem{Shuryak:2004pry}
E.~V. Shuryak, \emph{{The QCD vacuum, hadrons and the superdense matter}},
  vol.~71. World Scientific, 10, 2004,
  \href{https://doi.org/10.1142/5367}{10.1142/5367}.

\end{thebibliography}\endgroup
